\documentclass[AMA,STIX1COL]{WileyNJD-v2}
\usepackage{moreverb}
\usepackage{threeparttable, multicol, multirow, bbm}
\usepackage[algo2e]{algorithm2e}
\usepackage{booktabs}
\usepackage{comment}

\newcommand\BibTeX{{\rmfamily B\kern-.05em \textsc{i\kern-.025em b}\kern-.08em
T\kern-.1667em\lower.7ex\hbox{E}\kern-.125emX}}

\articletype{RESEARCH ARTICLE}%

\received{<day> <Month>, <year>}
\revised{<day> <Month>, <year>}
\accepted{<day> <Month>, <year>}

\begin{document}

\title{Constrained randomization and statistical inference for multi-arm parallel cluster randomized controlled trials}

\author[1,2]{Yunji Zhou}

\author[1,2]{Elizabeth L. Turner}

\author[1,2]{Ryan A. Simmons}

\author[3,4]{Fan Li$^{*,}$}

\authormark{ZHOU \textsc{et al.}}

\address[1]{\orgdiv{Department of Biostatistics and Bioinformatics}, \orgname{Duke University}, \orgaddress{\state{Durham, North Carolina}, \country{USA}}}

\address[2]{\orgdiv{Duke Global Health Institute}, \orgname{Duke University}, \orgaddress{\state{Durham, North Carolina}, \country{USA}}}

\address[3]{\orgdiv{Department of Biostatistics}, \orgname{Yale School of Public Health}, \orgaddress{\state{New Haven, Connecticut}, \country{USA}}}

\address[4]{\orgdiv{Center for Methods in Implementation and Prevention Science}, \orgname{Yale School of Public Health}, \orgaddress{\state{New Haven, Connecticut}, \country{USA}}}

\corres{$^*$Fan Li, Department of Biostatistics, Yale School of Public Health, New Haven, Connecticut, USA.\\
\email{fan.f.li@yale.edu}}

\fundingInfo{National Institute of Allergy and Infectious Diseases, Grant/Award Number: R01 AI141444; National Center for Advancing Translational Sciences, Grant/Award Numbers: UL1 TR002553 \& UL1 TR001863}

\abstract[Abstract]{A practical limitation of cluster randomized controlled trials (cRCTs) is that the number of available clusters may be small, resulting in an increased risk of baseline imbalance under simple randomization. Constrained randomization overcomes this issue by restricting the allocation to a subset of randomization schemes where sufficient overall covariate balance across comparison arms is achieved. \textcolor{black}{However, for multi-arm cRCTs, several design and analysis issues pertaining to constrained randomization have not been fully investigated.} Motivated by an ongoing multi-arm cRCT, we \textcolor{black}{elaborate the method of constrained randomization and} provide a comprehensive evaluation of the statistical properties of model-based and randomization-based tests under both simple and constrained randomization designs in multi-arm cRCTs, with varying combinations of design and analysis-based covariate adjustment strategies. In particular, as randomization-based tests have not been extensively studied in multi-arm cRCTs, we additionally develop most-powerful randomization tests under the linear mixed model framework for our comparisons. Our results indicate that under constrained randomization, both model-based and randomization-based analyses could gain power while preserving nominal type I error rate, given proper analysis-based adjustment for the baseline covariates. \textcolor{black}{Randomization-based analyses, however, are more robust against violations of distributional assumptions.} The choice of balance metrics and candidate set sizes and their implications on the testing of the pairwise and global hypotheses are also discussed. Finally, we caution against the design and analysis of multi-arm cRCTs with an extremely small number of clusters, due to insufficient degrees of freedom and the tendency to obtain an overly restricted randomization space.}


\jnlcitation{\cname{%
\author{Zhou Y}, 
\author{Turner EL}, 
\author{Simmons RA}, and 
\author{Li F}} (\cyear{2022}), 
\ctitle{Constrained randomization and statistical inference for multi-arm parallel cluster randomized controlled trials}, \cjournal{Statistics in Medicine}, \cvol{0}.}

\maketitle

\section{Introduction}\label{sec1}

Cluster randomized controlled trials (cRCTs) are designed to evaluate the effect of an intervention that is delivered to clusters of individuals.\cite{murray1998design,donner2000design,eldridge2012practical,campbell2014design} Examples of such clusters include schools, communities, factories, hospitals, and medical practices. This design is often used in practice when the intervention by its nature needs to be applied to an entire group of individuals or when treatment contamination might arise from the interaction between individuals in the same cluster. In addition, cRCTs can be used to capture the population-level direct and indirect effects of an intervention (for example, an intervention designed to reduce infectious disease transmission).\cite{hayes2017cluster} There are various types of cRCT designs, including parallel-arm and stepped wedge designs.\citep{turner2017review1} In this article, we focus on multi-arm parallel cRCTs, motivated by the design of TESTsmART, an ongoing trial evaluating interventions to improve antimalarial stewardship in the retail sector.\cite{woolsey2021incentivizing} In a multi-arm cRCT, each distinct arm can be defined by different combinations of interventions, or different “doses” of a single type of intervention,\cite{watson2020design} or it could be of interest to simultaneously examine the effects of multiple types of interventions.

A frequent practical limitation of cRCTs is the difficulty in recruiting a large number of randomization units (i.e., clusters) as compared to individually randomized trials.\cite{ivers2012allocation} With only a limited number of often heterogeneous clusters, simple randomization may fail to adequately balance important baseline prognostic covariates across arms.\cite{taljaard2016substantial} The lack of balance with respect to baseline covariates can lead to decreased statistical power and precision and may threaten the internal validity of the trial.\cite{raab2001balance, turner2017review1} In multi-arm cRCTs, the risk of chance imbalance in baseline covariates increases as the number of arms increases and the number of randomization units per arm decreases. Therefore, statistical methods controlling for baseline prognostic characteristics are particularly important for the appropriate design and analysis of such trials.

In parallel cRCTs, it has been well-known that design-based adjustment for baseline balance 
enhances comparability between clusters in different arms.\cite{ivers2012allocation} To minimize the risk of chance imbalance in small cRCTs with two arms, \textcolor{black}{a variety of restricted randomization methods, such as stratification and matching, have been proposed. However, these methods have inherent limitations. When there are multiple important baseline covariates available at the design phase, stratification may create too many sparse strata with incomplete fillings (overstratification) and the optimal number of strata is context dependent,\cite{kernan1999stratified} while matching introduces additional complexity to the estimation of the intracluster correlation coefficient (ICC), which hampers the reporting of ICC and planning of future trials, and limits the choice of statistical inference methods.\cite{donner2004pitfalls}} Covariate-based constrained randomization\cite{moulton2004covariate} has been developed as a promising design strategy that overcomes potential limitations of stratification and pair-matching.\cite{moulton2004covariate} 
In brief, constrained randomization involves (i) specifying important baseline cluster-level 
covariates to be balanced; (ii) enumerating all randomization schemes or simulating a large number of possible randomization schemes (duplicates should be removed if the schemes are randomly simulated); (iii) retaining a constrained randomization space with a subset of schemes where sufficient balance across baseline covariates is achieved according to some pre-specified balance metric; (iv) randomly selecting a scheme from the constrained space for implementation.\cite{gallis2018cvcrand,yu2019cvcrand} 
Despite this generic framework, implementation of constrained randomization strategies can differ with respect to the choice of balance metric as well as the size of the constrained space. Li et al.\cite{li2016evaluation,li2017evaluation} examined the impact of such design parameters under constrained randomization in 
two-arm parallel cRCTs. While the choice between the $l1$ and $l2$ balance metrics does not lead to substantial differences for statistical inference, they found that a smaller randomization space (such as 10\% of the simple randomization space with the smallest balance score) could improve the power of the model-based and randomization-based tests. Ciolino et al.\cite{ciolino2019choosing} and Watson et al.\cite{watson2020design} extended the balance metrics to multi-arm cRCTs. However, the consideration of alternative balance metrics as well as \textcolor{black}{the effect of constrained subspace sizes on the validity of analytical methods were} not fully articulated for multi-arm cRCTs in these prior studies (see Web Table 1 for comparison). 

Parallel to design-based adjustment, analysis-based adjustment for prognostic covariates often leads to improved precision of the estimated treatment effect; thus increasing the statistical power of the trial.\cite{pocock2002subgroup} Both the choice of covariates to be adjusted for and the choice of a proper primary analytical method are of crucial importance. Li et al.\cite{li2016evaluation,li2017evaluation} recommended that covariates used in design-based adjustment 
should also be accounted for in the subsequent analysis of a two-arm cRCT. 
The same recommendation has been enforced by Watson et al.\cite{watson2020design} in three-arm cRCTs for model-based tests to maintain a valid type I error rate as well as adequate power. While 
mixed-model regression has been routinely used 
to test for an intervention effect in cRCTs,\cite{turner2017review2}
the \textcolor{black}{randomization} test provides a flexible alternative that may be particularly attractive under the constrained randomization design. 
This is because the \textcolor{black}{randomization} test leads to exact inference with the nominal type I error rate and dispenses with any asymptotic approximation as in model-based inference. 
\cite{murray2006comparison} This unique feature helps alleviate concerns on potential small-sample biases from model-based inference. 
While model-based inference accompanied by small-sample degrees-of-freedom corrections have been studied in multi-arm cRCT designs by Watson et al.,\cite{watson2020design} the \textcolor{black}{randomization} test has not been extended to allow for valid inference under constrained randomization with multi-arm cRCTs. 
Furthermore, the relative performance of the model-based test and \textcolor{black}{randomization} test remains unexplored, and recommendations are needed to guide practice (see Web Table 1 for comparison). 

To address these knowledge gaps, we evaluate several statistical issues concerning the use of constrained randomization and the downstream statistical inference in the context of multi-arm cRCTs. Specifically, our contributions are three-fold: (i) we provide alternative balance metrics that could be used for constrained randomization with multi-arm cRCTs; (ii) we propose new \textcolor{black}{randomization} test statistics for efficient randomization-based inference with global and pairwise hypotheses of interest to multi-arm cRCTs; (iii) we clarify the relative performance between the model-based test and the \textcolor{black}{randomization} test under constrained randomization in multi-arm cRCTs, and detail key analytical considerations for each test (such as whether to adjust for a cluster-level aggregate or the corresponding individual-level covariate). The remainder of the paper is organized into five sections. In Section \ref{sec:motivating}, we use an ongoing study to motivate and illustrate the constrained randomization design. In Section \ref{sec:inference}, we describe the details of statistical approaches for testing hypotheses in multi-arm cRCTs. Our simulation study and results are presented in Sections \ref{sec:simulation_setup} and \ref{sec:simulation_results}. Section \ref{sec:discussion} gives concluding remarks.

\section{Constrained randomization in multi-arm cluster randomized controlled trials} \label{sec:motivating}

\subsection{Motivating example: the TESTsmART study}

The \textit{Malaria Diagnostic Testing and Conditional Subsidies to Target Artemisinin-Based Combination Therapies in the Retail Sector} (TESTsmART) study is an ongoing cRCT to evaluate strategies to increase appropriate treatment of malaria cases in the retail sector.\cite{woolsey2021incentivizing} In response to the over-consumption of artemisinin-based combination therapies (ACTs) in malaria-endemic countries, the study aims to target subsidized ACTs to those who receive a confirmatory diagnosis in private sector retail outlets. The study was originally planned as two four-arm cRCTs among registered retail outlets (clusters) conducted separately in two distinct study sites: western Kenya and Lagos, Nigeria. For the purpose of illustration, we focus on three of the four arms and the Nigerian study site in the current article. 

In this motivating example, forty-eight randomly selected retail outlets in Lagos will be allocated evenly across three intervention arms as described in Woolsey et al.\cite{woolsey2021incentivizing} Given the highly heterogeneous nature of the retail outlets, it is important to ensure balance on several baseline cluster-level factors of interest at the design stage. Three cluster-level variables can be obtained from the pre-randomization survey conducted by the study team, including daily patient volume of an outlet (continuous), which of the two geographical regions (A \& B) within the city of Lagos the outlet is located in (binary), and whether the outlet has malaria rapid diagnostic tests (mRDT) in stock prior to the intervention (binary). The actual randomization in the study considered only the second factor, but we will consider all three variables in the decision making process 
in this motivating example for demonstration purposes.

\subsection{Implementation of constrained randomization} \label{sec:balance_metric}

Covariate data for each cluster available before randomization can be used to improve overall balance through constrained randomization. This is achieved by restricting to treatment allocations that satisfy certain pre-specified criteria on overall balance. One allocation scheme will then be selected randomly from this constrained subspace. Whereas this general idea carries over from two-arm cRCTs to multi-arm cRCTs, the choice of balance metrics requires additional considerations because an overall balance is now defined based on all treatment arms. With two-arms, 
Raab and Butcher\cite{raab2001balance} introduced the $l2$ balance metric, and Li et al.\cite{li2016evaluation,li2017evaluation} developed the corresponding $l1$ balance metric. Little difference was found between the $l1$ and $l2$ balance metric in the study by Li et al.\cite{li2016evaluation,li2017evaluation}, hence both may be used interchangeably. 
With more than two arms, Ciolino et al.\cite{ciolino2019choosing} proposed a class of p-value based balance metrics, and concluded that Kruskal–Wallis test p-value with a threshold ($p>0.3$) leads to acceptable balance. Watson et al.\cite{watson2020design} extended the $l2$ balance metric in Raab and Butcher\cite{raab2001balance} for multi-arm cRCTs as the sum of the cluster-level standardized mean differences across all arms, and followed Li et al.\cite{li2017evaluation} to choose the lowest $10\%$ of the randomization space (in terms of the balance score) for constrained randomization. 
In the current study, we wish to control the maximum degree of the between-arm imbalance and therefore propose an alternative extension of the $l2$ metric of Raab and Butcher.\cite{raab2001balance} 
Specifically, our maximum pairwise $l2$ metric is given by
\begin{align}\label{eq:l2metric}
  B_{(l2)}=\max\limits_{i\neq i'}\left\{\sum_{l}\omega_{l}(\Bar{x}_{il}-\Bar{x}_{i'l})^2\right\},
\end{align}
where $\omega_{l}\geq0$ is the weight for the $l$th variable considered for balance and $l\in\{1,\ldots,L\}$, $\Bar{x}_{il},\Bar{x}_{i'l}$ denote the average of the $l$th cluster-level covariate from the $i$th and $i'$th arm, $i\neq i'$ and $i\text{, }i'\in \{1,\ldots,c\}$ with $c$ denoting the number of treatment arms. Both continuous and categorical variables are easily accommodated by the $l2$ balance score \eqref{eq:l2metric}. For categorical variables, a set of dummy variables can be used in the balance metric. Note that cluster-level data available for constrained randomization may also be aggregated from individual-level data. In practice, however, it is not always possible to obtain individual-level data at the design phase and cluster-level summaries from historical individual-level data may be used instead.\cite{gallis2018cvcrand} The weights $\omega_l$ represent the relative importance of each covariate to the balance score $B_{(l2)}$. One can choose to up-weight or down-weight certain covariates to reflect different priorities when prior knowledge of the strength of their associations with the outcome is available. Otherwise, a simple choice of the weights would be the inverse of the variance of the $l$th covariate $(\text{i.e., } \omega_{l}=1/\mathrm{var}(x_{l}))$ and this would be equivalent to standardizing the covariates such that their variances are unity. Finally, we notice that the $l2$ metric developed in Watson et al.\cite{watson2020design} defined balance in terms of the total imbalance of each arm with respect to the population mean, whereas our $l2$ balance metric achieves a similar purpose by bounding the maximum imbalance between any of all possible two-arm comparisons. In particular, both definitions are also connected with the balance metrics developed in the causal inference literature with multiple treatments. For example, the metric of Watson et al.\cite{watson2020design} resembles the so-called population standardized difference,\cite{mccaffrey2013tutorial} whereas our metric resembles the so-called pairwise standardized differences.\cite{li2019propensity} These two types of balance metrics often carry similar performance in empirical studies.\cite{li2019propensity}

The $l2$ balance metric \eqref{eq:l2metric} can be further extended to account for the covariances between the cluster-level factors. To this end, 
we additionally consider extending the the Mahalanobis distance metric discussed in Morgan and Rubin\cite{morgan2012rerandomization} to multi-arm cRCTs. Specifically, we define the maximum pairwise Mahalanobis distance metric as
\begin{align}\label{eq:mahalanobis}
B_{(M)}=\max\limits_{i\neq i'}\left\{(\Bar{\boldsymbol{X}}_{i}-\Bar{\boldsymbol{X}}_{i'})^{T}\boldsymbol{S}^{-1}(\Bar{\boldsymbol{X}}_{i}-\Bar{\boldsymbol{X}}_{i'})\right\}, 
\end{align}
where $\Bar{\boldsymbol{X}}_{i}$ is the $L\times1$ vector of the means of the covariates in the $i$th arm, and $\boldsymbol{S}$ is the $L\times L$ estimated sample covariance matrix of $\boldsymbol{X}$, the matrix of the covariates to balance on. It can be seen that $B_{(l2)}$ is a special case of $B_{(M)}$ when the off-diagonal covariance components of $\boldsymbol{S}$ are replace by zeros. 

Once a balance metric is specified, one can either enumerate all possible randomization schemes or randomly simulate a large number of randomization schemes within the simple randomization space (duplicates are removed if the schemes are randomly simulated) and calculate the corresponding balance score for each scheme. The next important aspect of constrained randomization is the cutoff value, by which we create the constrained randomization subspace. Let $q \in (0, 1]$ denote the cutoff value and $F_{B}$ denote the empirical cumulative distribution function of the balance scores calculated using any pre-specified balance metric. The cutoff value could be defined as the percentile such that the constrained space contains schemes with balance scores no larger than $F_{B}^{-1}(q)$, where $F_{B}^{-1}(\cdot)$ is the inverse empirical distribution function of the balance scores. The intuition is that the cutoff value approximately measures the proportion of schemes achieving sufficient balance on the covariates of interest. When $q = 1$, there is no constraint and simple randomization is implemented. When $0 < q < 1$, a subset of schemes with sufficient balance will be created and constrained randomization is implemented by selecting an allocation within the subset of schemes.

\subsection{Illustration based on the TESTsmART study} 

With a mix of binary and continuous variables available before randomization in the TESTsmART study, we consider the $l2$ balance metric, $B_{(l2)}$, and the Mahalanobis distance metric, $B_{(M)}$, to assess covariate balance and obtain the distributions of the metrics in Web Figure 1. We chose $q=0.1$ as the cutoff value for the constrained space. Table~\ref{TESTsmART_3arm} presents the comparison of the cluster-level variable means from three schemes selected by constrained randomization (CR) and simple randomization (SR). Unlike the realized schemes from CR, the scheme from SR has resulted in a highly imbalanced proportion of outlets from Region B across arms. At the same time, larger mean differences across arms with respect to the continuous group-level variables are also seen from the realized scheme under SR. Although this is only a single realization, it demonstrates the potential for observing covariate imbalance under SR, as well as advantages of CR in protecting from obtaining a scheme with large imbalance.


%
%

\begin{table}[htbp]
\caption{\label{TESTsmART_3arm} Average value of each covariate by treatment arm from three randomization schemes independently selected from (1) simple randomization (SR); (2) constrained randomization (CR) with $B_{(l2)}$ and candidate set size $q=0.1$; (3) CR with $B_{(M)}$ and candidate set size $q=0.1$.} \centering
\begin{tabular}{lllll}
  \toprule
\textbf{Randomization design}& \textbf{Cluster-level covariates}& \textbf{Arm 1} ($N_1=16$) & \textbf{Arm 2} ($N_2=16$) & \textbf{Arm 3} ($N_3=16$)\\ 
  \midrule
 SR &  \# in Region B & 6 (37.5\%)  &  2 (12.5\%)  & 8 (50\%)  \\ 
  & Patient volume & 17.75 & 19.31  & 26.12 \\ 
  & \# have mRDT in stock & 1 ( 6.2\%)  & 3 (18.8\%)  & 0 ( 0.0\%)  \\ \addlinespace
  
  CR with $B_{(l2)}$ metric &  \# in Region B & 6 (37.5\%)  & 5 (31.2\%)  & 5 (31.2\%) \\ 
  & Patient volume & 19.81  & 22.75  & 20.62  \\ 
  &\# have mRDT in stock &  1 ( 6.2\%)  & 2 ( 12.5\%)  & 1 ( 6.2\%) \\  \addlinespace
  
  CR with $B_{(M)}$ metric &  \# in Region B & 6 (37.5\%)  & 4 (25.0\%)  & 6 (37.5\%) \\ 
  & Patient volume &  22.56  & 21.50  & 19.12 \\  
  & \# have mRDT in stock & 1 ( 6.2\%)  & 2 (12.5\%)  & 1 ( 6.2\%) \\ 
   \bottomrule
\end{tabular}
\begin{tablenotes}
\item Abbreviation: SR, simple randomization; CR, constrained randomization; mRDT, malaria rapid diagnostic test.
\end{tablenotes}
\end{table}

\section{Statistical Inference under Constrained Randomization in multi-arm cluster randomized controlled trials} \label{sec:inference}

While we illustrated the application of constrained randomization to the TESTsmART study, appropriate statistical inference strategies under constrained randomization in multi-arm cRCTs need to to be further explored and could eventually inform the analysis of the TESTsmART study. Watson et al.\cite{watson2020design} have studied the performance of the mixed-model based test in their simulations, and we will review this approach in Section \ref{sec:modeltest}. In addition, because there is little prior discussion on how to carry out randomization-based inference in multi-arm cRCTs, and given randomization-based inference could be naturally coupled with constrained randomization,\cite{li2016evaluation,li2017evaluation} we develop most-powerful randomization tests in Section \ref{sec:randtest}, extending the approach of Braun and Feng \cite{braun2001optimal} to multiple arms and additional null hypotheses. 
Our evaluation assumes a cross-sectional design with only a single post-treatment outcome observation for each individual in each cluster, which resembles the TESTsmART study and represents the scenario where constrained randomization provides the maximum benefit for covariate balance.\cite{watson2020design} We do not evaluate repeated cross-sectional or cohort designs, but note that the horizontal before-after comparisons in these alternative designs already offer some protection against between-cluster imbalance. 

To proceed, we assume a single continuous outcome $Y_{jk}$ for each individual $k$ $( k = 1 , \ldots , m_j)$ nested within each cluster $j$ $( j = 1 , \ldots , G)$, where $G=\sum_{i=1}^{c}g_{i}$ with $g_{i}$ denoting the number of clusters in treatment arm $i$ ($i=1,\ldots,c$). \textcolor{black}{Let $T_{ij}$ denote the dichotomous treatment indicator for the treatment $i$ $(i=1,\ldots,c-1)$ where $T_{ij}=1$ if cluster $j$ is assigned to arm $i$ and $-1$ otherwise. We deviate from the standard notation of $T_{ij}\in\{0,1\}$ because re-parameterization, $T_{ij}\in\{-1,1\}$, is useful for elaborating the details of randomization-based inference in Section \ref{sec:randtest}.} Write $\boldsymbol{x}_{jk}$ as the $p$-dimensional vector of cluster-level or individual-level covariates. Of note, this vector may contain $p_1$ cluster-level and $p_2$ individual-level covariates. Further, if this vector only includes cluster-level covariates, we could simply replace $\boldsymbol{x}_{jk}$ by $\boldsymbol{x}_{j}$ and the rest follows.

\subsection{Model-based inference}\label{sec:modeltest}

Linear mixed model (LMM) regression is routinely used in the analyses of cRCTs. In this approach, estimates of both treatment and covariate effects at either the cluster or individual level, variance components, and the induced ICC can be obtained simultaneously. For multi-arm cRCTs with continuous outcomes, LMM with a single post-treatment outcome adjusting for baseline covariates can be expressed by
\begin{align}\label{eq:theoretical_model}
  Y_{jk}=\lambda+\delta_{1} T_{1j}+\ldots+\delta_{c-1} T_{c-1,j}+\boldsymbol{x}_{jk}'\boldsymbol{\beta}+\gamma_{j}+\epsilon_{jk}=\alpha_{jk}+\gamma_{j}+\epsilon_{jk},
\end{align}
where $\lambda$ is the overall intercept parameter, $\boldsymbol{\beta}$ is the parameter vector for covariates, $\boldsymbol{\delta}=(\delta_{1},\ldots,\delta_{c-1})'$ are the parameters associated with the effect of each treatment relative to the reference level. The within-cluster correlation is accounted for by a cluster-specific normally-distributed random effect $\gamma_{j} \sim \mathcal{N}(0,\,\sigma_{\gamma}^{2})$, and the individual-level error variance is $\epsilon_{jk}\sim \mathcal{N}(0,\sigma_\epsilon^2)$; independence is often assumed between $\gamma_j$ and $\epsilon_{jk}$. The choice of reference treatment arm depends on the context, and a natural reference arm in many multi-arm cRCTs is the control or usual care arm.
Conditional on estimates of random effect variance-covariance parameters, adjusted treatment effects are estimated based on maximum likelihood or restricted maximum likelihood.\cite{pinheiro2000mixed} Using these estimated adjusted treatment effects, a model-based hypothesis test of the treatment effects often uses the Wald statistic. 
In multi-arm cRCTs where not only the effect of each single treatment but also the joint comparison among all treatment effects may be of interest, the Wald test can be flexible enough to accommodate both purposes with appropriate considerations on the degrees of freedom. 

Specifically, the global null hypothesis of no treatment effect takes the form of $\mathcal{H}_0\text{: }\boldsymbol{\delta}=\boldsymbol{0}$, and proceeds with the Wald statistic, $R=\hat{\boldsymbol{\delta}}'\hat{\boldsymbol{\Sigma}}^{-1}\hat{\boldsymbol{\delta}}$, where $\boldsymbol{\Sigma}=\text{var}(\hat{\boldsymbol{\delta}})$ is the variance-covariance matrix of the treatment effect estimators. 
This statistic follows an asymptotic $\chi^2$-distribution with degrees of freedom $c-1$ under the global null. 
On the other hand, an alternative common practice in multi-arm trials is to investigate whether there is effect of each active intervention relative to the reference treatment in a pairwise fashion, and consider multiplicity adjustment to control for the family-wise error rate (FWER). In this case, each separate null for pairwise comparison $i$ is given by $\mathcal{H}_{0,i}:\delta_i=0$, for $i=1,\ldots,c-1$, and can be tested with a Wald $z$-statistic, $\hat{\delta}_i/\sqrt{\hat{\sigma}^2_i}$, where $\sigma_i^2=\text{var}(\hat{\delta}_i)$.
However, because cRCTs often only include a limited number of clusters (e.g. fewer than 20 clusters per arm), 
the actual sampling distribution of the test statistic may deviate from the Chi-squared or normal distribution. This renders the asymptotic approximation inaccurate and results in an inflated type I error rate.\cite{kahan2016increased} 

As a potential remedy to the inflated type I error rate, we consider the $F$-test as the alternative for the $\chi^2$-test when studying the global null, and $t$-test as an alternative for the $z$-test when studying the pairwise null.
\textcolor{black}{In addition, several small-sample corrections have been proposed to determine the appropriate denominator degrees of freedom (DoF) of the $F$-test.\cite{pinheiro2000mixed,kenward1997small,satterthwaite1946approximate,murray2007sizing} In this article, we consider the ``between-within'' correction, which sets the number of denominator DoF to the number of DoF at the cluster level.\cite{pinheiro2000mixed,murray2007sizing}} It is critical to determine whether the fixed effects in model \eqref{eq:theoretical_model} are at the cluster or individual level. For an individual-level fixed effect that changes within any cluster, within-cluster degrees of freedom should be assigned; otherwise, as for tests of intervention effects, the between-cluster degrees of freedom should be assigned. That is, when a joint test for the global null is performed, the $F$-statistic should be referenced to a (central) $F$-distribution with degrees of freedom equal to ($c-1$, $G-c-$ \# of cluster-level covariates). Similarly, for the separate null of each pairwise comparison ($\delta_i=0$), we use the $t$-test with the between-within degrees of freedom $=G-c-$ \# of cluster-level covariates.\cite{watson2020design} In two-arm cRCTs conducted under constrained randomization, Li et al.\cite{li2017evaluation} demonstrated that the $t$-test with the between-within degrees of freedom preserves the type I error rate, and similar findings (albeit in the absence of multiplicity adjustment) were observed in Watson et al.\cite{watson2020design} for three-arm cRCTs conducted under constrained randomization. 

\subsection{Randomization-based inference}\label{sec:randtest}

In two-arm parallel cRCTs, Li et al.\cite{li2016evaluation,li2017evaluation} have demonstrated that the \textcolor{black}{randomization} test provides a flexible framework for inference under constrained randomization. First introduced by Fisher,\cite{fisher1935design} the randomization test can often be more robust than the model-based methods in the analyses of cRCTs, because it is exact and does not rely on asymptotic approximation of the reference distribution. \textcolor{black}{In the literature of cRCTs, the terminologies "randomization test" and "permutation test" have been used interchangeably. However, they represent two distinct probability models such that subjects are randomly assigned to treatment in randomization tests while subjects are randomly sampled in permutation tests.\cite{ernst2004permutation} To acknowledge the fact that the test of interest in this article re-randomizes the treatment assignment, we henceforth refer to such tests as randomization tests. To carry out a randomization test in cRCTs}, one first defines a relevant test statistic. Then the outcome data are first analyzed based on the actual observed allocation of clusters to obtain an observed test statistic, which will be referenced against the randomization distribution calculated from shuffling or permuting the treatment labels according to the randomization space.\cite{good2000permutation,edgington1995randomization} 
The two-sided $p$-value is defined as the proportion of the test statistics obtained by permutation that are at least as extreme (in absolute values) as the observed one. 
While any sensible statistic may be considered for valid randomization-based inference, the power of the randomization test can depend on the choice of test statistics. 
Under two-arm cRCTs with simple randomization, Braun and Feng\cite{braun2001optimal} developed a 
uniformly most-powerful randomization (UMPR) test statistic in the sense of Lehmann and Stein\cite{lehmann1949theory} with a continuous outcome. The UMPR test statistic is derived from the joint marginal likelihood induced from a LMM and is invariant to the magnitude of effect size under the alternative. Their UMPR test can be favorable for three reasons: (i) it can achieve the highest statistical power when the LMM is correctly specified; (ii) it still maintains the nominal type I error rate when the LMM is incorrectly specified; and (iii) it is computationally efficient as a score-type test in which only one LMM is fit to estimate nuisance parameters; this is in sharp contrast to a computationally intensive Wald-type randomization test where the same LMM needs to be fit repeatedly under permutation. \textcolor{black}{Despite these advantages, randomization tests for cRCTs have so far only been considered when there are two arms, and we provide an extension to multi-arm cRCTs below.} 

\textcolor{black}{We extend randomization-based inference to multi-arm cRCTs under the specific case of normal alternatives where the random effect $\gamma_j\sim\mathcal{N}(0,\,\sigma_{\gamma}^{2})$, and the outcome $Y_{jk}|\gamma_{j} \sim\mathcal{N}(\alpha_{jk}+\gamma_{j},\sigma_{\epsilon}^2)$.} Again, we 
parameterize the treatment indicators in model (\ref{eq:theoretical_model}) such that $T_{ij}=1$ if cluster $j$ is assigned to arm $i$ and $-1$ otherwise. Under this model, we can write the joint marginal likelihood of the observed data as the product of the conditional distribution of $Y_{jk}$ integrated over the normally-distributed random effect, namely $\prod_{j=1}^G f(\boldsymbol{Y}_{j})$, where the $j${th} cluster's contribution to the marginal likelihood is
\begin{align}\label{eq:likelihood}
  f(\boldsymbol{Y}_{j})=\int \left\{ \prod_{k=1}^{m_j}f(Y_{jk}|\gamma_{j})\right\}f(\gamma_j)d\gamma_j.
\end{align}
In the above expression, $f(Y_{jk}|\gamma_{j})$ is the density of $\mathcal{N}(\alpha_{jk}+\gamma_{j},\sigma_{\epsilon}^2)$, and $f(\gamma_j)$ is $\mathcal{N}(0,\,\sigma_{\gamma}^{2})$. 
The most-powerful randomization test statistic is therefore the joint likelihood $\prod_{j=1}^G f(\boldsymbol{Y}_{j})$ under a specified alternative. 

For testing the separate null for each pairwise comparison, $\mathcal{H}_{0,i}\text{: }\delta_i=0$ ($i=1,\ldots,c-1$), we show in Web Appendix C that $\prod_{j=1}^G f(\boldsymbol{Y}_{j})$ depends on a simple kernel\cite{braun2001optimal}
\begin{align}\label{eq:umpp_single}
  S_i=\sum_{j=1}^{G}T_{ij}W_{j}\sum_{k=1}^{m_j}\left(Y_{jk}-
  \lambda-\boldsymbol{x}_{jk}'\boldsymbol{\beta}-\sum_{i'\neq i}\delta_{i'} T_{i'j}^{obs}\right),
\end{align}
regardless of the alternative $\mathcal{H}_{1,i}\text{: }\delta_i=\Delta_i$. In \eqref{eq:umpp_single}, 
$W_{j}=(\sigma_{\epsilon}^2 + m_j\sigma_{\gamma}^2)^{-1}$, $T_{i'j}^{obs}$ is the \emph{observed} treatment indicator for arm $i'\neq i$, and $T_{ij} \in \{1,-1\}$ is the treatment indicator dictating the randomization distribution of $S_i$. Because the form of the kernel is independent of $\Delta_i$, the test statistic $S_i$ is uniformly most-powerful for testing $\mathcal{H}_{0,i}$. Evidently, $S_i$ is a weighted sum of the cluster total errors, and depends on the nuisance parameters, $\sigma_\epsilon^2$, $\sigma_\gamma^2$, $\lambda$, $\boldsymbol{\beta}$ and $\delta_{i'}$ ($i'\neq i$). These nuisance model parameters are estimated by fitting the LMM (\ref{eq:theoretical_model}) on the observed data by setting $\delta_{i}=0$ \textcolor{black}{(i.e., removing the treatment indicator $T_{ij}$ from LMM (\ref{eq:theoretical_model}))}, and fixed across treatment permutations. 
To implement the pairwise test in multi-arm cRCTs, there is one important caveat for obtaining the randomization distribution of $S_i$. 
That is, the permutation of $T_{ij}$ should be a conditional permutation in which only the clusters in the arms evaluated in $\mathcal{H}_{0,i}$ (i.e., the reference arm and the arm receiving treatment $i$) are permuted while the clusters receiving other treatments are fixed. Operationally, we could first subset the (simple or constrained) randomization space such that $T_{i'j}=T_{i'j}^{obs}$ for all $i'\neq i$. Then we obtain the randomization distribution of $S_i$ only within this randomization subspace, where $T_{ij}$ is allowed to vary conditional on fixed treatment indicators for all other arms. This idea of conditional permutation was also explained in Wang et al.\cite{wang2020randomization} for multi-arm individual randomized trials, and applies to multi-arm cRCTs. Finally, because the randomization test requires a minimum of 20 allocation schemes to provide a $0.05$ level test, an overly constrained randomization space may lead to too few allocation schemes once we condition on $T_{i'j}=T_{i'j}^{obs}$ ($i'\neq i$) and fail to support a $0.05$ level test for $\mathcal{H}_{0,i}$. 
\textcolor{black}{Randomization-based confidence intervals for the treatment effect can be obtained based on the duality between interval estimation and hypothesis testing. For example, one can perform a grid search for $H_{0}$: $\delta_1=\Delta_{1,0},\ldots,\delta_{c-1}=\Delta_{c-1,0}$ with $\Delta_{1,0},\ldots,\Delta_{c-1,0}$ by carrying out a series of randomization tests. The confidence interval for $\delta_i$ is formed by the set of values for $\Delta_{i,0}$ not rejected by the randomization test. In addition, inverting randomization tests to obtain confidence intervals does not necessarily require a grid search. Faster algorithms to obtain randomization-based confidence intervals have been proposed for randomized controlled trials in general and studied for cRCTs.\cite{garthwaite1996confidence,garthwaite2009stochastic,rabideau2021randomization}}  

To test the global null $\mathcal{H}_0\text{: }\boldsymbol{\delta}=\boldsymbol{0}$, the likelihood-based statistic $\prod_{j=1}^G f(\boldsymbol{Y}_j)$ can still be used. However, this statistic does not simplify to a kernel as in \eqref{eq:umpp_multi} and will not lead to a UMPR because the statistic depends on the alternative $\boldsymbol{\delta}=\boldsymbol{\Delta}$. {\color{black} Therefore, we instead develop a locally most-powerful randomization (LMPR) test based on the efficient score statistic,\cite{cox1974theoretical} which is analytically tractable under the LMM marginal likelihood. Recall the marginal likelihood (\ref{eq:likelihood}) is
\begin{align*}
    \boldsymbol{Y}_j \sim \mathcal{N}\left(\boldsymbol{\alpha}_j = \boldsymbol{T}_j\boldsymbol{\delta}+\boldsymbol{Z}_j\boldsymbol{\eta},\boldsymbol{\Sigma}_j=\sigma_{\epsilon}^2\boldsymbol{I}_{m_j}+\sigma_{\gamma}^2\boldsymbol{J}_{m_j}\right)
\end{align*} 
where $\boldsymbol{T}_j=\boldsymbol{1}_{m_j}\otimes (T_{1j},T_{2j},\ldots,T_{c-1},j)$ is the $m_j \times (C-1)$ matrix of the treatment for cluster $j$, $\boldsymbol{Z}_j$ is the $m_j \times (1+p)$ design matrix including the column vector of intercepts and the $p$-dimensional covariates $\boldsymbol{X}_j$, $\boldsymbol{\delta}$ and $\boldsymbol{\eta}$ are the corresponding parameter vectors, $\boldsymbol{I}_{m_j}$ is the $m_j \times m_j$ identity matrix, and $\boldsymbol{J}_{m_j}=\boldsymbol{1}_{m_j}\boldsymbol{1}_{m_j}'$ is the $m_j \times m_j$ matrix of ones. The full score function is given by
\begin{align*}
    \boldsymbol{S}=\begin{pmatrix}
        \boldsymbol{S_\delta} \\ \boldsymbol{S_\eta}
    \end{pmatrix} = \begin{pmatrix}
        \frac{\partial}{\partial\boldsymbol{\delta}}\sum_{j=1}^G \log f(\boldsymbol{Y}_j) \\
        \frac{\partial}{\partial\boldsymbol{\eta}}\sum_{j=1}^G \log f(\boldsymbol{Y}_j)
    \end{pmatrix},
\end{align*}
with the expected information matrix given by 
\begin{align*}
    \boldsymbol{\mathcal{I}}=\begin{pmatrix}
        \boldsymbol{\mathcal{I}_{\delta\delta}} & \boldsymbol{\mathcal{I}_{\delta\eta}} \\ 
        \boldsymbol{\mathcal{I}_{\eta\delta}}  & \boldsymbol{\mathcal{I}_{\eta\eta}}
    \end{pmatrix} = -E\begin{pmatrix}
        \frac{\partial^2}{\partial\boldsymbol{\delta}\partial\boldsymbol{\delta'}}\sum_{j=1}^G \log f(\boldsymbol{Y}_j)  & 
        \frac{\partial^2}{\partial\boldsymbol{\delta}\partial\boldsymbol{\eta'}}\sum_{j=1}^G \log f(\boldsymbol{Y}_j)  \\
        \frac{\partial^2}{\partial\boldsymbol{\eta}\partial\boldsymbol{\delta'}}\sum_{j=1}^G \log f(\boldsymbol{Y}_j)  & 
        \frac{\partial^2}{\partial\boldsymbol{\eta}\partial\boldsymbol{\eta'}}\sum_{j=1}^G \log f(\boldsymbol{Y}_j) 
    \end{pmatrix}.
\end{align*}
Since we need to estimate the nuisance parameter $\eta$, we summarize $\boldsymbol{S}$ using the efficient score statistic and define the locally most-powerful randomization (LMPR) test statistic as 
\begin{align}\label{eq:umpp_multi}
    {Q}=\boldsymbol{S_\delta}'(\boldsymbol{\mathcal{I}_{\delta\delta}-\mathcal{I}_{\delta\eta}\mathcal{I}_{\eta\eta}^{-1}\mathcal{I}_{\eta\delta}})^{-1}\boldsymbol{S_\delta},
\end{align}
where each term is evaluated under the global null. Specifically, we show in Web Appendix C that the efficient score is 
\begin{equation*}
\boldsymbol{S_\delta}
=\sum_{j=1}^{G} \boldsymbol{T}_j'\boldsymbol{\Sigma}_j^{-1}(\boldsymbol{Y}_j-\boldsymbol{Z}_j\boldsymbol{\eta})=
\begin{pmatrix} \sum_{j=1}^{G} T_{1j}\boldsymbol{1}_{m_j}'\boldsymbol{\Sigma}_j^{-1}(\boldsymbol{Y}_j-\boldsymbol{Z}_j\boldsymbol{\eta}) & \ldots & \sum_{j=1}^{G}T_{(C-1)j}\boldsymbol{1}_{m_j}'\boldsymbol{\Sigma}_j^{-1}(\boldsymbol{Y}_j-\boldsymbol{Z}_j\boldsymbol{\eta})
 \end{pmatrix}'
\end{equation*}
where $\boldsymbol{\Sigma}_j^{-1}=({1}/{\sigma_\epsilon^2})\boldsymbol{I}_{m_j}-({m_j\sigma_\gamma^2})/\left\{\sigma_\epsilon^2(\sigma_\epsilon^2+m_j\sigma_\gamma^2)\right\}\boldsymbol{J}_{m_j}$ is the analytical inverse of the compound symmetric matrix.\cite{li2018sample} Furthermore, the upper left component of the information matrix is derived as
\begin{align*}
\boldsymbol{\mathcal{I}_{\delta\delta}}=
\sum_{j=1}^G E\left(\boldsymbol{T}'_j\boldsymbol{\Sigma}_j^{-1}\boldsymbol{T}_j\right)
=\sum_{j=1}^{G} \frac{m_j}{\sigma_\epsilon^2+m_j\sigma_\gamma^2}
	\begin{pmatrix}
    q_{11}     & q_{12} & \cdots & q_{1,c-1} \\  
    q_{21}     & q_{22} & \cdots & q_{2,c-1} \\  
    \vdots   & \vdots     &  \ddots & \vdots     \\ 
    q_{c-1,1} & q_{c-1,2} & \cdots & q_{c-1,c-1}  
    \end{pmatrix},
\end{align*}
with the diagonal components $q_{ii}=1$, and off-diagonal components $q_{ii'}=(1-2\pi_{i}-2\pi_{i'})$, and $\pi_{i}$ is the allocation proportion to treatment arm $i$. The lower right component of the information matrix is derived as $\boldsymbol{\mathcal{I}_{\eta\eta}}= 
\sum_{j=1}^G \boldsymbol{Z}'_j\boldsymbol{\Sigma}_j^{-1}\boldsymbol{Z}_j$, and the off-diagonal component is given by
\begin{equation*}
\boldsymbol{\mathcal{I}_{\eta\delta}}=\boldsymbol{\mathcal{I}'_{\delta\eta}}=E\left( \sum_{j=1}^G \boldsymbol{Z}'_j\boldsymbol{\Sigma}_j^{-1}\boldsymbol{T}_j\right)
=\sum_{j=1}^{G} \frac{1}{\sigma_\epsilon^2+m_j\sigma_\gamma^2}
\begin{pmatrix}
(2\pi_1-1)m_j & (2\pi_2-1)m_j & \ldots & (2\pi_{C-1}-1)m_j\\
(2\pi_1-1)\sum_{k=1}^{m_j}\boldsymbol{x}_{jk} & 
(2\pi_2-1)\sum_{k=1}^{m_j}\boldsymbol{x}_{jk} & 
\ldots & 
(2\pi_{C-1}-1)\sum_{k=1}^{m_j}\boldsymbol{x}_{jk}
\end{pmatrix}.
\end{equation*}
The detailed derivation are provided in Web Appendix C.} 
Similar to the UMPP for testing the pairwise null, the nuisance parameters in the LMPP is estimated by fitting the LMM \eqref{eq:theoretical_model} once to the observed data (setting $\delta_1 =\ldots= \delta_{c-1} = 0$) and fixed across treatment permutations. Finally, unlike the UMPR test statistic $S_i$, the randomization distribution of $Q$ is dictated by the joint distribution of $(T_{1j},\ldots,T_{c-1,j})'$, and therefore the randomization distribution of $Q$ can be calculated according to the randomization space for the entire treatment vector. To facilitate the operation of the above randomization tests, we provide detailed execution steps in Algorithm \ref{alg:p-test}.

\begin{algorithm} 
\SetAlgoLined

 Specify the null hypothesis\;
 
  \If{$\mathcal{H}_0\text{: }\boldsymbol{\delta}=\boldsymbol{0}$ 
  }{
    Fit LMM (\ref{eq:theoretical_model}) assuming $\delta_{1}=\ldots=\delta_{c-1}=0$ \; \\
    Obtain the observed test statistic $S^{*}=Q$ using equation (\ref{eq:umpp_multi})\; \\
    \If{simple randomization of the $G$ clusters is performed in design phase}{
     1. Enumerate all possible randomization schemes of $G$ clusters within $c$ arms, or randomly sample $20,000$ randomization schemes within $c$ arms and remove replicates \; \\
     2. Suppose there are $R$ schemes in the randomization space \; \\
     3. Calculate the test statistic $S^{(r)}$ ($r=1,\ldots,R$) under each of the $R$ schemes using equation (\ref{eq:umpp_multi})\;    
    }
    \If{constrained randomization of the $G$ clusters is performed in design phase}{
     1. Enumerate all possible randomization schemes of $G$ clusters within $c$ arms, or randomly sample $20,000$ randomization schemes within $c$ arms and remove replicates \; \\
     2. Calculate overall covariate balance score $B$ across all arms \; \\
     3. Remove those randomization schemes with $B>B^{*}$, where $B^{*}$ is the corresponding balance score used in design phase \; \\
     4. Suppose there are ${R}$ schemes left in the randomization space \; \\
     5. Calculate the test statistic $S^{(r)}$ ($r=1,\ldots,R$) under each of the ${R}$ schemes using equation (\ref{eq:umpp_multi})\;
    }
   }
   \If{$\mathcal{H}_{0,i}\text{: }\delta_i=0$}{
   Fit LMM (\ref{eq:theoretical_model}) assuming $\delta_{i}=0$, where $i$ takes a value between $1$ and $c-1$  \; \\
    Obtain the observed test statistic $S^{*}=S_i$ using equation (\ref{eq:umpp_single})\; \\
    \If{simple randomization of the $G$ clusters is performed in design phase}{
     1. Enumerate all possible randomization schemes where only clusters in the $i${th} arm and the reference arm are permuted, holding the other assignment fixed. Or randomly sample $20,000$ randomization schemes where only clusters in the $i${th} arm and the reference arm are permuted and remove replicates, holding the other assignment fixed \; \\
     2. Suppose there are $R$ schemes in the randomization space \; \\
     3. Calculate the test statistic $S^{(r)}$ ($r=1,\ldots,R$) under each of the $R$ schemes using equation (\ref{eq:umpp_single})\;    
    }
    \If{constrained randomization of the $G$ clusters is performed in design phase}{
     1. Enumerate all possible randomization schemes where only clusters in the $i${th} arm and the reference arm are permuted, holding the other assignment fixed. Or randomly sample $20,000$ randomization schemes where only clusters in the $i${th} arm and the reference arm are permuted and remove replicates, holding the other assignment fixed \; \\
     2. Calculate overall covariate balance score $B$ across all arms \; \\
     3. Remove those randomization schemes with $B>B^{*}$, where $B^{*}$ is the corresponding balance score used in design phase \; \\
     4. Suppose there are ${R}$ schemes left in the randomization space \; \\
     5. Calculate the test statistic $S^{(r)}$ ($r=1,\ldots,{R}$) under each of the ${R}$ schemes using equation (\ref{eq:umpp_single})\;
    }
  }
  Calculate the two-sided $p$-value $=\frac{1}{R}\sum_{r=1}^{R}\mathbb{1}(|S^{(r)}|\geq|S^{*}|)$, where $\mathbb{1}(|S^{(r)}|\geq|S^{*}|)=1$ if $|S^{(r)}|\geq|S^{*}|$ and $=0$ otherwise

 \caption{Executing the randomization test in multi-arm cRCTs}\label{alg:p-test}
\end{algorithm}

\section{Methods for the simulation studies} \label{sec:simulation_setup}

We conduct a simulation study to assess the impact of the choice of the candidate set sizes and the balance metrics for constrained randomization, the choice of adjusted versus unadjusted analysis, as well as the use of model-based versus randomization-based inference in a multi-arm parallel cRCT setting. Wherever applicable, we followed and extended the simulation design described in Li et al.\cite{li2016evaluation} 

Overall, we designed and reported our simulation study according to the ``ADEMP'' structure of key steps and decisions in simulation studies described by Morris et al.\cite{morris2019using} We described the \textbf{A}im in the previous paragraph, \textbf{D}ata generation process in Section \ref{sec:DGP}, \textbf{E}stimand (we interpret this in our case as the target hypothesis because the goal here is for testing instead of estimation) in Section \ref{sec:H0}, \textbf{M}ethods in Section \ref{sec:sim_methods}, and chose Type I error rate and power as the main \textbf{P}erformance measures reported in Section \ref{sec:simulation_results}. We conducted a series of simulations based on a parallel multi-arm cRCT with cross-sectional design with a single post-treatment continuous outcome $Y_{jk}$ for each individual $k$ $( k = 1 , \ldots , m)$ nested within each cluster $j$ $( j = 1 , \ldots , c\times g )$ where $c$ is the number of treatment arms and $g$ is the number of clusters nested within each treatment arm. We therefore considered balanced designs with equal number of clusters ($g$) in each arm and equal number of individuals ($m$) in each cluster. Our choice of the number of clusters $(g)$ and individuals $(m)$ was motivated by the TESTsmART trial. We varied $g$ using values of $3\text{, }5\text{, and } 10$ to evaluate the performance of the randomization and analysis methods with different (but small) numbers of clusters. \textcolor{black}{We considered a small number of clusters because a recent systematic review\cite{ivers2012allocation} suggested that the median number of clusters in published cRCTs is only 21 (IQR: 12 - 52).} Each cluster was assumed to contain $m=150$ individuals, resembling the TESTsmART trial. This number was not varied since it is well-known that the effective sample size of a cRCT is largely driven by the number of clusters rather than the cluster size.\cite{donner2000design} We assumed a three-arm design $(c=3)$, with one arm receiving ‘standard of care’ and serving as the control arm. Findings from three-arm simulation studies can be informative and easily applied to other multi-arm parallel cRCT settings with more arms, thus a three-arm design was selected for the simulation studies due to computational efficiency. To ensure stable estimates for the type I error rate and power, we ran $10,000$ Monte Carlo iterations for each combination of the parameters. That is, the acceptable bounds for $5\%$ Type I error rate are $(4.57\%, 5.43\%)$. 

\subsection{Data generation process (DGP)} \label{sec:DGP}
Let $Y_{jk}$ be the outcome for each subject $k$ $(k=1,\ldots,m)$, nested within each cluster $j$ $(j=1,\ldots,3g)$. We generated the outcome data with two cluster-level binary covariates and two individual-level continuous covariates from the following linear mixed model:

\begin{align}\label{eq:outcome}
  Y_{jk}=\boldsymbol{z}_{jk}'\boldsymbol{\beta}_z+\boldsymbol{x}_{j}'\boldsymbol{\beta}_x+\delta_{1} T_{1j}+\delta_{2} T_{2j}+\gamma_{j}+\epsilon_{jk} \\
  j=1,\ldots,3g, k=1,\ldots,m.\nonumber
\end{align}

In this model, $\boldsymbol{z}_{jk}$ is the $2\times1$ vector of individual-level continuous covariates; $\boldsymbol{x}_{j}$ is the $2\times1$ vector of cluster-level binary covariates. Each of the two individual-level covariates was independently generated from $\mathcal{N}(\mu_{j},\,\sigma_{z}^{2})$. The cluster-specific means $\mu_{j} $ were randomly generated from a uniform distribution with support $(-2, 2)$. Each of the two cluster-level covariates was independently simulated from a Bernoulli distribution with probability 0.3 (a modest probability of being either 1 or 0). The strength of the association of each covariate and the outcome was fixed at a value of 1, so that   $\boldsymbol{\beta}_z=\boldsymbol{\beta}_x=\boldsymbol{1}_{2\times1} $. For a three-arm design, we need two dichotomous treatment indicators $T_{1j}$ and $T_{2j}$, \textcolor{black}{which take the values of -1 and 1}, to contrast the two treatment conditions to the control condition. The error term $\epsilon_{jk}$ was independently generated from $\mathcal{N}(0,\,\sigma_{\epsilon}^{2})$, where $\sigma_{\epsilon}^2=4$ and the cluster-specific random effect $\gamma_{j}$ was generated from $\mathcal{N}(\mu_\gamma,\,\sigma_{\gamma}^{2})$, where $\mu_\gamma=1$, $\sigma_{\gamma}^{2}=\rho\sigma_{\epsilon}^{2}/(1-\rho)$, and $\rho$ is the intraclass correlation coefficient (ICC). Three ICC values were considered for each level of $g$: $0.01$, $0.05$, and $0.10$. \textcolor{black}{The intervention effects $2\delta_1$ and $2\delta_2$ were fixed at zero for studying type I error and were specified such that the standardized effect size is approximately 0.5 or 0.75 
for studying power.}

\textcolor{black}{To further inform the comparison, we considered 2 alternative data generation processes. First, we reduced the effect sizes to 75\%, 50\%, and 25\% of the original magnitude for the case of $g=10$ to avoid the ceiling effect on power. Second, we generated non-normal data to evaluate the robustness of the analytical methods under violations of distributional assumptions. Following Small et al.\cite{small2008randomization}, we specified the random cluster effect $\gamma_j$ and error term $\epsilon_{jk}$ to follow the standard Cauchy distribution, which has a heavier tail than the normal distribution.} \textcolor{black}{Note that test statistics (\ref{eq:umpp_single}) and (\ref{eq:umpp_multi}) are no longer UMPR and LMPR test statistics under non-normal data generation process as they are derived from full likelihood. However, it would still be of interest to clarify the relative performances of statistics (\ref{eq:umpp_single}) and (\ref{eq:umpp_multi}), compared to model-based inference.}

\subsection{Null hypotheses and multiplicity adjustment} \label{sec:H0}

As noted in Section \ref{sec:inference}, different types of hypotheses could be of interest in multi-arm parallel cRCTs.\cite{juszczak2019reporting} We specified a set of three hypotheses (see Table~\ref{hypotheses}) based on model \eqref{eq:outcome} 
where there is one control condition and two treatment conditions. We adopted a hierarchical approach and specified a global hypothesis comparing all three arms at once as the first step. Then we compared the two arms receiving active treatments with the control arm separately. To adjust for multiple testing, we considered two scenarios and two adjustment strategies accordingly. In the first scenario, we were interested in each of the treatment comparisons individually and the two active treatments were evaluated distinctly. In this case, adjustment of family-wise error rate (FWER) is not needed\cite{li2017introduction,parker2020non} and we therefore fixed the significance level at $5\%$ for each pairwise hypothesis. In the second scenario, we considered a setting whereby the two active treatment arms consisted of the same treatment given at different doses and if either one of the treatment doses showed a statistically significant effect relative to ‘standard of care’, we would conclude that there was evidence of treatment effect. 
It would be recommended to control for FWER when interventions are related and findings are summarized into one single conclusion.\cite{li2017introduction,parker2020non,khan2020prevalence} Therefore, we performed a conservative Bonferroni adjustment\cite{aickin1996adjusting} for the two pairwise hypotheses of the treatment effects (i.e., alpha level $ = 2.5\%$). \textcolor{black}{Given that the adjustment of FWER is context dependent and there is no consensus in current literature, the purpose of performing such adjustment in our simulation studies is to evaluate whether the analytical methods carry the appropriate error rate under multiplicity corrections. Providing theoretical guidance on when to perform multiplicity adjustment is beyond the scope of this article.}

\begin{table}[htbp]
\caption{\label{hypotheses} Null and alternative hypotheses tested in the simulation studies. \textcolor{black}{For the pairwise hypothesis, we consider two cases: one where no multiplicity corrections are performed and the other where multiplicity is controlled.} Abbreviation: FWER, family-wise error rate.}\centering
\begin{tabular*}{500pt}{@{\extracolsep\fill}lcccc@{\extracolsep\fill}}
\toprule
\textbf{Comparison} & $\boldsymbol{\mathcal{H}_0}$  & $\boldsymbol{\mathcal{H}_1}$  & \textbf{Alpha level}  & \textbf{Alpha level controlling for FWER} \\
\midrule
Global hypothesis & $\delta_{1}=\delta_{2}=0$  & $\delta_{1}\neq0$ or $\delta_{2}\neq0$ & $5\%$  & Not Applicable  \\
Pairwise hypothesis & $\delta_{1}=0$  & $\delta_{1}\neq0$  & $5\%$  & $2.5\%$   \\
Pairwise hypothesis & $\delta_{2}=0$  & $\delta_{2}\neq0$  & $5\%$  & $2.5\%$   \\
\bottomrule
\end{tabular*}
\end{table}

\subsection{Design-based versus analysis-based adjustment} \label{sec:sim_methods}
Design-based adjustment was implemented through constrained randomization in contrast to simple randomization. For simple randomization, the final allocation scheme was selected from the full randomization space, regardless of the covariate balance. For constrained randomization, the final allocation scheme was selected from the randomization subspace where sufficient balance on the covariates of interest among all treatment arms was achieved with respect to the balance metrics described in Section \ref{sec:balance_metric}. To compare the choice of covariates in the design stage, we considered different combinations of the covariates to balance in the design phase via constrained randomization: (i) only the two cluster-level binary covariates ($\boldsymbol{x}_j$) were adjusted for in the design; (ii) only the two individual-level covariates aggregated at the cluster level ($\bar{\boldsymbol{z}}_{j}$) were adjusted for in the design; (iii) all four covariates were adjusted for in the design. Precisely, the use of individual-level covariates aggregated at cluster level in the design phase implies that the individuals in each cluster should be already recruited at the time of randomization, which may not always be the case in cRCTs. However, we consider this scenario to mimic the practice when there are high-quality estimates on the cluster-level means of the individual-level covariates from historical data available prior to randomization. To implement constrained randomization, we first enumerated all 1680 possible random assignments when $g=3$, and, for $g=5 \text{ or } 10$, we randomly sampled 20,000 assignments and removed duplicates as an approximation of the complete randomization space since full enumeration can quickly become computationally intractable with larger numbers of clusters. Then we calculated the balance scores for all the randomization schemes we enumerated $(g=3)$ or randomly sampled $(g=5 \text{ or } 10)$. The constrained space was defined as the subspace in which the corresponding balance scores of the randomization schemes were lower than a cutoff value. Since the absolute magnitude of the balance scores has no intuitive interpretation, we chose the cutoff such that the candidate set size varies from exactly 100 to $10\%$ and $50\%$ of the full randomization space. 

\begin{table}[htbp]
\caption{\label{combination} Combinations of design-based and analysis-based adjustment strategies (indicated by `$\checkmark$') evaluated in the simulation studies. Unadj: unadjusted; Adj-C: analysis adjusted for cluster-level covariates/aggregates; Adj-I, analysis adjusted for individual-level covariates; Fully Adj-C, analysis adjusted for all covariates with cluster-level aggregates; Fully Adj-I, analysis adjusted for all covariates with individual-level covariates (when available).}
\centering
\begin{tabular}{p{4cm}cccccccc}
\toprule
&  & \multicolumn{6}{c}{\textbf{Analysis-based adjustment}} \\
\cmidrule(lr){3-8}
&  & Unadj & Adj-C
& Adj-C  & Adj-I & Fully Adj-C & Fully Adj-I  \\
\multicolumn{2}{c}{\textbf{Design-based adjustment}}  & $\emptyset$ & $\boldsymbol{x}_{j}$ & $\bar{\boldsymbol{z}}_{j}$  & $\boldsymbol{z}_{jk}$ & $\{\boldsymbol{x}_{j}, \bar{\boldsymbol{z}}_{j}\}$ & $\{\boldsymbol{x}_{j}, \boldsymbol{z}_{jk}\}$  \\
\cmidrule(lr){1-2}\cmidrule(lr){3-8}
Simple randomization  & $\emptyset$ & $\checkmark$ & $\checkmark$ & $\checkmark$ & $\checkmark$ & $\checkmark$ & $\checkmark$   \\
Constrained randomization & $\boldsymbol{x}_{j}$ & $\checkmark$ & $\checkmark$ &   &  & $\checkmark$ & $\checkmark$ \\
Constrained randomization & $\bar{\boldsymbol{z}}_{j}$ & $\checkmark$ &  & $\checkmark$  & $\checkmark$ & $\checkmark$ & $\checkmark$  \\
Constrained randomization & $\{\boldsymbol{x}_{j}, \bar{\boldsymbol{z}}_{j}\}$ & $\checkmark$ &  & & & $\checkmark$ & $\checkmark$\\
\bottomrule
\end{tabular}
\end{table}

In the analysis phase, we compared the model-based and the randomization-based tests in Section \ref{sec:inference} for both simple and constrained randomization. For each test, we compared three types of covariate adjustment strategies (analysis-based adjustment): (i) no adjustment; (ii) adjustment of covariates according to the covariates used in the design stage; and (iii) fully adjusted with all four covariates. Whenever applicable, we also compared the adjustment for the actual individual-level covariates ($\boldsymbol{z}_{jk}$) versus for their cluster-level aggregates ($\bar{\boldsymbol{z}}_{j}$); this latter strategy perfectly conforms to the design, where the cluster-level aggregates of individual-level covariates are used under constrained randomization. Under simple randomization, we examined all possible covariate adjustment strategies considered for constrained randomization to compare the net benefit of design-based adjustment in the presence of analysis-based adjustment. 
To further elucidate our simulation design, Table~\ref{combination} provides a summary of design-based and analysis-based adjustment we considered (with acronyms defined and explained). In Table~\ref{combination}, the comparison of each column within a specific row reveals the benefit of analysis-based adjustment, whereas the comparison of each row within a specific column reveals the benefit of design-based adjustment. 
All analyses were conducted in R 3.6.0\cite{Rsoftware} with randomization programmed with the package \verb"randomizr"\cite{coppock2019randomizr} and linear mixed analysis performed with the package \verb"nlme"\cite{pinheiro2020nlme}.

\section{RESULTS FROM THE SIMULATION STUDIES} \label{sec:simulation_results}

\begin{figure*}[h]
	\begin{center}
	\includegraphics[trim=5 15 5 10, clip, width=0.75\linewidth]{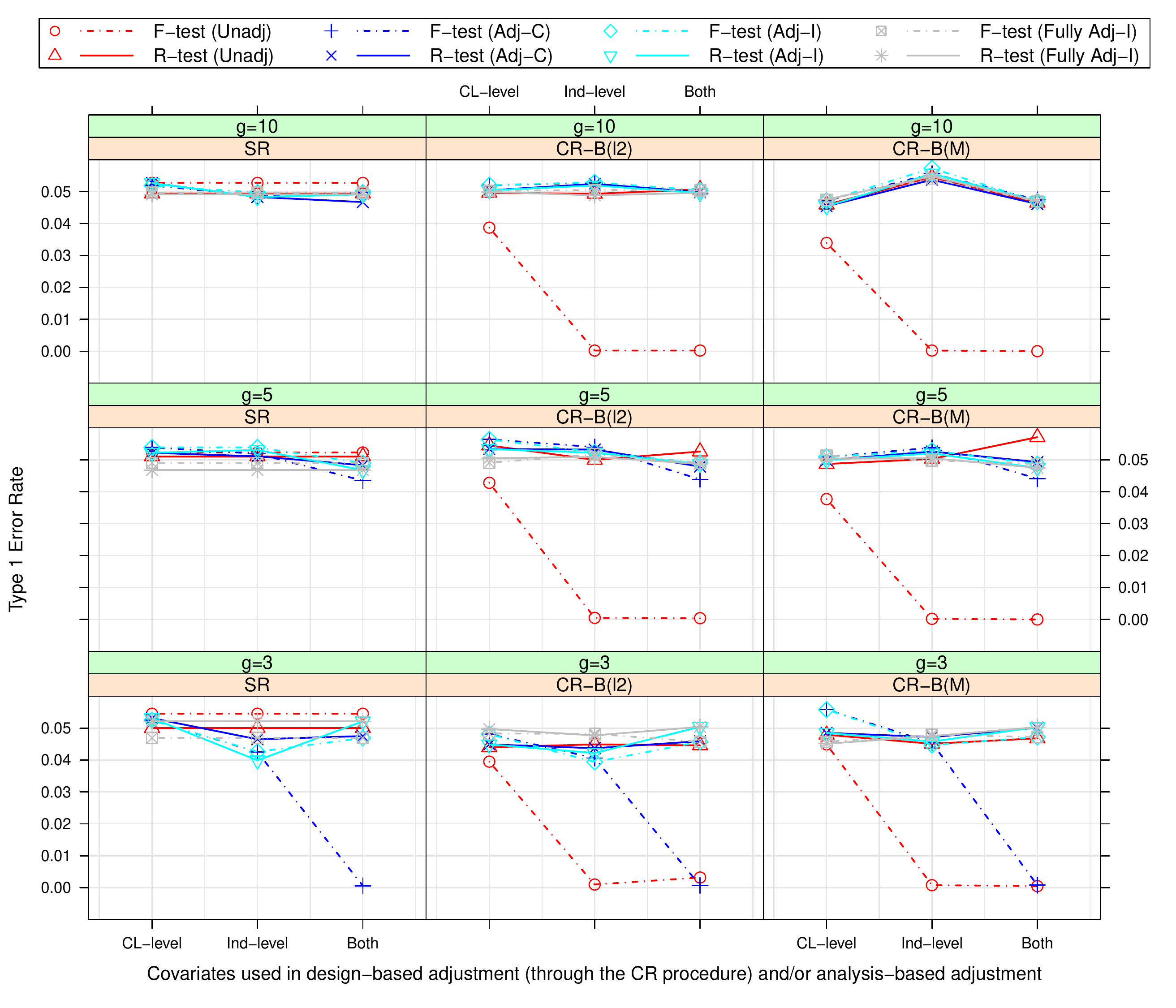}
	\end{center}
	\caption{\label{global_alpha_fig} Type I error rates for the global hypothesis ($\mathcal{H}_{0}\text{: }\delta_{1}=\delta_{2}=0$) under simple randomization (SR) versus constrained randomization (CR) with 2 balance metrics $B_{(l2)}$ and $B_{(M)}$. CR implemented using covariates indicated on the horizontal axis; candidate set size $=10\%$ under CR; ICC $=0.05$; alpha level = $5\%$; R-test: randomization test; CL-level: cluster-level covariates, $\boldsymbol{x}_j$; Ind-level: individual-level covariates, $\boldsymbol{z}_{jk}$; Unadj: unadjusted test; Adj-C: test adjusted for the covariates on the horizontal axis (with individual-level covariates aggregated at the cluster level); Adj-I: test adjusted for the covariates on the horizontal axis (with actual individual-level covariates); Fully Adj-I: test adjusted for all four covariates (with actual individual-level covariates).}
\end{figure*}

In Table~\ref{global_alpha} and Figure \ref{global_alpha_fig}, we summarized the Monte Carlo type I error rates for the global hypothesis comparing all three arms ($\mathcal{H}_{0}\text{: }\delta_{1}=\delta_{2}=0$) under simple randomization (SR) and constrained randomization (CR); in Table~\ref{global_power} and Figure~\ref{global_power_fig}, we summarized the corresponding results for power. The results for the pairwise hypotheses ($\mathcal{H}_{0}\text{: }\delta_{1}=0$ and $\mathcal{H}_{0}\text{: }\delta_{2}=0$) are presented in Web Appendix D. 
To simplify the presentation, we held the ICC fixed at $0.05$ throughout  
and compared results for $g=$ 3, 5, and 10. Results with other ICC values are qualitatively similar and presented in Web Appendix E. The tables and figures focus on the comparisons with a slightly different emphasis. In the tables, we focused on comparing the model-based and randomization-based tests under SR and CR (using $B_{(l2)}$) with the candidate set size ranging from $50\%$ to $10\%$ of the full randomization space as well as with the candidate set size is exactly $100$. That is, we study the consequence of constrained randomization with an increasing level of constraint. All CR scenarios considered all four covariates in the balance metric; the adjusted tests therefore controlled for all four covariates accordingly, with either the actual individual-level covariates or their cluster-level aggregates.
In the figures, we fixed the candidate set size at $10\%$ and compared the tests under SR and CR (using both $B_{(l2)}$ and $B_{(M)}$), across different design-based and analysis-based adjustment strategies. For design-based adjustment using CR, we controlled for three combinations of the covariates, as in Table~\ref{combination}.
For analysis-based adjustment, we chose to adjust for no covariates, the covariates used in the design phase, and all covariates, again as outlined in Table~\ref{combination}. Additional results and discussions on multiplicity adjustment 
were presented in Web Appendix F, \textcolor{black}{results under reduced effect sizes were presented in Web Appendix G, and results under non-normal DGP were presented in Web Appendix H.}

\begin{table}[htbp]
\caption{\label{global_alpha} Type I error rates for the global hypothesis ($\mathcal{H}_{0}\text{: }\delta_{1}=\delta_{2}=0$) under simple randomization (SR) versus constrained randomization (CR) \textcolor{black}{with candidate set sizes = 50\%, 10\%, and 100 of the randomization space}. All covariates were used in constrained randomization and the adjusted tests; constrained randomization was implemented using the $l2$ metric; ICC $=0.05$. The nominal type I error rate is $0.05$, and the acceptable range for nominal type I error rate with $10,000$ replicates is $(0.0457,0.0543)$.} \centering
\begin{tabular}{p{2cm}p{2.15cm}  p{0.62cm}p{0.62cm}p{0.62cm}p{0.8cm} p{0.62cm}p{0.62cm}p{0.62cm}p{0.8cm} p{0.62cm}p{0.62cm}p{0.62cm}p{0.62cm} }
  \toprule
  & &  \multicolumn{4}{c}{\textbf{$\chi^2$-test}} & \multicolumn{4}{c}{\textbf{$F$-test}} & \multicolumn{4}{c}{\textbf{Randomization test}} \\
  \cmidrule(lr){3-6}\cmidrule(lr){7-10}\cmidrule(lr){11-14}
 \# of clusters per arm  & Analysis-based adjustment & SR & CR (50$\%$) & CR (10$\%$) & CR (100) & SR & CR (50$\%$) & CR (10$\%$) & CR (100) & SR & CR (50$\%$) & CR (10$\%$) & CR (100) \\ 
  \hline
$g=10$ & Unadj & 0.070 & 0.009 & 0.000 & 0.000 & 0.053 & 0.005 & 0.000 & 0.000 & 0.049 & 0.050 & 0.051 & 0.042 \\ 
 & Adj-C & 0.068 & 0.066 & 0.072 & 0.067 & 0.050 & 0.047 & 0.049 & 0.048 & 0.047 & 0.048 & 0.050 & 0.038\\
 & Adj-I & 0.066 & 0.068 & 0.069 & 0.065 & 0.050 & 0.049 & 0.050 & 0.048 & 0.049 & 0.050 & 0.050 & 0.039 \\
 \addlinespace
  
$g=5$ & Unadj & 0.091 & 0.022 & 0.003 & 0.000 & 0.052 & 0.008 & 0.000 & 0.000 & 0.051 & 0.052 & 0.053 & 0.040 \\ 
 & Adj-C & 0.102 & 0.096 & 0.104 & 0.103 & 0.043 & 0.044 & 0.044 & 0.047 & 0.048 & 0.049 & 0.048 & 0.042 \\ 
 & Adj-I & 0.090 & 0.096 & 0.093 & 0.095 & 0.049 & 0.050 & 0.049 & 0.052 & 0.047 & 0.048 & 0.049 & 0.042 \\  
  \addlinespace
  
$g=3$ & Unadj & 0.132 & 0.050 & 0.012 & 0.006 & 0.054 & 0.014 & 0.003 & 0.001 & 0.050 & 0.049 & 0.045 & 0.040 \\ 
 & Adj-C & 0.164 & 0.174 & 0.164 & 0.167 & 0.001 & 0.000 & 0.001 & 0.001 & 0.048 & 0.049 & 0.046 & 0.043 \\ 
 & Adj-I & 0.159 & 0.165 & 0.157 & 0.158 & 0.047 & 0.049 & 0.046 & 0.049 & 0.052 & 0.048 & 0.050 & 0.040 \\

   \bottomrule
\end{tabular}
\end{table}

\subsection{Type I error rate} \label{sec:alpha_results}

With regard to type I error rate of the model-based tests, three distinct patterns were observed. First, for the global hypothesis, the model-based tests  are conservative under CR if no analysis-based adjustment was performed in accordance with the design-based adjustment. The type I error rates of the unadjusted model-based tests decrease as the candidate set size decreases (Table~\ref{global_alpha}). In addition, the conservative performance was observed regardless of the balance metrics being used and the combinations of covariates being constrained on (Figure~\ref{global_alpha_fig}). Second, appropriate analysis-based adjustment under CR brings the type I error rates of the test with appropriate degrees of freedom (i.e., $F$-test for the global hypothesis) back to the nominal level. The performance of the adjusted $F$-test is similar to that under simple randomization when $g=$ 10 or 5 (Table~\ref{global_alpha}). Similar patterns were observed for the pairwise hypotheses (Web Appendix D). Third, the $\chi^2$-test for the global hypothesis is anti-conservative with type I error rates higher than $0.0543$ (the acceptable upper bound accounting for Monte Carlo error) with $g=10$ and are inflated to over $15\%$ when $g=3$ (Table~\ref{global_alpha}). For this reason, the $z$-test, which is the counterpart of the $\chi^2$-test for the pairwise hypotheses ($\mathcal{H}_{0}\text{: }\delta_{1}=0$ and $\mathcal{H}_{0}\text{: }\delta_{2}=0$), was not considered further in our simulations. 
On the other hand, the $F$-test and $t$-test provide adequate small-sample correction and maintain the correct size of type I error rates after analysis-based adjustment except for $g=3$, as shown in Table~\ref{global_alpha} and Web Table 2, as well as Figure~\ref{global_alpha_fig} and Web Figures 2-3. This is because the small-sample correction is achieved by modifying the (denominator) degrees of freedom and each cluster-level covariate counts towards those degrees of freedom. As a result, the $F$-test and $t$-test adjusted for all covariates aggregated at the cluster level fail to achieve the correct size of type I error rate due to the lack of (denominator) degrees of freedom when the number of clusters is very small (e.g., $g=3$). 

\textcolor{black}{Different patterns were observed for the randomization test. For most scenarios, the proposed randomization test maintains the nominal type I error rate for the global hypothesis under both SR and CR, regardless of analysis-based adjustments. Throughout this article, we refer to analysis-based adjustment as the explicit adjustment of covariates in LMM (\ref{eq:theoretical_model}) for either model-based or randomization-based inference. Note that the covariates used in the design phase are already implicitly adjusted for in the randomization tests through the constrained randomization space used to create the empirical distribution of the test statistics.} The unadjusted randomization test has similar performance to that of the adjusted versions, indicating the validity of a randomization test under CR does not rely on analysis-based adjustment (Table~\ref{global_alpha} and Figure~\ref{global_alpha_fig}). Similar results were observed for the pairwise hypotheses (Web Appendix D). This is in sharp contrast to the model-based tests, which heavily depend on adequate analysis-based adjustment for validity. However, two other important factors have an impact on the performance of the randomization test, both of which are related to the number of randomization schemes used to construct the exact randomization distribution of the test statistics. First, the size of the constrained subspace should not be extremely small. When the candidate set size is only 100, the type I error rates for testing the global hypothesis become lower than 0.05. The pairwise hypothesis was observed to be even more sensitive to the candidate set size. For example, when $g=5$, the type I error rates for the pairwise hypothesis start to drop substantially below the nominal level even when the candidate set is $10\%$ of the full randomization space (Web Table 2). This is because the randomization distribution for the pairwise hypothesis depends on the clusters in the control arm and only one of the active treatment arms, resulting in a limited number of possible schemes, if the constrained randomization space is already small. 
Second, the number of clusters per arm is a key determinant of the size of the full randomization space and the constrained randomization space, and therefore can compromise the rejection rate of the randomization tests with a very small number of clusters. An extreme example arises from the pairwise hypothesis when $g=3$. The full randomization space for a test comparing two of the three arms is $\binom{6}{3}=20$. With such a limited randomization space, the rejection of any null hypothesis at the $5\%$ significance level is not feasible, let alone further reduction of the number of acceptable randomization schemes by constrained randomization. 

The combinations of the analysis-based and design-based adjustment strategies listed in Table~\ref{combination} were compared with respect to type I error rate. \textcolor{black}{When constrained randomization is used in the design phase, adjustment of the corresponding covariates in the LMM is required for the validity of model-based tests, but not for randomization-based tests.} Further adjustment of the covariates beyond those used in CR does not affect type I error rate. Adjustment of covariates using individual level data (i.e., Adj-I \& Fully Adj-I) or cluster level aggregates (i.e., Adj-C \& Fully Adj-C) has little impact on type I error rate, except for $g=3$ where there is an insufficient number of clusters to support the between-within denominator degrees of freedom in the model-based tests. For this reason, the fully adjusted tests using cluster level aggregates were excluded from the figures. Last but not least, the two balance metrics has little impact on both the model-based and randomization-based tests under constrained randomization (see the CR panels of Figure~\ref{global_alpha_fig}and Web Figures 2-3).

\subsection{Power} \label{sec:power_results}

\begin{table}[h]
\caption{\label{global_power} Power for the global hypothesis ($\mathcal{H}_{0}\text{: }\delta_{1}=\delta_{2}=0$) under simple randomization (SR) versus constrained randomization (CR) \textcolor{black}{with candidate set sizes = 50\%, 10\%, and 100 of the randomization space}. All covariates were used in constrained randomization and the adjusted tests; constrained randomization was implemented using the $l2$ metric; ICC $=0.05$; alpha level $=5\%$; \textcolor{black}{power values corresponding to non-nominal type I errors are shaded out.}} \centering
\begin{tabular}{p{2.0cm}p{2.15cm}  p{0.62cm}p{0.62cm}p{0.62cm}p{0.8cm} p{0.62cm}p{0.62cm}p{0.62cm}p{0.8cm} p{0.62cm}p{0.62cm}p{0.62cm}p{0.62cm} }
  \toprule
  & &  \multicolumn{4}{c}{\textbf{$\chi^2$-test}} & \multicolumn{4}{c}{\textbf{$F$-test}} & \multicolumn{4}{c}{\textbf{Randomization test}} \\
  \cmidrule(lr){3-6}\cmidrule(lr){7-10}\cmidrule(lr){11-14}
 \# of clusters per arm  & Analysis-based adjustment & SR & CR (50$\%$) & CR (10$\%$) & CR (100) & SR & CR (50$\%$) & CR (10$\%$) & CR (100) & SR & CR (50$\%$) & CR (10$\%$) & CR (100) \\ 
  \midrule
$g=10$ & Unadj & \textcolor{gray}{0.525} & \textcolor{gray}{0.489} & \textcolor{gray}{0.433} & \textcolor{gray}{0.385} & 0.464 & \textcolor{gray}{0.425} & \textcolor{gray}{0.355} & \textcolor{gray}{0.293} & 0.469 & 0.642 & 0.826 & \textcolor{gray}{0.960} \\ 
 & Adj-C & \textcolor{gray}{1.000} & \textcolor{gray}{1.000} & \textcolor{gray}{1.000} & \textcolor{gray}{1.000} & 1.000 & 1.000 & 1.000 & 1.000 & 1.000 & 1.000 & 1.000 & \textcolor{gray}{1.000} \\ 
 & Adj-I & \textcolor{gray}{1.000} & \textcolor{gray}{1.000} & \textcolor{gray}{1.000} & \textcolor{gray}{1.000} & 1.000 & 1.000 & 1.000 & 1.000 & 1.000 & 1.000 & 1.000 & \textcolor{gray}{1.000} \\ 
 \addlinespace
  
$g=5$ & Unadj & \textcolor{gray}{0.322} & \textcolor{gray}{0.240} & \textcolor{gray}{0.159} & \textcolor{gray}{0.076} & 0.224 & \textcolor{gray}{0.152} & \textcolor{gray}{0.084} & \textcolor{gray}{0.028} & 0.221 & 0.314 & 0.429 & \textcolor{gray}{0.551} \\ 
 & Adj-C & \textcolor{gray}{0.982} & \textcolor{gray}{0.994} & \textcolor{gray}{0.997} & \textcolor{gray}{0.998} & 0.951 & 0.975 & 0.989 & 0.993 & 0.850 & 0.923 & 0.970 & \textcolor{gray}{0.956} \\ 
 & Adj-I & \textcolor{gray}{0.996} & \textcolor{gray}{0.998} & \textcolor{gray}{0.999} & \textcolor{gray}{0.999} & 0.989 & 0.995 & 0.996 & 0.997 & 0.958 & 0.984 & 0.992 & \textcolor{gray}{0.967} \\ 
   \addlinespace
  
$g=3$ & Unadj & \textcolor{gray}{0.266} & \textcolor{gray}{0.186} & \textcolor{gray}{0.102} &\textcolor{gray}{ 0.091} & 0.128 & \textcolor{gray}{0.074} & \textcolor{gray}{0.030} & \textcolor{gray}{0.027} & 0.121 & 0.149 & 0.177 & \textcolor{gray}{0.147} \\ 
 & Adj-C & \textcolor{gray}{0.761} & \textcolor{gray}{0.815} & \textcolor{gray}{0.872} & \textcolor{gray}{0.882} & \textcolor{gray}{0.180} & \textcolor{gray}{0.226} & \textcolor{gray}{0.282} & \textcolor{gray}{0.294} & 0.267 & 0.273 & 0.339 & \textcolor{gray}{0.284} \\ 
 & Adj-I & \textcolor{gray}{0.908} & \textcolor{gray}{0.933} & \textcolor{gray}{0.947} & \textcolor{gray}{0.950} & 0.653 & 0.703 & 0.742 & 0.739 & 0.538 & 0.560 & 0.566 & \textcolor{gray}{0.453} \\

   \bottomrule
\end{tabular}
\end{table}

\begin{figure*}[h]
	\begin{center}
	\includegraphics[trim=5 15 5 10, clip, width=0.75\linewidth]{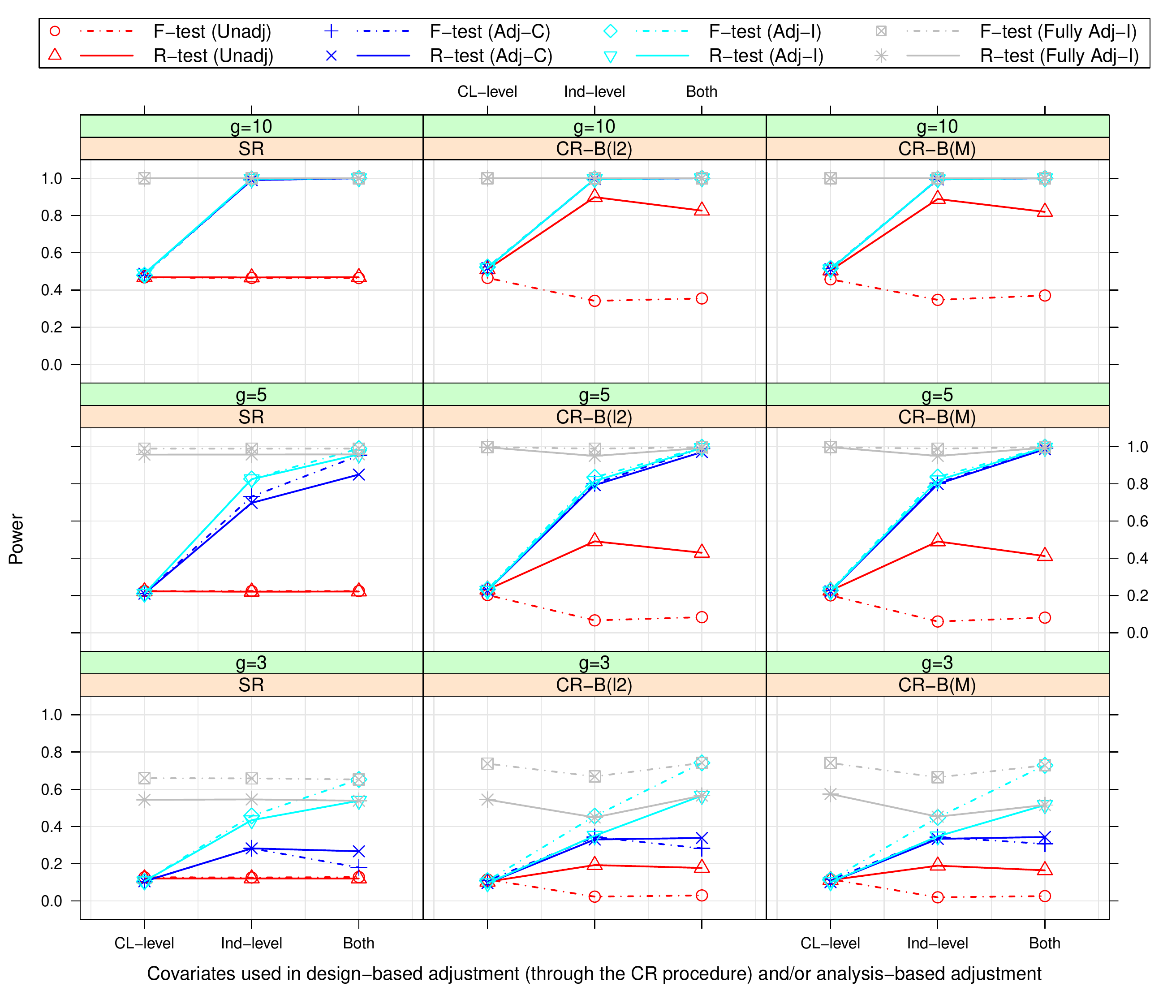}
	\end{center}
	\caption{\label{global_power_fig} Power for the global hypothesis ($\mathcal{H}_{0}\text{: }\delta_{1}=\delta_{2}=0$) under simple randomization (SR) versus constrained randomization (CR) with 2 balance metrics $B_{(l2)}$ and $B_{(M)}$. CR implemented using covariates indicated on the horizontal axis; candidate set size $=10\%$ under CR; ICC $=0.05$; alpha level $=5\%$; R-test: randomization test; CL-level: cluster-level covariates, $\boldsymbol{x}_j$; Ind-level: individual-level covariates, $\boldsymbol{z}_{jk}$; Unadj: unadjusted test; Adj-C: test adjusted for the covariates on the horizontal axis (with individual-level covariates aggregated at the cluster level); Adj-I: test adjusted for the covariates on the horizontal axis (with actual individual-level covariates); Fully Adj-I: test adjusted for all four covariates (with actual individual-level covariates).}
\end{figure*}

In this section, we describe the performance of the methods under comparison in terms of power, with a particular interest in methods that maintain nominal Type I error rates. With regard to power for the global hypothesis, the performance of model-based and randomization-based tests differ in many of the scenarios considered. First, power for the unadjusted model-based tests (i.e., $\chi^2$-test and $F$-test) decreases as the candidate set size decreases under CR (see Table~\ref{global_power}), indicating that the design-based adjustment alone does not achieve optimal power for cRCTs if results are analyzed using the model-based tests. Moreover, these tests have already been shown to be overly conservative 
in Section \ref{sec:alpha_results}. Second, after appropriate analysis-based adjustment of the covariates in accordance with the design-based adjustment, the adjusted model-based tests can achieve higher power than the adjusted randomization-based tests, when the number of clusters per arm is fairly limited (say $g=$ 3 or 5). With more clusters per arm ($g=10$), the adjusted model-based tests and the randomization-based test demonstrate similar level of power, confirming the properties of the UMPR and LMPR derived in Section \ref{sec:randtest}. \textcolor{black}{Moreover, given the power for the global hypothesis ($\mathcal{H}_{0}\text{: }\delta_{1}=\delta_{2}=0$) under $g=10$ reaches the maximum, we present power under $g=10$ with reduced effect sizes in Web Appendix G to further support this argument.} Relatedly, adjustment with prognostic covariates in the model-based tests can lead to considerably increased power even under SR (see Table~\ref{global_power} and the left panel of Figure~\ref{global_power_fig}). In this case, constrained randomization can mildly improve power over simple randomization for the adjusted model-based tests. The results for the pairwise hypotheses ($\mathcal{H}_{0}\text{: }\delta_{1}=0$ and $\mathcal{H}_{0}\text{: }\delta_{2}=0$) are similar (See Web Appendix D).

In contrast to the model-based tests, increased power is observed for randomization tests under CR versus SR regardless of analysis-based adjustment, as long as there is a sufficient number of randomization schemes to ensure a valid randomization test. For example, great improvement in power is seen in the unadjusted randomization test, for which the power under CR increases as the candidate set size decreases in most scenarios, and can be almost three times higher than that under SR in the case where $g=10$ (see top right of Table~\ref{global_power}). This implies that the design-based adjustment alone could achieve adequate power with randomization-based inference. However, this power gain for the randomization test from the design-based adjustment does not reach the level attained by the analysis-based adjustment. Despite the improvement in power under CR, our results demonstrate the importance of the validity of the randomization test, which highly depends on whether an adequate number of randomization schemes are available. This is jointly determined by the number of clusters and the size of the constrained randomization space. In particular, power for the randomization tests does not always increase monotonically as the candidate set size decreases, especially when $g \le 10$. This is because overly constraining can reduce the number of acceptable randomization schemes to construct the exact distribution of the randomization test statistic. Moreover, when the number of clusters is small, for example $g=3$, the power for randomization tests becomes unacceptably low under either SR and CR, as there are insufficient clusters to enumerate a randomization space that is large enough to make valid inference. Importantly, the pairwise hypothesis is particularly sensitive to the number of clusters and the size of the constrained randomization space, compared with the global hypothesis (see Web Appendix D). 

Besides the differences in model-based and randomization-based methods, similar patterns for the design and analysis-based adjustment strategies in Table~\ref{combination} can be found and are applied to all tests. In general, constrained randomization improves power, given proper analysis-based adjustment of the covariates used in constrained randomization. In addition, when CR was performed, further adjustment of the covariates associated with the outcome in the analysis phase beyond those used in constrained randomization (i.e., Fully Adj-I vs Adj-I, Fully Adj-C vs Adj-C) could gain additional power. Noticeably, the adjusted tests are more powerful than their unadjusted version for both SR and CR and power improves with an increasing number of covariates that are known to be predictive of the outcome being adjusted. Moreover, among the analysis-based adjustment strategies, the adjustment of individual-level continuous covariates using their aggregated version (i.e., Adj-C and Fully Adj-C) is less powerful than keeping them at the individual level (i.e., Adj-I and Fully Adj-I). For this reason, the fully adjusted tests using cluster level aggregates were excluded from the figures to simplify presentation. Finally, there is little impact of the balance metrics on power as shown in Figure~\ref{global_power_fig} and Web Figures 4-5.

\subsection{Results under different ICCs and multiplicity adjustment} \label{sec:ICC_FWER_results}

Results under an ICC of $0.10$ and $0.01$ were presented in Web Appendix E. We focused on the comparison of the performance of the design-based and analysis-based adjustment strategies within each level of ICC. Results under multiplicity adjustment (alpha $=2.5\%$) for the tests of the two pairwise hypotheses ($\mathcal{H}_{0}\text{: }\delta_{1}=0$ and $\mathcal{H}_{0}\text{: }\delta_{2}=0$) were summarized in Web Appendix F. Results in these settings were similar to those in Sections \ref{sec:alpha_results} and \ref{sec:power_results}. That is, when constrained randomization is performed, model-based tests depend on the adjustment of the covariates used in constrained randomization in the models to maintain nominal type I error rate, while the randomization tests do not require this adjustment. Constrained randomization, as compared to simple randomization, improves power for properly adjusted model based tests and for randomization tests regardless of analysis-based adjustment. Further adjustment of prognostic covariates in the analysis phase improves power for both model-based and randomization tests. Analysis-based adjustment with the individual level data whenever available is preferred over using the cluster level aggregates, even though the latter perfectly matches the design-based adjustment through constrained randomization. The two balance metrics have little impact on the results. 

\subsection{Results under alternative DGP}

\textcolor{black}{Results under non-normal DGP were presented in Web Appendix H. In the case where either the random cluster effect $\gamma_j$ or error term $\epsilon_{jk}$ follows the standard Cauchy distribution instead of the normal distribution, the randomization test has superior performance over the model-based test. In particular, when the LMM assumption fails to hold, the model-based test fails to maintain the nominal type I error rate, regardless of adjustment strategies under CR, and yields consistently smaller power than the randomization test. These results hold for both the UMPR and LMPR, depending on the hypothesis of interest, demonstrating that the randomization test is in general more robust against violations of the LMM assumptions, compared to the model-based tests.}

\section{Discussion} \label{sec:discussion}

Constrained randomization is a useful tool in cRCTs to balance multiple baseline covariates; thus protecting the internal validity of a trial. Previous simulation studies have investigated the considerations on constrained randomization and subsequent statistical analysis in two-arm parallel cRCT settings.\cite{li2016evaluation,li2017evaluation} While extensions of constrained randomization to three-arm parallel cRCTs were recently pursued,\cite{ciolino2019choosing,watson2020design} investigations on the choice of design parameters are currently limited, and the statistical tests previously considered were restricted to linear mixed models.
\cite{watson2020design} \textcolor{black}{Motivated by a recent multi-arm cRCT,\cite{woolsey2021incentivizing} we investigated the application of constrained randomization and provided a comprehensive evaluation of its statistical properties. For the design of multi-arm cRCTs, we proposed two alternative balance metrics (the maximum pairwise $l2$ metric and the maximum Mahalanobis distance metric) for the implementation of constrained randomization, and provided a detailed discussion of the impact of these alternative balance metrics as well as different sizes of the constrained space. For statistical inference under multi-arm cRCTs, we extended the theory of optimal randomization test by developing new randomization test statistics and articulating the different randomization procedures and spaces required for testing the global and pairwise hypotheses.} A comparison between the model-based and randomization-based tests was carried out to elucidate the performance of each statistical analysis approach under constrained randomization to generate practical recommendations.

Our simulation study demonstrated that when the baseline covariates are balanced through design-based adjustment via constrained randomization, both the model-based and randomization-based analyses could potentially gain power and maintain the nominal type I error rate. However, this desirable property can depend on proper adjustment in the analyses for the baseline covariates being constrained upon, especially for the model-based tests. For the model-based analyses, constrained randomization without corresponding analysis-based adjustment leads to conservative inference with no improvement in power. With appropriate adjustment of the covariates used in constrained randomization, model-based analyses were able to achieve greater power compared to simple randomization, while maintaining nominal type I error rate. \textcolor{black}{This finding is consistent with that in previous cRCTs simulations\cite{li2016evaluation,li2017evaluation,watson2020design} as well as that in individually randomized trials.\cite{lewis1999statistical,kernan1999stratified,kahan2012reporting}} Among the model-based analyses, the Wald $F$-test (for global hypothesis) and $t$-test (for pairwise hypothesis) are preferred over the $\chi^{2}$-test and $z$-test for their ability to carry the correct test size, especially in multi-arm cRCTs with a limited number of clusters per arm. However, the model-based test fails to maintain nominal type I error rate when normality assumption does not hold. For randomization-based analysis (randomization test), we provide compelling evidence to show that substantial power could be obtained and nominal type I error rate can be well maintained under constrained randomization. Notably, the power of the randomization test given sufficient randomization space is similar to that of the model-based test coupled with the correct modeling assumptions. \textcolor{black}{In contrast to the model-based test, the randomization-based test does not depend on correct distributional assumptions and analysis-based adjustment in the LMM to maintain valid inference in terms of type I error rate.} Although analysis-based adjustment is not necessary for the randomization test under constrained randomization, the statistical power could be further improved if the covariates being constrained on are adjusted in the analyses. Moreover, when covariates are adjusted for in the analysis phase, the improvement in power as a result of constrained randomization becomes relatively modest. In addition, further improvement of power could be achieved through the analysis-based adjustment of additional covariates that are not used in the design but predict the outcome. Finally, for both model-based and randomization-based analyses, we have advocated for analysis-based adjustment of individual-level covariates instead of their cluster-level aggregates, whenever applicable, to achieve higher power. \textcolor{black}{This is particularly appealing in small cRCTs, because the adjustment of cluster-level surrogates further consumes the already highly limited cluster-level degrees of freedom, which may lead to non-nominal error rate and reduced power. However, individual-level adjustment requires the exclusion or imputation of missing covariate data, while cluster-level adjustment may be less prone to this issue.\cite{hooper2018analysis}} \textcolor{black}{In general, analysis-based adjustment of strong prognostic covariates is recommended, even after design-based adjustment via constrained randomization. However, we also recognize that it may be challenging to pre-specify the “correct” set of prognostic covariates to include in the analysis. In individually randomized trials analyzed by analysis of covariance models, previous work has shown that adjusting for baseline covariates generally does not lead to asymptotic efficiency loss for estimating the average treatment effect.\cite{zeng2021propensity,yang2001efficiency,wang2019analysis} In two-arm cRCTs analyzed by linear mixed models, however, the efficiency implications of covariate adjustment may depend on the correct specification of the covariance structure and are generally more complicated.\cite{wang2021robustness} Additional research and guidance are also needed on optimal ways to select baseline covariates for adjustment in multi-arm cRCTs.} 

Compared to two-arm cRCTs, multi-arm cRCTs may embody a more diverse combination of hypotheses for treatment effects, including but not limited to the global hypothesis comparing all treatments and the pairwise hypothesis evaluating the effects of each treatment arm compared to the usual care. For model-based analyses, these hypotheses can be tested with the readily available output of the linear mixed model fit with mainstream software. In contrast, the randomization test can proceed differently according to the type of hypothesis being tested. Specifically, the test for the global hypothesis should be referenced against the randomization space where all treatment arms are permuted, whereas the test for the pairwise hypothesis should be carried out against the subspace where only the treatments under comparison are permuted (i.e., a conditional permutation). Therefore, as demonstrated by our simulation results, the randomization test for the pairwise hypothesis is more sensitive to small cRCTs and a tight constrained randomization space, because of the limited allocation schemes after conditioning on the observed treatment assignments for clusters receiving other treatments. On the other hand, with a sufficient number of clusters such as $g=10$ and as long as the constrained randomization space is not too tight, the randomization test we developed carries similar power to the model-based test. This highlights one caveat for using randomization test in small multi-arm RCTs with CR: even though they have better control of type I error rates, they may fail to provide a powerful test due to insufficient number of permutations supported by the constrained randomization design. 

These findings led us to believe that appropriate statistical inference can be challenging in (multi-arm) cRCTs with a very small number of clusters per arm, even after constrained randomization that balances the baseline covariates. \textcolor{black}{Such trials are not uncommon, as 8 studies with at most 3 clusters per condition were identified in the systematic review by Varnell et al.\cite{varnell2004design} and 8 studies with at most 5 clusters per condition were identified in a recent review of cRCTs with binary outcomes by Turner et al.\cite{turner2021completeness} We caution against the design and analysis of such trials for the following reasons.} \textcolor{black}{First, randomization-based inference may not be feasible under the 0.05  significance level or be sufficiently powerful} due to an insufficient number of possible allocation schemes to construct the randomization distribution. An example would be the case where $g=3$, the type I error rate and power for the pairwise comparison is exactly zero, \textcolor{black}{because the rejection of any null hypothesis at the $5\%$ significance level based on p-value $< 0.05$ is infeasible with a full randomization space of size $\binom{6}{3}=20$.} Although the adjusted $F$-test and $t$-test provided acceptable power and carried the desired type I error rate when $g=3$, this evidence may not be generalizable to other small cRCTs, \textcolor{black}{as the data generation process in our simulation studies fully satisfy the distributional assumptions of linear mixed models. Importantly, small sample corrections for the degrees of freedom can be used to ensure reasonable type I error rate and the randomization test can be considered as a flexible randomization-based alternative. However, theoretical guidance on the choices of small sample correction methods varies with multiple factors and the current methods may not always achieve the appropriate test size,\cite{watson2020design} while the randomization test may be invalid when the number of clusters is extremely small.} 
Furthermore, a small number of clusters limits the ability to adjust for cluster-level covariates in the analyses. The between-within denominator degree of freedom of the $F$-test and the degree of freedom of the $t$-test is calculated as the difference between the total number of clusters and the total number of cluster-level parameters, including the treatment arms and cluster-level covariates. 
In such situations, any model-based test adjusting for an excessive number of covariates is expected to be invalid due to insufficient degrees of freedom. With insufficient analysis-based adjustment, the model-based test becomes overly conservative. Because of these reasons, even though constrained randomization could achieve balance in very small cRCTs, we caution against the use of multi-arm cRCTs with a very limited number of clusters.

The choice of the two balance metrics considered in our study show little difference in terms of type I error and power. This similarity is anticipated since the two balance metrics under comparison only differ in whether the correlation among the baseline covariates is incorporated in the balance score. Compared to the choice of a balance metric, the size of the constrained randomization subspace has a more crucial impact on the analysis. Typically, power gain is achieved with a smaller constrained subspace, but this does not suggest a monotone inverse relationship. In fact, we observed in our study that overly constrained randomization can decrease the power of the tests. In addition, this impact manifests more in the pairwise hypothesis compared with the global hypothesis. In our simulations with $g = 5$ and $g=10$, a constrained randomization space $q = 0.1$ works fairly well in terms of type I error and power for the global hypothesis. For the pairwise hypothesis, however, a larger $q=0.5$ is necessary when $g=5$ to provide sufficient allocation schemes, especially when it comes to the application of the randomization test. For conducting multi-arm cRCTs, it would be useful to develop an algorithm to check the validity of the constrained randomization and to avoid an overly constrained randomization space, 
perhaps along the lines of 
Moulton\cite{moulton2004covariate}.

There are a few possible limitation of our study. First, 
our simulations are based on a balanced design. We assumed the same number of clusters per arm and the same number of individuals per cluster. Design balance at the cluster level is common with cRCTs and is essential to ensure the validity of the randomization test.\cite{braun2001optimal,gail1996design} In addition, the mixed-model methods are generally robust to unbalanced design at the cluster level.\cite{murray1998design} 
On the other hand, variable cluster size can be easily incorporated into our randomization tests without affecting its validity, whereas the appropriate choice of model-based tests and small-sample corrections may depend on the degree of cluster size variability in multi-arm cRCTs, which is an open question for further research. 
Second, we did not incorporate heterogeneous ICC structures across participating clusters. In some cases, the intervention effect may not be constant and will likely lead to higher ICC values for groups receiving intervention, giving rise to a random intervention model.\citep{murray1998design,hemming2018modeling} \textcolor{black}{Third, we only considered the $F-$ and $t-$tests with cluster-level degrees of freedom as the small sample correction method in our simulation studies. Other correction methods have been thoroughly studied in Watson et al.\cite{watson2020design} As a complement, we demonstrated that the randomization test can also be a flexible alternative for controlling type 1 error rate because it does not dependent on further analysis-based adjustment nor any degree of freedom corrections. Finally, we limited the simulation studies to continuous outcomes. However, the relative performance of different design and analytical methods, the covariate adjustment strategies, and other practical suggestions should be largely generalizable to multi-arm cRCTs with binary outcomes. More research is needed to evaluate constrained randomization and the subsequent analytical issues in multi-arm cRCTs with more complex ICC structures and different types of outcomes, in order to better guide statistical practice where balance is actively sought for in the design phase.}

\vspace{-0.2in}

\section*{Acknowledgements} 

The research presented in the manuscript was partially funded by the National Institute
of Allergy and Infectious Diseases of the National Institutes of Health (NIH) under Award Number R01 AI141444 (PI: Dr. Wendy Prudhomme O'Meara). YZ and RAS were supported in part by the National Center For Advancing Translational Sciences of the NIH under Award Number UL1 TR002553. FL was supported in part by the National Center For Advancing Translational Sciences of the NIH under Award Number UL1 TR001863. The content is solely the responsibility of the authors and does not necessarily represent the official views of the NIH. We thank Dr. Wendy Prudhomme O'Meara and Theodoor Visser for sharing the data set from the TESTsmART trial. We thank Xueqi Wang and John A. Gallis for helpful discussions and computational assistance. The authors also thank the Editor, Associate Editor and two anonymous reviewers for constructive comments and suggestions, which have greatly improved the paper.

\section*{Data availability statement}
The data that motivate and inform the simulation studies in this article are available upon request from the corresponding author. The data are not publicly available due to privacy or ethical restrictions.

\section*{References}

\begingroup
\renewcommand{\section}[2]{} 

\endgroup

\section*{Supporting information}
Additional supporting information may be found online in the Supporting Information section at the end of this article.

\begingroup
\renewcommand{\section}[2]{} 
\appendix 
\endgroup

\renewcommand{\figurename}{WEB FIGURE}
\setcounter{figure}{0}
\renewcommand{\tablename}{WEB TABLE}
\setcounter{table}{0}

\newpage 
\begin{center}
	\begin{LARGE}
		\textbf{
			Web Appendix for ``Constrained randomization and statistical inference for multi-arm parallel cluster randomized controlled trials''} \\
	\end{LARGE}
	\vspace{5mm} 
	\begin{large}
		Yunji Zhou, Elizabeth L. Turner, Ryan A. Simmons, Fan Li$^{*}$ \\
	\end{large}
\end{center}

\section*{Appendix A: Comparison of previous studies and this article}

Web Table~\ref{Comparison} shows the comparison between this article and other studies that investigated the extensions of constrained randomization to three-arm parallel cluster randomized controlled trials (cRCTs).\cite{ciolino2019choosing,watson2020design} Specifically, this article provided additional discussion of the implementation of constrained randomization with more design choices, including \textcolor{black}{more extreme} cutoff values of acceptable covariate balance and \textcolor{black}{alternative} balance metrics (the maximum pairwise $l2$ metric and the maximum Mahalanobis distance metric) for multi-arm cRCTs. For statistical inference under constrained randomization, new randomization tests for the global hypothesis and pairwise hypotheses in the context of multiple treatments were developed and evaluated in addition to linear mixed models. A comparison between the model-based and randomization-based tests was carried out to provide recommendations on the relative performance of each statistical analysis approach under simple and constrained randomization.

\begin{table}[ht]
	\centering\caption{\label{Comparison}  Comparison of studies examining constrained randomization for multi-arm parallel cluster randomized controlled trials (cRCTs). } 
	\begin{tabular}{rlll}
		\toprule
		& \textbf{Ciolino et al.}\cite{ciolino2019choosing} & \textbf{Watson et al.}\cite{watson2020design} & \textbf{This article}  \\
		\midrule
		Type of study & Parallel cRCT & Parallel cRCT  & Parallel cRCT\\ 
		Number of arms & 3 & 3 & 3 \\
		Study design & Not Applicable & \vtop{\hbox{\strut Post only}\hbox{\strut Repeated cross-section}\hbox{\strut Cohort}} & Post only \\
		$\#$ of clusters per arm & \vtop{\hbox{\strut $10\colon10\colon10$}\hbox{\strut $6\colon18\colon18$}}  & \vtop{\hbox{\strut $3\colon3\colon3$}\hbox{\strut $\vdots$}\hbox{\strut $11\colon11\colon11$}}  & \vtop{\hbox{\strut $3\colon3\colon3$}\hbox{\strut $5\colon5\colon5$}\hbox{\strut $10\colon10\colon10$}} \\
		Cluster sizes & Not Applicable & 10 to 40 & 150 \\
		Unequal cluster sizes & Not Applicable & TRUE & FALSE \\
		Type of covariates & Continuous & Continuous & \vtop{\hbox{\strut Continuous}\hbox{\strut Categorical}} \\ 
		Imbalance metrics & \vtop{\hbox{\strut min(ANOVA p-values)}\hbox{\strut min(KW p-values)}\hbox{\strut MANOVA p-value}\hbox{\strut min(WRS test p-values)}\hbox{\strut min($t$-test p-values)}} & $l2$ metric & \vtop{\hbox{\strut $l2$ metric}\hbox{\strut Mahalanobis distance}} \\
		Cutoff of acceptable balance & P-value $> 0.30$ & \textcolor{black}{Best $90\%$ to $10\%$} & Best $50\%$, $10\%$, $100$\\
		Type of outcome & Not Applicable & Continuous & Continuous \\ 
		Non-normal data & Not Applicable & FALSE & TRUE \\
		Analytical methods & Not Applicable & Linear mixed model & \vtop{\hbox{Linear mixed model}\hbox{\strut Randomization test}}  \\
		ICC & Not Applicable  & 0.001, 0.05 & 0.01, 0.05, 0.10 \\
		Multiplicity adjustment & Not Applicable & FALSE & TRUE \\
		\bottomrule
	\end{tabular}
	\begin{tablenotes}
		\item Abbreviations: ICC, intraclass correlation coefficient; ANOVA, analysis of variance; KW, Kruskal–Wallis; MANOVA, multivariate analysis of variance; WRS, Wilcoxon rank-sum. 
	\end{tablenotes}
	
\end{table}

\section*{Appendix B: Additional analysis of the TESTsmART trial}

\begin{figure*}[htbp]
	\begin{center}
		\includegraphics[trim=0 20 25 50, clip, width=0.75\linewidth]{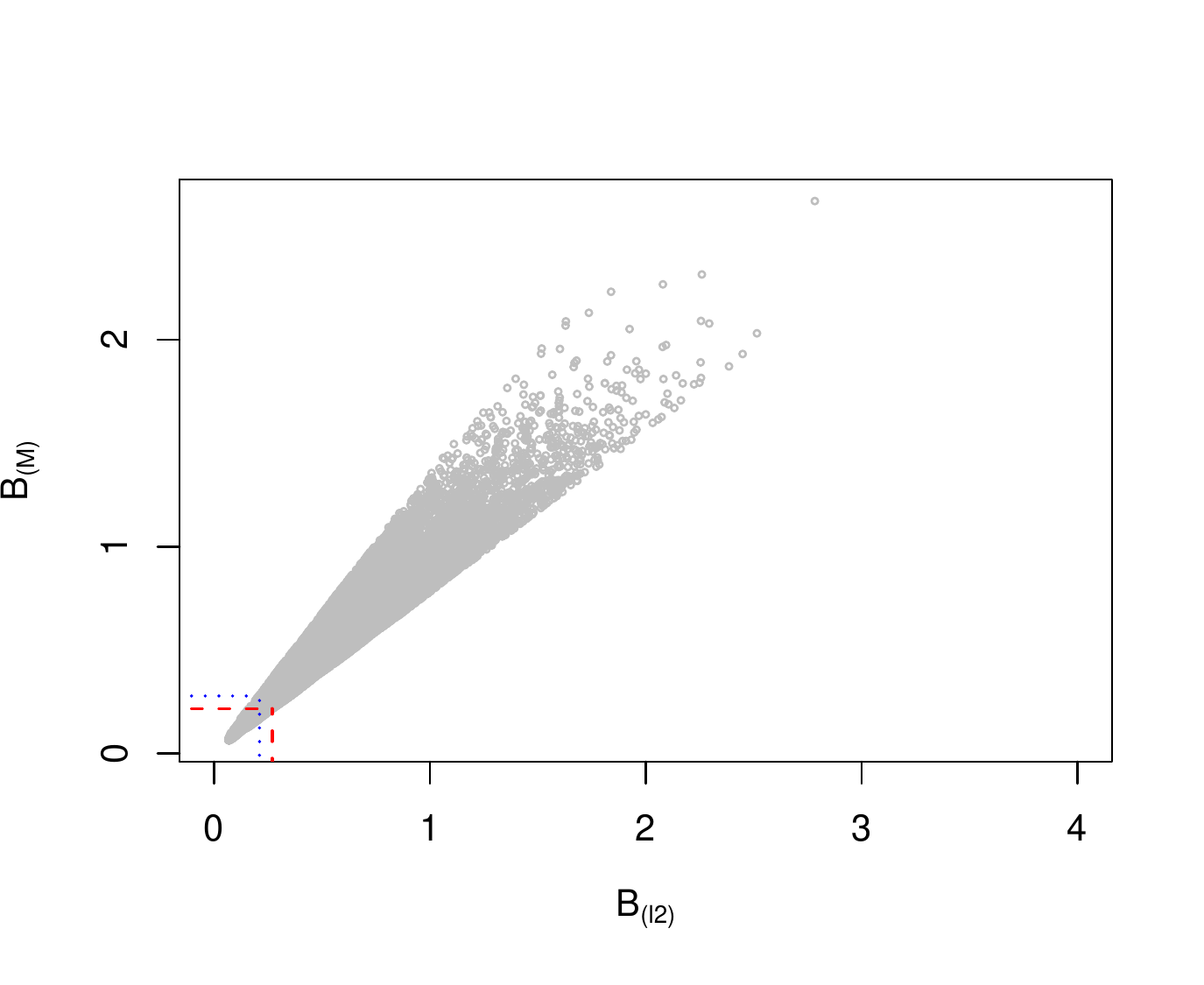}
	\end{center}
	\caption{\label{balance_score}  Plot of balance scores from the $l2$ metric ($B_{(l2)}$) against balance scores from the Mahalanobis distance ($B_{(M)}$) in the TESTsmART study. The two constrained randomization spaces ($q = 0.1$) with $B_{(M)}$ and with $B_{(l2)}$ are marked by the long-dashed red lines and the dotted blue lines, respectively. 
	}
\end{figure*}

Web Figure~\ref{balance_score} is a scatter plot of the balance scores based on the $l2$ metric ($B_{(l2)}$) against those based on the Mahalanobis distance ($B_{(M)}$) with default weights. The constrained randomization space with $q = 0.1$ under each balance metric is indicated in the plot. Each constrained space reduces the likelihood of outlet/cluster-level covariate imbalance by selecting the most balanced allocation schemes from the complete space.

\section*{Appendix C: Derivation of test statistic (6)}
From LMM (3) and the cluster likelihood defined in (4), we have the uniformly most-powerful randomization (UMPR) test statistic for the pairwise hypothesis ($\mathcal{H}_{0,i}\text{: }\delta_i=0$) as the joint likelihood $\prod_{j=1}^G f(\boldsymbol{Y}_{j})$. This test statistic corresponds to the UMPR test because it is independent of the alternative hypothesis $\delta_i=\Delta_i$. The UMPR test statistic can be simplified to test statistic (5) in the special case where $f(Y_{jk}|\gamma_{j})$ is $\mathcal{N}(\alpha_{jk}+\gamma_{j},\sigma_{\epsilon}^2)$, and $f(\gamma_j)$ is $\mathcal{N}(0,\,\sigma_{\gamma}^{2})$.\cite{braun2001optimal} The proof can be found in the Appendix of Braun and Feng \cite{braun2001optimal} and we summarize it here for completeness. In the case where $f(Y_{jk}|\gamma_{j})$ is $\mathcal{N}(\alpha_{jk}+\gamma_{j},\sigma_{\epsilon}^2)$, and $f(\gamma_j)$ is $\mathcal{N}(0,\,\sigma_{\gamma}^{2})$, the cluster likelihood (4) can be written as 
\begin{align*}
&\int^{\infty}_{-\infty} \prod_{k=1}^{m_j} (2\pi\sigma_{\epsilon}^2)^{-1/2}\text{exp}\left\{ -\frac{1}{2\sigma_{\epsilon}^2}(y_{jk}-\alpha_{jk}-\gamma)^2 \right\} \times (2\pi\sigma_{\gamma}^2)^{-1/2}\text{exp}\left\{-\frac{1}{2\sigma_{\gamma}^2}(\gamma-0)^2\right\}d\gamma \\
=&\int^{\infty}_{-\infty} (2\pi\sigma_{\epsilon}^2)^{-m_{j}/2}\text{exp}\left\{ -\frac{1}{2\sigma_{\epsilon}^2}\sum_{k=1}^{m_j}(y_{jk}-\alpha_{jk}-\gamma)^2 \right\} \times (2\pi\sigma_{\gamma}^2)^{-1/2}\text{exp}\left\{-\frac{1}{2\sigma_{\gamma}^2}(\gamma-0)^2\right\}d\gamma \\
\propto&\int^{\infty}_{-\infty} \text{exp}\left\{ -\frac{1}{2\sigma_{\epsilon}^2}\sum_{k=1}^{m_j}(y_{jk}-\alpha_{jk}-\gamma)^2-\frac{1}{2\sigma_{\gamma}^2}\gamma^2\right\}d\gamma \\
\propto&\int^{\infty}_{-\infty}\text{exp}\left\{  -\frac{1}{2\sigma_{\epsilon}^2}\sum_{k=1}^{m_j}(y_{jk}-\alpha_{jk})^2 + \frac{1}{\sigma_{\epsilon}^2}\sum_{k=1}^{m_j}(y_{jk}-\alpha_{jk})\gamma-\frac{m_j\sigma_{\gamma}^2+\sigma_{\epsilon}^2}{2\sigma_{\epsilon}^2\epsilon_{\gamma}^2}\gamma^2 \right\}d\gamma \\
\propto& \int^{\infty}_{-\infty}\text{exp}\left\{-\frac{1}{2\sigma_{\epsilon}^2}\left[\sum_{k=1}^{m_j}(y_{jk}-\alpha_{jk})^2 -\frac{\sigma_{\gamma}^2}{m_j\sigma_{\gamma}^2+\sigma_{\epsilon}^2}\left(\sum_{k=1}^{m_j}(y_{jk}-\alpha_{jk})\right)^2 + \frac{m_j\sigma_{\gamma}^2+\sigma_{\epsilon}^2}{\sigma_{\gamma}^2}\left(\gamma-\frac{\sigma_{\gamma}^2}{m_j\sigma_{\gamma}^2+\sigma_{\epsilon}^2}\sum_{k=1}^{m_j}(y_{jk}-\alpha_{jk}) \right)^2 \right]  \right\}d\gamma \\
\propto & \text{ exp}\left\{-\frac{1}{2\sigma_{\epsilon}^2}\left[\sum_{k=1}^{m_j}(y_{jk}-\alpha_{jk})^2 -\frac{\sigma_{\gamma}^2}{m_j\sigma_{\gamma}^2+\sigma_{\epsilon}^2}\left(\sum_{k=1}^{m_j}(y_{jk}-\alpha_{jk})\right)^2\right] \right\}\int^{\infty}_{-\infty} \text{exp}\left\{-\frac{\left(\gamma-\frac{\sigma_{\gamma}^2}{m_j\sigma_{\gamma}^2+\sigma_{\epsilon}^2}\sum_{k=1}^{m_j}(y_{jk}-\alpha_{jk}) \right)^2}{2\frac{\sigma_{\epsilon}^2\sigma_{\gamma}^2}{m_j\sigma_{\gamma}^2+\sigma_{\epsilon}^2}} \right\}d\gamma \\
\propto& \text{ exp}\left\{-\frac{1}{2\sigma_{\epsilon}^2}\left[\sum_{k=1}^{m_j}(y_{jk}-\alpha_{jk})^2 -\frac{\sigma_{\gamma}^2}{m_j\sigma_{\gamma}^2+\sigma_{\epsilon}^2}\left(\sum_{k=1}^{m_j}(y_{jk}-\alpha_{jk})\right)^2\right] \right\}
\end{align*}
\textcolor{black}{Following Braun and Feng\cite{braun2001optimal}, we focus on the exponent due to monotinicity and rewrite the function as
	\begin{align*}
	& -\frac{1}{2\sigma_{\epsilon}^2}\left[\sum_{k=1}^{m_j}(y_{jk}-\alpha_{jk})^2 -\frac{\sigma_{\gamma}^2}{m_j\sigma_{\gamma}^2+\sigma_{\epsilon}^2}\left(\sum_{k=1}^{m_j}(y_{jk}-\alpha_{jk})\right)^2\right] \\
	= & -\frac{1}{2\sigma_{\epsilon}^2}\left\{\sum_{k=1}^{m_j}\left[y_{jk}-(\lambda+\boldsymbol{x}_{jk}'\boldsymbol{\beta}+\sum_{i'\neq i}\delta_{i'} T_{i'j})-\delta_iT_{ij}\right]^2 -\frac{\sigma_{\gamma}^2}{m_j\sigma_{\gamma}^2+\sigma_{\epsilon}^2}\left[\sum_{k=1}^{m_j}y_{jk}-(\lambda+\boldsymbol{x}_{jk}'\boldsymbol{\beta}+\sum_{i'\neq i}\delta_{i'} T_{i'j})-\delta_iT_{ij}\right]^2\right\} \\
	= & C_1+C_2+\frac{\delta_iT_{ij}}{\sigma_{\epsilon}^2}(1-m_{j}c_{j})\sum_{k=1}^{m_j}\left[y_{jk}-\left(\lambda+\boldsymbol{x}_{jk}'\boldsymbol{\beta}+\sum_{i'\neq i}\delta_{i'} T_{i'j}\right)\right],
	\end{align*}
	where 
	\begin{align*}
	& c_j=\sigma_{\gamma}^2/(\sigma_{\epsilon}^2+m_j\sigma_{\gamma}^2), \\
	& C_1=-\left\{ \sum_{k=1}^{m_j}\left[y_{jk}-\left(\lambda+\boldsymbol{x}_{jk}'\boldsymbol{\beta}+\sum_{i'\neq i}\delta_{i'} T_{i'j}\right)\right]^2+m_j(T_{ij}\delta_i)^2 \right\} /2\sigma_{\epsilon}^2, \\
	& C_2=c_j\left\{ \left(\sum_{k=1}^{m_j}\left[y_{jk}-\left(\lambda+\boldsymbol{x}_{jk}'\boldsymbol{\beta}+\sum_{i'\neq i}\delta_{i'} T_{i'j}\right)\right] \right)^2 +(m_jT_{ij}\delta_i)^2 \right\}/2\sigma_{\epsilon}^2,
	\end{align*}
	$T_{i'j}$ is replaced by the \emph{observed} treatment indicator $T_{i'j}^{obs}$ for arm $i'\neq i$, and $T_{ij} \in \{1,-1\}$ is the treatment indicator.} By ignoring $C_1$ and $C_2$, which are invariant to the treatment assignment because $T_{ij}^2=1$, we can obtain the cluster specific statistic as
\begin{align*}
\frac{T_{ij}}{\sigma_{\epsilon}^2+m_j\sigma_{\gamma}^2}\sum_{k=1}^{m_j}\left(Y_{jk}-
\lambda-\boldsymbol{x}_{jk}'\boldsymbol{\beta}-\sum_{i'\neq i}\delta_{i'} T_{i'j}^{obs}\right).
\end{align*}
By summing over all clusters, we can obtain the overall statistic as
\begin{align*}
\sum_{j=1}^{G}T_{ij}W_{j}\sum_{k=1}^{m_j}\left(Y_{jk}-
\lambda-\boldsymbol{x}_{jk}'\boldsymbol{\beta}-\sum_{i'\neq i}\delta_{i'} T_{i'j}^{obs}\right),
\end{align*}
where $W_j=(\sigma_{\epsilon}^2+m_j\sigma_{\gamma}^2)^{-1}$.

For the global hypothesis ($\mathcal{H}_0\text{: }\boldsymbol{\delta}=\boldsymbol{0}$), the joint likelihood $\prod_{j=1}^G f(\boldsymbol{Y}_{j})$ can still be used, but the resulting statistic does not lead to a UMPR test because it depends on the alternative $\boldsymbol{\delta}=\boldsymbol{\Delta}$. \textcolor{black}{Therefore, we additionally develop a locally most-powerful randomization (LMPR) test based on the marginal likelihood (4), which is rewritten as
	\begin{align*}
	\boldsymbol{Y}_j \sim \mathcal{N}(\boldsymbol{\alpha}_j = \boldsymbol{T}_j\boldsymbol{\delta}+\boldsymbol{Z}_j\boldsymbol{\eta},\boldsymbol{\Sigma}_j=\sigma_{\epsilon}^2\boldsymbol{I}+\sigma_{\gamma}^2\boldsymbol{J})
	\end{align*} 
	where $\boldsymbol{T}_j$ is the $m_j \times (C-1)$ matrix of the treatment assignment parameterized as $\{-1,1\}$, $\boldsymbol{Z}_j$ is the $m_j \times (1+p)$ design matrix including the column vector of ones and the $p$-dimensional covariates $\boldsymbol{X}_j$, $\boldsymbol{\delta}$ and $\boldsymbol{\eta}$ are the corresponding parameter vectors, $\boldsymbol{I}$ is the $m_j \times m_j$ identity matrix, and $\boldsymbol{J}$ is the $m_j \times m_j$ matrix of ones. The full score function is given by
	\begin{align*}
	\boldsymbol{S}=\begin{pmatrix}
	\boldsymbol{S_\delta} \\ \boldsymbol{S_\eta}
	\end{pmatrix} = \begin{pmatrix}
	\frac{\partial}{\partial\boldsymbol{\delta}}\sum_{j=1}^G \log f(\boldsymbol{Y}_j) \\
	\frac{\partial}{\partial\boldsymbol{\eta}}\sum_{j=1}^G \log f(\boldsymbol{Y}_j)
	\end{pmatrix},
	\end{align*}
	with full information matrix given by 
	\begin{align*}
	\boldsymbol{\mathcal{I}}=\begin{pmatrix}
	\boldsymbol{\mathcal{I}_{\delta\delta}} & \boldsymbol{\mathcal{I}_{\delta\eta}} \\ 
	\boldsymbol{\mathcal{I}_{\eta\delta}}  & \boldsymbol{\mathcal{I}_{\eta\eta}}
	\end{pmatrix} = -E\begin{pmatrix}
	\frac{\partial^2}{\partial\boldsymbol{\delta}\partial\boldsymbol{\delta'}}\sum_{j=1}^G \log f(\boldsymbol{Y}_j)  & 
	\frac{\partial^2}{\partial\boldsymbol{\delta}\partial\boldsymbol{\eta'}}\sum_{j=1}^G \log f(\boldsymbol{Y}_j)  \\
	\frac{\partial^2}{\partial\boldsymbol{\eta}\partial\boldsymbol{\delta'}}\sum_{j=1}^G \log f(\boldsymbol{Y}_j)  & 
	\frac{\partial^2}{\partial\boldsymbol{\eta}\partial\boldsymbol{\eta'}}\sum_{j=1}^G \log f(\boldsymbol{Y}_j) 
	\end{pmatrix}.
	\end{align*}
	Since we need to estimate the nuisance parameter $\eta$, we summarize $\boldsymbol{S}$ using the efficient score statistic and define the locally most-powerful randomization (LMPR) test statistic as 
	\begin{align*}
	{Q}=\boldsymbol{S_\delta}'(\boldsymbol{\mathcal{I}_{\delta\delta}-\mathcal{I}_{\delta\eta}\mathcal{I}_{\eta\eta}^{-1}\mathcal{I}_{\eta\delta}})^{-1}\boldsymbol{S_\delta}
	\end{align*}
	To operationalize the LMPR test statistic, we calculate both $\boldsymbol{S}$ and $\boldsymbol{\mathcal{I}}$ evaluated under the global null in the following steps. \\
	(1) We first calculate $\boldsymbol{S_\delta}|_{\boldsymbol{\delta}=\boldsymbol{0}}$
	\begin{flalign*}
	\boldsymbol{S_\delta}|_{\boldsymbol{\delta}=\boldsymbol{0}} = & \frac{\partial}{\partial\boldsymbol{\delta}}\sum_{j=1}^G \log f(\boldsymbol{Y}_j)|_{\boldsymbol{\delta}=\boldsymbol{0}}  \\ 
	= & \sum_{j=1}^{G} \boldsymbol{T}'\boldsymbol{\Sigma}_j^{-1}(\boldsymbol{Y}_j-\boldsymbol{T}_j\boldsymbol{\delta}-\boldsymbol{Z}_j\boldsymbol{\eta})|_{\boldsymbol{\delta}=\boldsymbol{0}}  \\
	= & \sum_{j=1}^{G} \boldsymbol{T}_j'\boldsymbol{\Sigma}_j^{-1}(\boldsymbol{Y}_j-\boldsymbol{Z}_j\boldsymbol{\eta}),
	\end{flalign*} 
	which can be written in a $(C-1) \times 1$ matrix as
	\begin{align*}
	\begin{pmatrix} \sum_{j=1}^{G} T_{1j}\boldsymbol{1}_{m_j}'\boldsymbol{\Sigma}_j^{-1}(\boldsymbol{Y}_j-\boldsymbol{Z}_j\boldsymbol{\eta}) & \ldots & \sum_{j=1}^{G}T_{(C-1)j}\boldsymbol{1}_{m_j}'\boldsymbol{\Sigma}_j^{-1}(\boldsymbol{Y}_j-\boldsymbol{Z}_j\boldsymbol{\eta})
	\end{pmatrix}'
	\end{align*}
	(2) We then derive $\boldsymbol{\mathcal{I}_{\delta\delta}}$
	\begin{align*}
	\boldsymbol{\mathcal{I}_{\delta\delta}}= & -E\left\{\frac{\partial^2}{\partial\boldsymbol{\delta}\partial\boldsymbol{\delta'}}\sum_{j=1}^G \log f(\boldsymbol{Y}_j)\right\} \\
	= & E\left( \sum_{j=1}^G \boldsymbol{T}'_j\boldsymbol{\Sigma}_j^{-1}\boldsymbol{T}_j\right) \\
	= & \sum_{j=1}^G E\left(\boldsymbol{T}'_j\boldsymbol{\Sigma}_j^{-1}\boldsymbol{T}_j\right)
	\end{align*}
	The diagonal element of $\boldsymbol{\mathcal{I}_{\delta\delta}}$ is derived as follows
	\begin{align*}
	\sum_{j=1}^G E\left(T_{ij}\boldsymbol{1}'\boldsymbol{\Sigma}_j^{-1}T_{ij}\boldsymbol{1}\right) = & \sum_{j=1}^{G} E\left(\boldsymbol{1}'\boldsymbol{\Sigma}_j^{-1}\boldsymbol{1} \right) \\
	= & \sum_{j=1}^{G} \frac{m_j}{\sigma_{\epsilon}^2}-\frac{m_j^2\sigma_\gamma^2}{\sigma_\epsilon^2(\sigma_\epsilon^2+m_j\sigma_\gamma^2)} \\
	= & \sum_{j=1}^{G} \frac{m_j}{\sigma_\epsilon^2+m_j\sigma_\gamma^2} \\
	=&\sum_{j=1}^{G} m_{j}W_{j}
	\end{align*}
	The off-diagonal element of $\boldsymbol{\mathcal{I}_{\delta\delta}}$ is derived as follows 
	\begin{align*}
	\sum_{j=1}^G E\left(T_{ij}\boldsymbol{1}'\boldsymbol{\Sigma}_j^{-1}T_{i'j}\boldsymbol{1}\right) = & \sum_{j=1}^{G} E\left(T_{ij}T_{i'j} \right)\left(\boldsymbol{1}'\boldsymbol{\Sigma}_j^{-1}\boldsymbol{1} \right) \\
	= & \sum_{j=1}^{G} \frac{m_j E(T_{ij}T_{i'j})}{\sigma_\epsilon^2+m_j\sigma_\gamma^2} \\
	=&\sum_{j=1}^{G} m_{j}W_{j} E(T_{ij}T_{i'j})
	\end{align*}
	To calculate $E(T_{ij}T_{i'j})$, we need to obtain the joint probability density function of $T_{ij}$ and $T_{i'j}$ with respect to $\pi_{ij}$ and $\pi_{i'j}$, which are the known probabilities of a cluster $j$ being assigned to arm $i$ and $i'$, respectively:
	\begin{align*}
	f(T_{ij},T_{i'j};\pi_{ij},\pi_{i'j})=\begin{cases}
	0 & \text{if } T_{ij}=1 \text{ \& } T_{i'j}=1\\
	\pi_{ij} & \text{if } T_{ij}=1 \text{ \& } T_{i'j}=-1\\
	1-\pi_{ij}-\pi_{i'j} & \text{if } T_{ij}=-1 \text{ \& } T_{i'j}=-1\\
	\pi_{i'j} & \text{if } T_{ij}=-1 \text{ \& } T_{i'j}=1
	\end{cases}  
	\end{align*}
	Therefore, $E(T_{ij}T_{i'j})=\sum T_{ij}T_{i'j}f(T_{ij},T_{i'j};\pi_{ij},\pi_{i'j})=0-\pi_{ij}+1-\pi_{ij}-\pi_{i'j}-\pi_{i'j}=1-2\pi_{ij}-2\pi_{i'j}$ \\
	(3) Next, we derive $\boldsymbol{\mathcal{I}_{\eta\eta}}$
	\begin{align*}
	\boldsymbol{\mathcal{I}_{\eta\eta}}= & -E\left\{\frac{\partial^2}{\partial\boldsymbol{\eta}\partial\boldsymbol{\eta'}}\sum_{j=1}^G \log f(\boldsymbol{Y}_j)\right\} \\
	= &  \sum_{j=1}^G \boldsymbol{Z}'_j\boldsymbol{\Sigma}_j^{-1}\boldsymbol{Z}_j
	\end{align*}
	where $\boldsymbol{\Sigma}_j^{-1}$ can be shown to be equal to $\frac{1}{\sigma_\epsilon^2}\boldsymbol{I}_{m_j}-\frac{m_j\sigma_\gamma^2}{\sigma_\epsilon^2(\sigma_\epsilon^2+m_j\sigma_\gamma^2)}\boldsymbol{J}_{m_j}$ following Appendix A of Li et al.\cite{li2018sample} \\
	(4) Next, we derive $\boldsymbol{\mathcal{I}_{\eta\delta}}$
	\begin{align*}
	\boldsymbol{\mathcal{I}_{\eta\delta}}= & -E\left\{\frac{\partial^2}{\partial\boldsymbol{\eta}\partial\boldsymbol{\delta'}}\sum_{j=1}^G \log f(\boldsymbol{Y}_j)\right\} \\
	= & E\left( \sum_{j=1}^G \boldsymbol{Z}'_j\boldsymbol{\Sigma}_j^{-1}\boldsymbol{T}_j\right) \\
	= & \sum_{j=1}^G \boldsymbol{Z}'_j\boldsymbol{\Sigma}_j^{-1}E\left(\boldsymbol{T}_j\right) \\
	= & \sum_{j=1}^G \boldsymbol{Z}'_j \left( \frac{1}{\sigma_\epsilon^2}E(\boldsymbol{T}_j)-\frac{m_j\sigma_\gamma^2}{\sigma_\epsilon^2(\sigma_\epsilon^2+m_j\sigma_\gamma^2)}\boldsymbol{1} \begin{pmatrix}2\pi_1-1 & \cdots & 2\pi_{(C-1)}-1 \end{pmatrix}  \right) \\
	= & \sum_{j=1}^G \boldsymbol{Z}'_j \left(\frac{1}{\sigma_\epsilon^2}E(\boldsymbol{T}_j)- \frac{m_j\sigma_\gamma^2}{\sigma_\epsilon^2(\sigma_\epsilon^2+m_j\sigma_\gamma^2)}E(\boldsymbol{T}_j)\right) \\
	= & \sum_{j=1}^G \boldsymbol{Z}'_j \left( \frac{1}{\sigma_\epsilon^2+m_j\sigma_\gamma^2}E(\boldsymbol{T}_j) \right).
	\end{align*}
	Since each column of $E(\boldsymbol{T}_j)$ corresponds to a treatment variable $T_{ij}$ and $E(T_{ij})=\pi_i-(1-\pi_i)=2\pi_i-1$, then the $i$th column of $E(\boldsymbol{T}_j)$ is written by $(2\pi_i-1)\boldsymbol{1}_{m_j}$, and we therefore can write
	\begin{equation*}
	\boldsymbol{\mathcal{I}_{\eta\delta}}=\sum_{j=1}^{G} \frac{1}{\sigma_\epsilon^2+m_j\sigma_\gamma^2}
	\begin{pmatrix}
	(2\pi_1-1)m_j & (2\pi_2-1)m_j & \ldots & (2\pi_{C-1}-1)m_j\\
	(2\pi_1-1)\sum_{k=1}^{m_j}\boldsymbol{x}_{jk} & 
	(2\pi_2-1)\sum_{k=1}^{m_j}\boldsymbol{x}_{jk} & 
	\ldots & 
	(2\pi_{C-1}-1)\sum_{k=1}^{m_j}\boldsymbol{x}_{jk}
	\end{pmatrix}.
	\end{equation*}
	\\
	(5) Finaly, we obtain $\boldsymbol{\mathcal{I}_{\delta\eta}} = \boldsymbol{\mathcal{I}_{\eta\delta}}'$   
}

\clearpage

\section*{Appendix D: Results for the pairwise hypotheses}

\begin{figure*}[htbp]
	\begin{center}
		\includegraphics[trim=5 15 5 10, clip, width=0.85\linewidth]{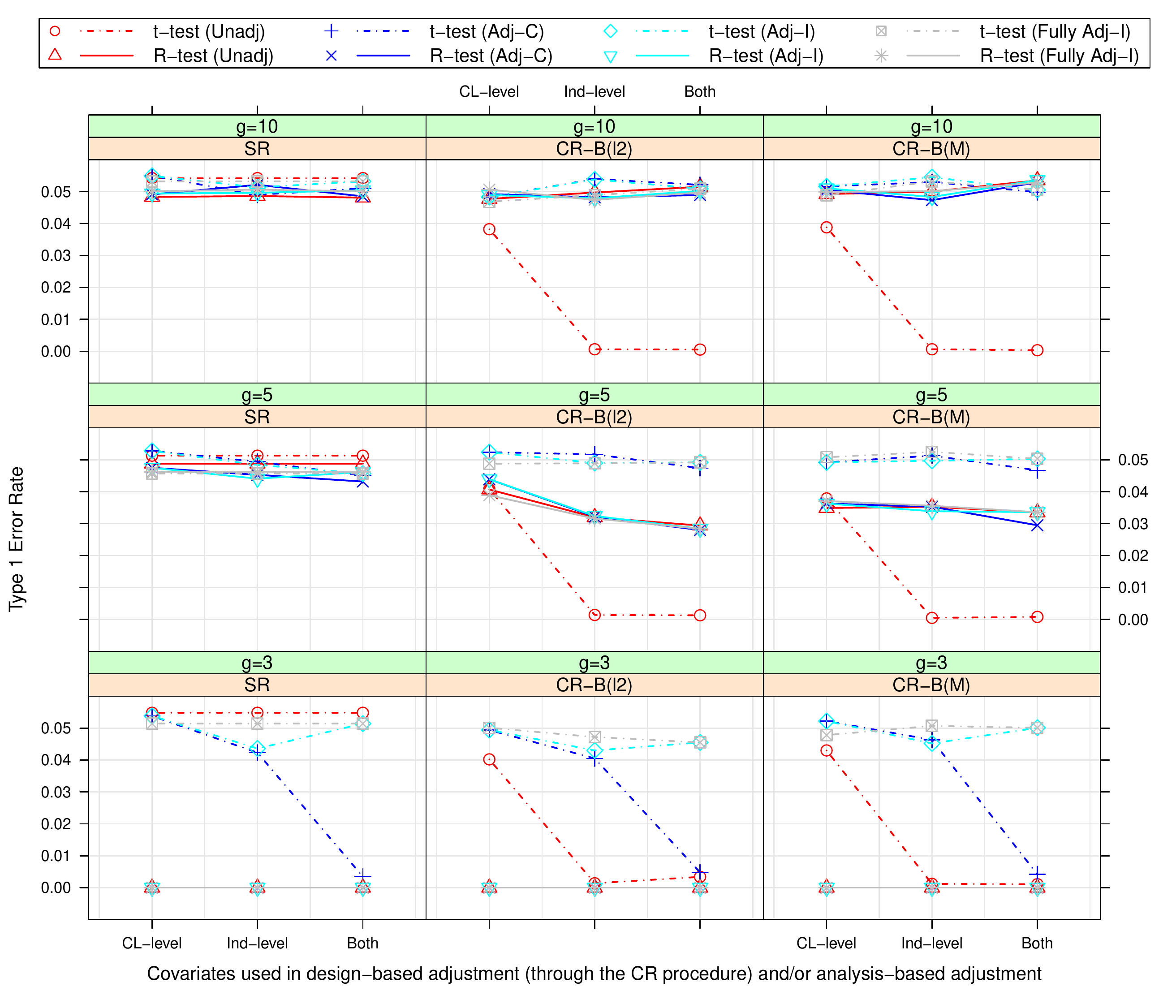}
	\end{center}
	\caption{\label{fig:indi_alpha_fig}  Type I error rates for the pairwise hypothesis ($\mathcal{H}_{0}\text{: }\delta_{1}=0$) under simple randomization (SR) versus constrained randomization (CR) with 2 balance metrics $B_{(l2)}$ and $B_{(M)}$. CR implemented using covariates indicated on the horizontal axis; candidate set size $=10\%$ under CR; ICC $=0.05$; alpha level $=5\%$; R-test: randomization test; CL-level: cluster-level covariates, $\boldsymbol{x}_j$; Ind-level: individual-level covariates, $\boldsymbol{z}_{jk}$; Unadj: unadjusted test; Adj-C: test adjusted for the covariates on the horizontal axis (with individual-level covariates aggregated at the cluster level); Adj-I: test adjusted for the covariates on the horizontal axis (with actual individual-level covariates); Fully Adj-I: test adjusted for all four covariates (with actual individual-level covariates).}%
	%
\end{figure*}

\begin{figure*}[htbp]
	\begin{center}
		\includegraphics[trim=5 15 5 10, clip, width=0.85\linewidth]{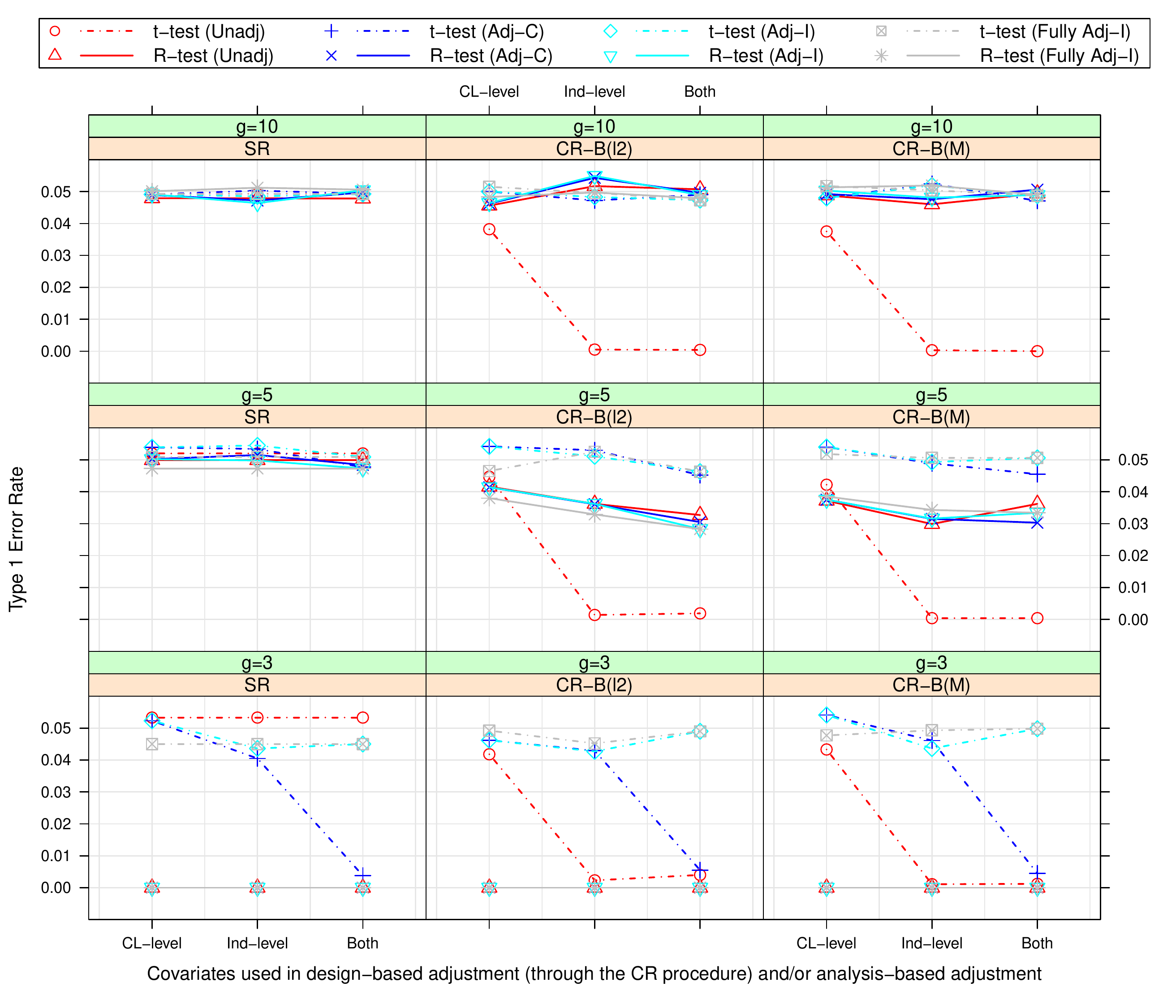}
	\end{center}
	\caption{\label{fig:indi2_alpha_fig}  Type I error rates for the pairwise hypothesis ($\mathcal{H}_{0}\text{: }\delta_{2}=0$) under simple randomization (SR) versus constrained randomization (CR) with 2 balance metrics $B_{(l2)}$ and $B_{(M)}$. CR implemented using covariates indicated on the horizontal axis; candidate set size $=10\%$ under CR; ICC $=0.05$; alpha level $=5\%$; R-test: randomization test; CL-level: cluster-level covariates, $\boldsymbol{x}_j$; Ind-level: individual-level covariates, $\boldsymbol{z}_{jk}$; Unadj: unadjusted test; Adj-C: test adjusted for the covariates on the horizontal axis (with individual-level covariates aggregated at the cluster level); Adj-I: test adjusted for the covariates on the horizontal axis (with actual individual-level covariates); Fully Adj-I: test adjusted for all four covariates (with actual individual-level covariates).}%
	%
\end{figure*}

\begin{figure*}[htbp]
	\begin{center}
		\includegraphics[trim=5 15 5 10, clip, width=0.85\linewidth]{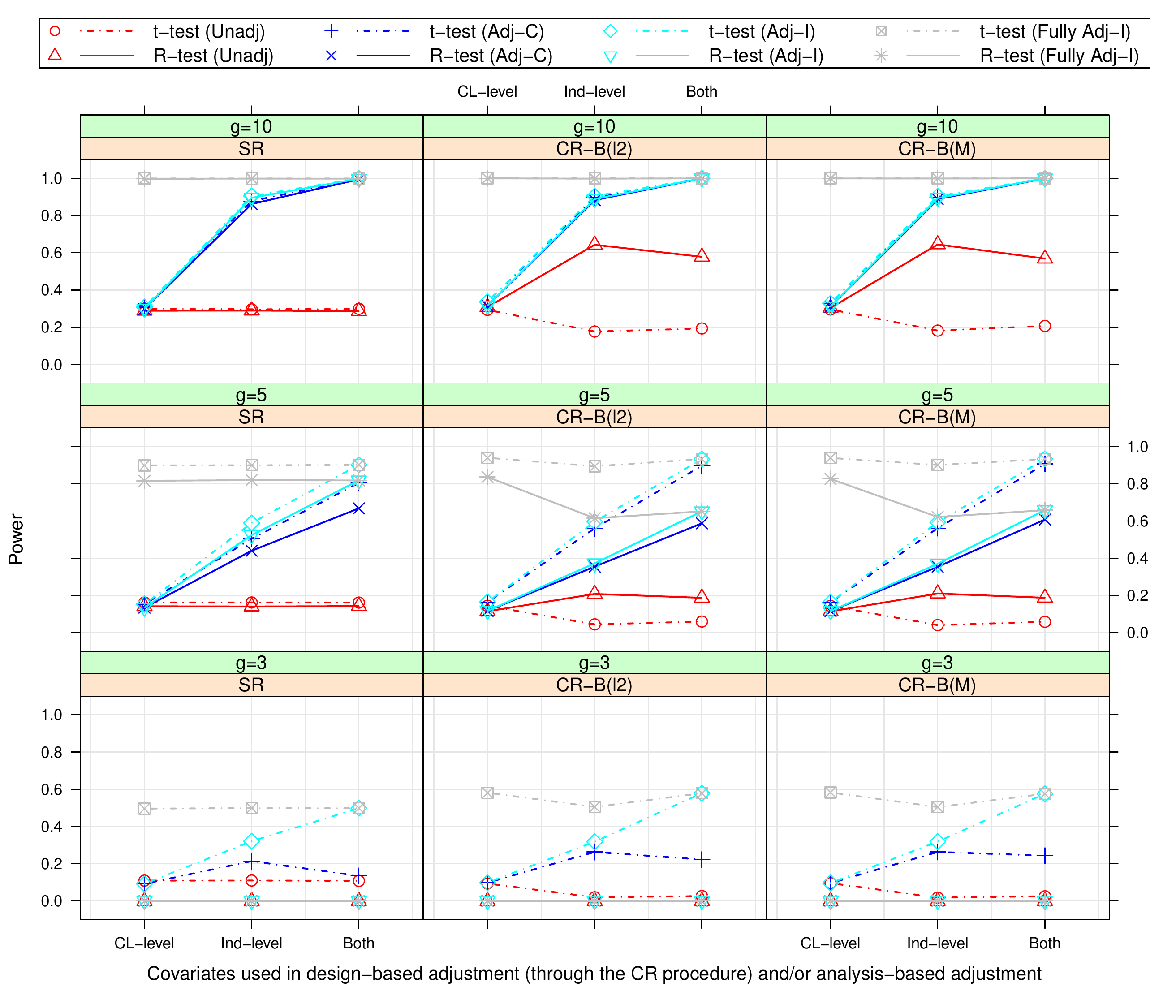}
	\end{center}
	\caption{\label{indi_power_fig}  Power for the pairwise hypothesis ($\mathcal{H}_{0}\text{: }\delta_{1}=0$) under simple randomization (SR) versus constrained randomization (CR) with 2 balance metrics $B_{(l2)}$ and $B_{(M)}$. CR implemented using covariates indicated on the horizontal axis; candidate set size $=10\%$ under CR; ICC $=0.05$; alpha level $=5\%$; R-test: randomization test; CL-level: cluster-level covariates, $\boldsymbol{x}_j$; Ind-level: individual-level covariates, $\boldsymbol{z}_{jk}$; Unadj: unadjusted test; Adj-C: test adjusted for the covariates on the horizontal axis (with individual-level covariates aggregated at the cluster level); Adj-I: test adjusted for the covariates on the horizontal axis (with actual individual-level covariates); Fully Adj-I: test adjusted for all four covariates (with actual individual-level covariates). 
	}
\end{figure*}

\begin{figure*}[htbp]
	\begin{center}
		\includegraphics[trim=5 15 5 10, clip, width=0.85\linewidth]{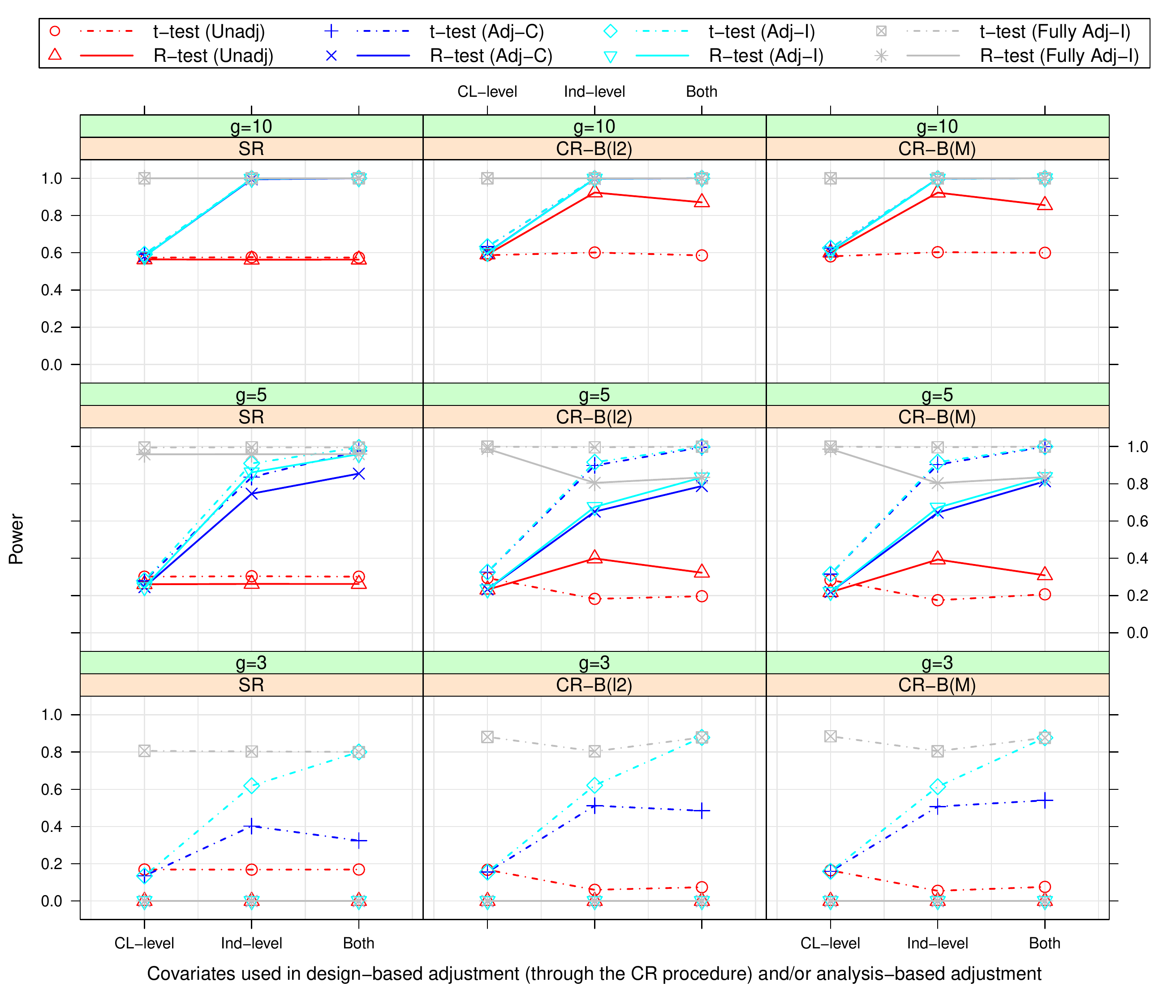}
	\end{center}
	\caption{\label{indi2_power_fig}  Power for the pairwise hypothesis ($\mathcal{H}_{0}\text{: }\delta_{2}=0$) under simple randomization (SR) versus constrained randomization (CR) with 2 balance metrics $B_{(l2)}$ and $B_{(M)}$. CR implemented using covariates indicated on the horizontal axis; candidate set size $=10\%$ under CR; ICC $=0.05$; alpha level $=5\%$; R-test: randomization test; CL-level: cluster-level covariates, $\boldsymbol{x}_j$; Ind-level: individual-level covariates, $\boldsymbol{z}_{jk}$; Unadj: unadjusted test; Adj-C: test adjusted for the covariates on the horizontal axis (with individual-level covariates aggregated at the cluster level); Adj-I: test adjusted for the covariates on the horizontal axis (with actual individual-level covariates); Fully Adj-I: test adjusted for all four covariates (with actual individual-level covariates). 
	}
\end{figure*}

We presented the results for the two pairwise hypotheses ($\mathcal{H}_{0}\text{: }\delta_{1}=0$ and $\mathcal{H}_{0}\text{: }\delta_{2}=0$) in this section. In Web Table~\ref{indi_alpha} and Web Figures~\ref{fig:indi_alpha_fig}-\ref{fig:indi2_alpha_fig}, we  summarized the Monte Carlo type I error rates under simple randomization (SR) and constrained randomization (CR), while in Web Table~\ref{indi_power} and Web Figures~\ref{indi_power_fig}-\ref{indi2_power_fig}, we summarized the corresponding results for power. We held the ICC fixed at $0.05$ throughout this section 
and compared results for $g=$ 3, 5, and 10.

\begin{table}[htbp]
	\caption{\label{indi_alpha}  Type I error rates for the pairwise hypotheses ($\mathcal{H}_{0}\text{: }\delta_{1}=0$ and $\mathcal{H}_{0}\text{: }\delta_{2}=0$) under simple randomization (SR) versus constrained randomization (CR) \textcolor{black}{with candidate set sizes = 50\%, 10\%, and 100 of the randomization space}. All covariates were used in constrained randomization and the adjusted tests; constrained randomization was implemented using the $l2$ metric; ICC $=0.05$; alpha level $=5\%$. The nominal type I error rate is $0.05$, and the acceptance range for nominal type I error rate with $10,000$ replicates is $(0.0457,0.0543)$. } \centering
	\begin{tabular}{p{1.5cm}p{2.0cm}p{2.15cm}  p{0.62cm}p{0.62cm}p{0.62cm}p{0.8cm} p{0.62cm}p{0.62cm}p{0.62cm}p{0.8cm} }
		\toprule
		& & &  \multicolumn{4}{c}{\textbf{$t$-test}} & \multicolumn{4}{c}{\textbf{Randomization test}} \\
		\cmidrule(lr){4-7}\cmidrule(lr){8-11}
		$\mathcal{H}_{0}$ & \# of clusters per arm  & Analysis-based adjustment & SR & CR (50$\%$) & CR (10$\%$) & CR (100) & SR & CR (50$\%$) & CR (10$\%$) & CR (100) \\ 
		\midrule
		$\delta_{1}=0$ &$g=10$ & Unadj & 0.054 & 0.009 & 0.000 & 0.000 & 0.048 & 0.050 & 0.051 & 0.050 \\ 
		& & Adj-C & 0.051 & 0.052 & 0.052 & 0.046 & 0.048 & 0.052 & 0.049 & 0.048 \\ 
		& & Adj-I & 0.053 & 0.052 & 0.051 & 0.047 & 0.050 & 0.051 & 0.050 & 0.049 \\  \addlinespace
		
		& $g=5$ & Unadj & 0.051 & 0.012 & 0.001 & 0.000 & 0.049 & 0.045 & 0.029 & 0.000 \\ 
		& & Adj-C & 0.045 & 0.043 & 0.047 & 0.050 & 0.043 & 0.044 & 0.028 & 0.001 \\ 
		& & Adj-I & 0.046 & 0.050 & 0.049 & 0.050 & 0.046 & 0.041 & 0.028 & 0.000 \\ \addlinespace
		
		& $g=3$ & Unadj & 0.055 & 0.018 & 0.003 & 0.002 & 0.000 & 0.000 & 0.000 & 0.000 \\ 
		& & Adj-C & 0.004 & 0.005 & 0.005 & 0.004 & 0.000 & 0.000 & 0.000 & 0.000 \\ 
		& & Adj-I & 0.051 & 0.050 & 0.045 & 0.050 & 0.000 & 0.000 & 0.000 & 0.000 \\  \addlinespace
		
		$\delta_{2}=0$ & $g=10$ & Unadj & 0.049 & 0.010 & 0.000 & 0.000 & 0.048 & 0.052 & 0.051 & 0.051 \\ 
		& & Adj-C & 0.049 & 0.048 & 0.049 & 0.050 & 0.050 & 0.049 & 0.050 & 0.050 \\ 
		& & Adj-I & 0.049 & 0.049 & 0.047 & 0.049 & 0.050 & 0.052 & 0.049 & 0.051 \\  \addlinespace
		
		& $g=5$ & Unadj & 0.052 & 0.013 & 0.002 & 0.000 & 0.050 & 0.043 & 0.033 & 0.001 \\ 
		& & Adj-C & 0.048 & 0.046 & 0.045 & 0.048 & 0.048 & 0.042 & 0.030 & 0.000 \\ 
		& & Adj-I & 0.051 & 0.048 & 0.046 & 0.051 & 0.047 & 0.043 & 0.028 & 0.000 \\  \addlinespace
		
		& $g=3$ & Unadj & 0.053 & 0.019 & 0.004 & 0.002 & 0.000 & 0.000 & 0.000 & 0.000 \\ 
		& & Adj-C & 0.004 & 0.006 & 0.005 & 0.004 & 0.000 & 0.000 & 0.000 & 0.000 \\ 
		& & Adj-I & 0.045 & 0.048 & 0.049 & 0.045 & 0.000 & 0.000 & 0.000 & 0.000 \\  
		
		\bottomrule
	\end{tabular}
\end{table}

\clearpage 

\clearpage

\clearpage 

\begin{table}[htbp]
	\caption{\label{indi_power}  Power for the pairwise hypotheses ($\mathcal{H}_{0}\text{: }\delta_{1}=0$ and $\mathcal{H}_{0}\text{: }\delta_{2}=0$) under simple randomization (SR) versus constrained randomization (CR) \textcolor{black}{with candidate set sizes = 50\%, 10\%, and 100 of the randomization space}. All covariates were used in constrained randomization and the adjusted tests; constrained randomization was implemented using the $l2$ metric; ICC $=0.05$; alpha level $=5\%$; \textcolor{black}{power values corresponding to non-nominal type I errors are shaded out.} } \centering
	\begin{tabular}{p{1.5cm}p{2.0cm}p{2.15cm}  p{0.62cm}p{0.62cm}p{0.62cm}p{0.8cm} p{0.62cm}p{0.62cm}p{0.62cm}p{0.8cm} }
		\toprule
		& & &  \multicolumn{4}{c}{\textbf{$t$-test}} & \multicolumn{4}{c}{\textbf{Randomization test}} \\
		\cmidrule(lr){4-7}\cmidrule(lr){8-11}
		$\mathcal{H}_{0}$ & \# of clusters per arm  & Analysis-based adjustment & SR & CR (50$\%$) & CR (10$\%$) & CR (100) & SR & CR (50$\%$) & CR (10$\%$) & CR (100) \\ 
		\midrule
		$\delta_{1}=0$ & $g=10$ & Unadj & 0.298 & \textcolor{gray}{0.258} & \textcolor{gray}{0.193} & \textcolor{gray}{0.132} & 0.286 & 0.408 & 0.578 & 0.800 \\ 
		& & Adj-C & 0.998 & 0.999 & 1.000 & 1.000 & 0.994 & 0.998 & 0.999 & 0.999 \\
		& & Adj-I & 0.999 & 0.999 & 1.000 & 1.000 & 0.998 & 0.999 & 0.999 & 0.999 \\ 
		\addlinespace
		
		& $g=5$ & Unadj & 0.162 & \textcolor{gray}{0.108} & \textcolor{gray}{0.060} & \textcolor{gray}{0.018} & 0.143 & 0.175 & \textcolor{gray}{0.188} & \textcolor{gray}{0.004} \\ 
		& & Adj-C & 0.805 & 0.853 & 0.897 & 0.914 & 0.668 & 0.701 & \textcolor{gray}{0.588} & \textcolor{gray}{0.007} \\ 
		& & Adj-I & 0.901 & 0.923 & 0.933 & 0.937 & 0.818 & 0.810 & \textcolor{gray}{0.652} & \textcolor{gray}{0.007} \\ 
		\addlinespace
		
		& $g=3$ & Unadj & 0.108 & \textcolor{gray}{0.064} & \textcolor{gray}{0.027} & \textcolor{gray}{0.024} & \textcolor{gray}{0.000} & \textcolor{gray}{0.000} & \textcolor{gray}{0.000} & \textcolor{gray}{0.000} \\ 
		& & Adj-C & \textcolor{gray}{0.135} & \textcolor{gray}{0.176} & \textcolor{gray}{0.223} & \textcolor{gray}{0.235} & \textcolor{gray}{0.000} & \textcolor{gray}{0.000} & \textcolor{gray}{0.000} & \textcolor{gray}{0.000} \\ 
		& & Adj-I & 0.499 & 0.546 & 0.579 & 0.590 & \textcolor{gray}{0.000} & \textcolor{gray}{0.000} & \textcolor{gray}{0.000} & \textcolor{gray}{0.000} \\ \addlinespace
		
		$\delta_{2}=0$ & $g=10$ & Unadj & 0.574 & \textcolor{gray}{0.578} & \textcolor{gray}{0.586} & \textcolor{gray}{0.606} & 0.563 & 0.712 & 0.871 & 0.982 \\ 
		& & Adj-C & 1.000 & 1.000 & 1.000 & 1.000 & 1.000 & 1.000 & 1.000 & 1.000 \\ 
		& & Adj-I & 1.000 & 1.000 & 1.000 & 1.000 & 1.000 & 1.000 & 1.000 & 1.000 \\
		\addlinespace
		
		& $g=5$ & Unadj & 0.301 & \textcolor{gray}{0.256} & \textcolor{gray}{0.196} & \textcolor{gray}{0.120} & 0.262 & 0.323 & \textcolor{gray}{0.322} & \textcolor{gray}{0.005} \\ 
		& & Adj-C & 0.976 & 0.993 & 0.996 & 0.998 & 0.855 & 0.908 & \textcolor{gray}{0.788} & \textcolor{gray}{0.008} \\ 
		& & Adj-I & 0.995 & 0.999 & 0.999 & 1.000 & 0.959 & 0.970 & \textcolor{gray}{0.834} & \textcolor{gray}{0.008} \\ 
		\addlinespace
		
		& $g=3$ & Unadj & 0.169 & \textcolor{gray}{0.131} & \textcolor{gray}{0.074} & \textcolor{gray}{0.064} & \textcolor{gray}{0.000} & \textcolor{gray}{0.000} & \textcolor{gray}{0.000} & \textcolor{gray}{0.000} \\ 
		& & Adj-C & \textcolor{gray}{0.324} & \textcolor{gray}{0.394} & \textcolor{gray}{0.485} & \textcolor{gray}{0.500} & \textcolor{gray}{0.000} & \textcolor{gray}{0.000} & \textcolor{gray}{0.000} & \textcolor{gray}{0.000} \\ 
		& & Adj-I & 0.800 & 0.848 & 0.878 & 0.882 & \textcolor{gray}{0.000} & \textcolor{gray}{0.000} & \textcolor{gray}{0.000} & \textcolor{gray}{0.000} \\  
		\bottomrule
	\end{tabular}
\end{table}

\clearpage 

\clearpage 

\clearpage 

\section*{Appendix E: Results under different intraclass correlation coefficient}

We presented the results under an intraclass correlation coefficient (ICC) of $0.05$ in the main text. Two alternative ICCs were considered: $0.10$ and $0.01$, and presented the results under alternative ICCs in this section. We avoided the comparison of results across different ICCs. Instead, we compared the performance of the design-based and analysis-based adjustment strategies within each level of ICC. In Web Tables~\ref{alpha_10}-\ref{power_10}, we summarized the type I error rate and power under ICC $=0.10$, while in Web Tables~\ref{alpha_01}-\ref{power_01}, we summarized the results under ICC $=0.01$. The number of clusters per arm $g$ is held at 5 and the alpha level is held at 5\% throughout the comparison in this section. As a counterpart of the $\chi^2$-test, the $z$-test for the pairwise hypotheses ($\mathcal{H}_{0}\text{: }\delta_{1}=0$ and $\mathcal{H}_{0}\text{: }\delta_{2}=0$) was not considered further in our simulations because it will carry inflated type I error rate.

\begin{table}[htbp]
	\caption{\label{alpha_10}  Results under ICC $=0.10$: Type I error rates under simple randomization (SR) versus constrained randomization (CR) \textcolor{black}{with candidate set sizes = 50\%, 10\%, and 100 of the randomization space}. All covariates were used in constrained randomization and the adjusted tests; constrained randomization was implemented using the $l2$ metric; $g=5$. The nominal type I error rate is $0.05$, and the acceptable range for nominal type I error rate with $10,000$ replicates is $(0.0457,0.0543)$.} \centering
	\begin{tabular}{p{1.5cm}p{2.15cm}  p{0.62cm}p{0.62cm}p{0.62cm}p{0.8cm} p{0.62cm}p{0.62cm}p{0.62cm}p{0.8cm} p{0.62cm}p{0.62cm}p{0.62cm}p{0.62cm} }
		\toprule
		& &  \multicolumn{4}{c}{\textbf{$\chi^2$-test}} & \multicolumn{4}{c}{\textbf{$F$-test/$t$-test}} & \multicolumn{4}{c}{\textbf{Randomization test}} \\
		\cmidrule(lr){3-6}\cmidrule(lr){7-10}\cmidrule(lr){11-14}
		$\mathcal{H}_{0}$ & Analysis-based adjustment & SR & CR (50$\%$) & CR (10$\%$) & CR (100) & SR & CR (50$\%$) & CR (10$\%$) & CR (100) & SR & CR (50$\%$) & CR (10$\%$) & CR (100) \\ 
		\hline
		$\delta_{1}$=$\delta_{2}$=0 & Unadj & 0.090 & 0.025 & 0.004 & 0.000 & 0.053 & 0.010 & 0.001 & 0.000 & 0.052 & 0.051 & 0.051 & 0.043 \\ 
		&  Adj-C & 0.112 & 0.108 & 0.106 & 0.107 & 0.053 & 0.052 & 0.050 & 0.050 & 0.053 & 0.050 & 0.050 & 0.041 \\ 
		&  Adj-I & 0.100 & 0.099 & 0.095 & 0.095 & 0.054 & 0.052 & 0.049 & 0.050 & 0.054 & 0.050 & 0.048 & 0.040 \\ 
		\addlinespace
		
		$\delta_{1}$=0 & Unadj & --- & --- & --- & --- & 0.052 & 0.014 & 0.002 & 0.000 & 0.046 & 0.039 & 0.035 & 0.000 \\ 
		&Adj-C & --- & --- & --- & --- & 0.053 & 0.052 & 0.051 & 0.050 & 0.047 & 0.042 & 0.030 & 0.000 \\ 
		& Adj-I & --- & --- & --- & --- & 0.053 & 0.053 & 0.049 & 0.050 & 0.050 & 0.041 & 0.031 & 0.000 \\  
		\addlinespace
		
		\bottomrule
	\end{tabular}
	\begin{flushleft}
	\end{flushleft}
\end{table}

\begin{table}[htbp]
	\caption{\label{power_10} Results under ICC $=0.10$: Power under simple randomization (SR) versus constrained randomization (CR) \textcolor{black}{with candidate set sizes = 50\%, 10\%, and 100 of the randomization space}. All covariates were used in constrained randomization and the adjusted tests; constrained randomization was implemented using the $l2$ metric; $g=5$; \textcolor{black}{power values corresponding to non-nominal type I errors are shaded out.} } \centering
	\begin{tabular}{p{1.5cm}p{2.15cm}  p{0.62cm}p{0.62cm}p{0.62cm}p{0.8cm} p{0.62cm}p{0.62cm}p{0.62cm}p{0.8cm} p{0.62cm}p{0.62cm}p{0.62cm}p{0.62cm} }
		\toprule
		& &  \multicolumn{4}{c}{\textbf{$\chi^2$-test}} & \multicolumn{4}{c}{\textbf{$F$-test/$t$-test}} & \multicolumn{4}{c}{\textbf{Randomization test}} \\
		\cmidrule(lr){3-6}\cmidrule(lr){7-10}\cmidrule(lr){11-14}
		$\mathcal{H}_{0}$ & Analysis-based adjustment & SR & CR (50$\%$) & CR (10$\%$) & CR (100) & SR & CR (50$\%$) & CR (10$\%$) & CR (100) & SR & CR (50$\%$) & CR (10$\%$) & CR (100) \\ 
		\hline
		$\delta_{1}$=$\delta_{2}$=0 & Unadj & \textcolor{gray}{0.341} & \textcolor{gray}{0.273} & \textcolor{gray}{0.201} & \textcolor{gray}{0.123} & 0.240 & \textcolor{gray}{0.175} & \textcolor{gray}{0.111} & \textcolor{gray}{0.056} & 0.239 & 0.327 & 0.428 & 0.529  \\ 
		& Adj-C & \textcolor{gray}{0.914} & \textcolor{gray}{0.945} & \textcolor{gray}{0.964} & \textcolor{gray}{0.978} & 0.814 & 0.858 & 0.899 & 0.923 & 0.719 & 0.804 & 0.870 & 0.835 \\ 
		& Adj-I & \textcolor{gray}{0.961} & \textcolor{gray}{0.972} & \textcolor{gray}{0.978} & \textcolor{gray}{0.981} & 0.914 & 0.928 & 0.941 & 0.951 & 0.869 & 0.903 & 0.927 & 0.863\\  
		\addlinespace
		
		$\delta_{1}$=0 & Unadj & --- & --- & --- & --- & 0.171 & \textcolor{gray}{0.125} & \textcolor{gray}{0.077} & \textcolor{gray}{0.036} & 0.152 & \textcolor{gray}{0.184} & \textcolor{gray}{0.184} & \textcolor{gray}{0.003} \\ 
		& Adj-C & --- & --- & --- & --- &  0.604 & 0.650 & 0.699 & 0.737 & 0.500 & \textcolor{gray}{0.508} & \textcolor{gray}{0.423} & \textcolor{gray}{0.005} \\ 
		& Adj-I & --- & --- & --- & --- & 0.718 & 0.742 & 0.758 & 0.776 & 0.631 & \textcolor{gray}{0.615} & \textcolor{gray}{0.483} & \textcolor{gray}{0.006} \\  
		\addlinespace
		
		\bottomrule
	\end{tabular}
\end{table}

\clearpage

\begin{table}[htbp]
	\caption{\label{alpha_01} Results under ICC $=0.01$: Type I error rates under simple randomization (SR) versus constrained randomization (CR) \textcolor{black}{with candidate set sizes = 50\%, 10\%, and 100 of the randomization space}. All covariates were used in constrained randomization and the adjusted tests; constrained randomization was implemented using the $l2$ metric; $g=5$. The nominal type I error rate is $0.05$, and the acceptable range for nominal type I error rate with $10,000$ replicates is $(0.0457,0.0543)$.} \centering
	\begin{tabular}{p{1.5cm}p{2.15cm}  p{0.62cm}p{0.62cm}p{0.62cm}p{0.8cm} p{0.62cm}p{0.62cm}p{0.62cm}p{0.8cm} p{0.62cm}p{0.62cm}p{0.62cm}p{0.62cm} }
		\toprule
		& &  \multicolumn{4}{c}{\textbf{$\chi^2$-test}} & \multicolumn{4}{c}{\textbf{$F$-test/$t$-test}} & \multicolumn{4}{c}{\textbf{Randomization test}} \\
		\cmidrule(lr){3-6}\cmidrule(lr){7-10}\cmidrule(lr){11-14}
		$\mathcal{H}_{0}$ & Analysis-based adjustment & SR & CR (50$\%$) & CR (10$\%$) & CR (100) & SR & CR (50$\%$) & CR (10$\%$) & CR (100) & SR & CR (50$\%$) & CR (10$\%$) & CR (100) \\ 
		\hline
		$\delta_{1}$=$\delta_{2}$=0 & Unadj & 0.093 & 0.019 & 0.001 & 0.000 & 0.056 & 0.007 & 0.000 & 0.000 & 0.055 & 0.053 & 0.049 & 0.038 \\ 
		& Adj-C & 0.024 & 0.023 & 0.024 & 0.022 & 0.004 & 0.003 & 0.004 & 0.003 & 0.048 & 0.046 & 0.051 & 0.040 \\ 
		& Adj-I & 0.086 & 0.087 & 0.090 & 0.096 & 0.041 & 0.044 & 0.043 & 0.050 & 0.045 & 0.050 & 0.048 & 0.042 \\ \addlinespace
		
		$\delta_{1}$=0 & Unadj & --- & --- & --- & --- & 0.054 & 0.012 & 0.001 & 0.000 & 0.050 & 0.046 & 0.030 & 0.000 \\ 
		& Adj-C& --- & --- & --- & --- &  0.010 & 0.010 & 0.011 & 0.007 & 0.046 & 0.047 & 0.032 & 0.001 \\ 
		& Adj-I & --- & --- & --- & --- &  0.042 & 0.049 & 0.049 & 0.050 & 0.042 & 0.043 & 0.032 & 0.000 \\  
		\addlinespace
		
		\bottomrule
	\end{tabular}
\end{table}

\begin{table}[htbp]
	\caption{\label{power_01} Results under ICC $=0.01$: Power under simple randomization (SR) versus constrained randomization (CR) \textcolor{black}{with candidate set sizes = 50\%, 10\%, and 100 of the randomization space}. All covariates were used in constrained randomization and the adjusted tests; constrained randomization was implemented using the $l2$ metric; $g=5$; \textcolor{black}{power values corresponding to non-nominal type I errors are shaded out.} } \centering
	\begin{tabular}{p{1.5cm}p{2.15cm}  p{0.62cm}p{0.62cm}p{0.62cm}p{0.8cm} p{0.62cm}p{0.62cm}p{0.62cm}p{0.8cm} p{0.62cm}p{0.62cm}p{0.62cm}p{0.62cm} }
		\toprule
		& &  \multicolumn{4}{c}{\textbf{$\chi^2$-test}} & \multicolumn{4}{c}{\textbf{$F$-test/$t$-test}} & \multicolumn{4}{c}{\textbf{Randomization test}} \\
		\cmidrule(lr){3-6}\cmidrule(lr){7-10}\cmidrule(lr){11-14}
		$\mathcal{H}_{0}$ & Analysis-based adjustment & SR & CR (50$\%$) & CR (10$\%$) & CR (100) & SR & CR (50$\%$) & CR (10$\%$) & CR (100) & SR & CR (50$\%$) & CR (10$\%$) & CR (100) \\ 
		\hline
		$\delta_{1}$=$\delta_{2}$=0 & Unadj & \textcolor{gray}{0.283} & \textcolor{gray}{0.209} & \textcolor{gray}{0.111} & \textcolor{gray}{0.035} & 0.192 & \textcolor{gray}{0.119} & \textcolor{gray}{0.052} & \textcolor{gray}{0.011} & 0.188 & 0.293 & 0.409 & \textcolor{gray}{0.571} \\ 
		& Adj-C & \textcolor{gray}{1.000} & \textcolor{gray}{1.000} & \textcolor{gray}{1.000} & \textcolor{gray}{1.000} & \textcolor{gray}{0.999} & \textcolor{gray}{1.000} & \textcolor{gray}{1.000} & \textcolor{gray}{1.000} & 0.941 & 0.989 & 0.999 & \textcolor{gray}{0.999} \\ 
		& Adj-I & \textcolor{gray}{1.000} & \textcolor{gray}{1.000} & \textcolor{gray}{1.000} & \textcolor{gray}{1.000} & \textcolor{gray}{1.000} & \textcolor{gray}{1.000} & \textcolor{gray}{1.000} & 1.000 & 0.992 & 1.000 & 1.000 & \textcolor{gray}{1.000} \\   
		\addlinespace
		
		$\delta_{1}$=0 & Unadj & --- & --- & --- & --- & 0.143 & \textcolor{gray}{0.100} & \textcolor{gray}{0.041} & \textcolor{gray}{0.007} & 0.125 & 0.171 & \textcolor{gray}{0.182} & \textcolor{gray}{0.003} \\ 
		& Adj-C & --- & --- & --- & --- & \textcolor{gray}{0.985} & \textcolor{gray}{0.997} & \textcolor{gray}{0.999} & \textcolor{gray}{1.000} & 0.887 & 0.936 & \textcolor{gray}{0.823} & \textcolor{gray}{0.008} \\ 
		& Adj-I & --- & --- & --- & --- & \textcolor{gray}{0.999} & 1.000 & 1.000 & 1.000 & \textcolor{gray}{0.974} & \textcolor{gray}{0.976} & \textcolor{gray}{0.853} & \textcolor{gray}{0.008}  \\   
		\addlinespace
		
		\bottomrule
	\end{tabular}
\end{table}

\clearpage

\section*{Appendix F: Results under multiplicity adjustment}

\begin{figure*}[htbp]
	\begin{center}
		\includegraphics[trim=5 15 5 10, clip, width=0.85\linewidth]{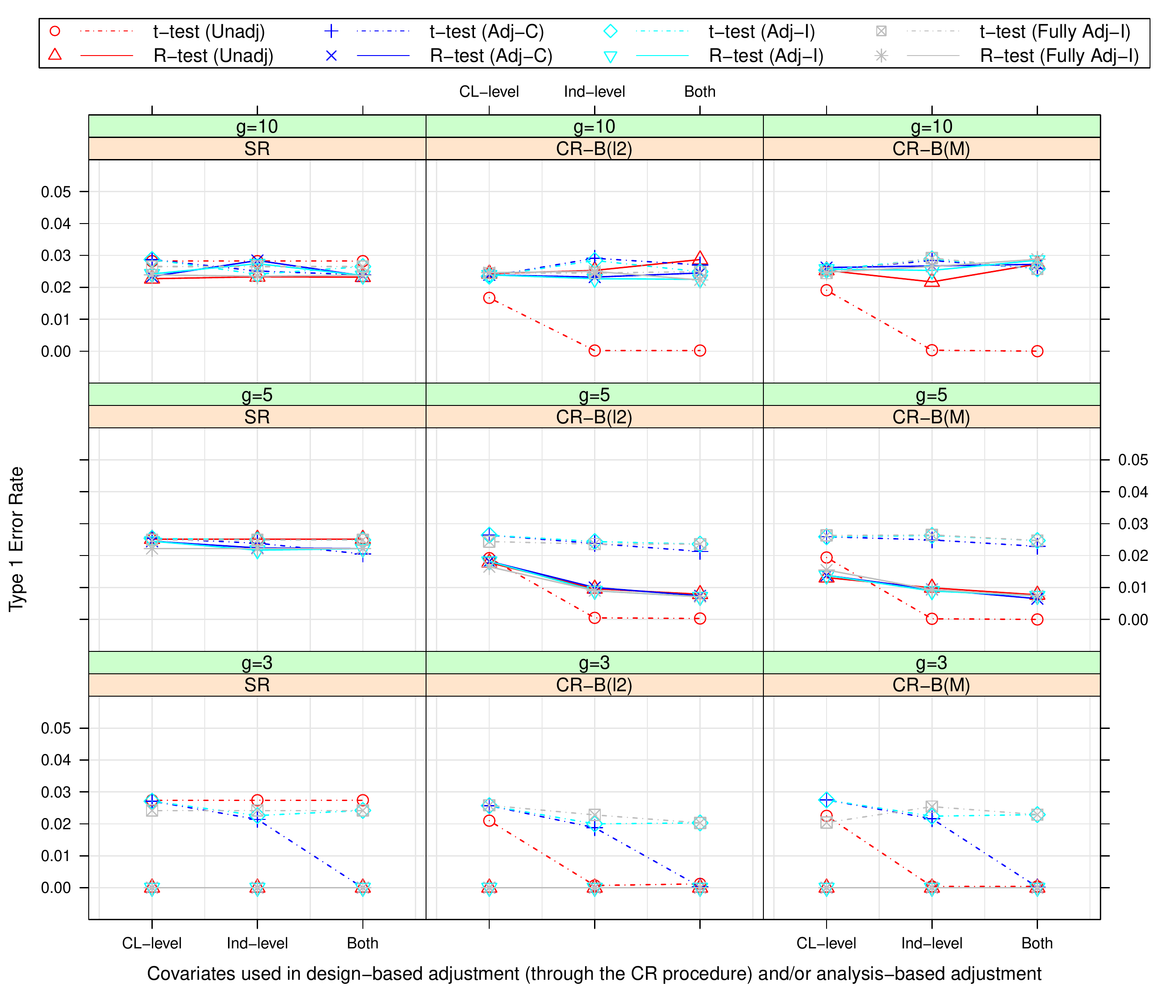}
	\end{center}
	\caption{\label{indi_alpha_FWER_fig} Results under  multiplicity adjustment: type I error rates for the pairwise hypothesis ($\mathcal{H}_{0}\text{: }\delta_{1}=0$) under simple randomization (SR) versus constrained randomization (CR) with 2 balance metrics $B_{(l2)}$ and $B_{(M)}$. CR implemented using covariates indicated on the horizontal axis; candidate set size $=10\%$ under CR; ICC $=0.05$; alpha level $=2.5\%$; R-test: randomization test; CL-level: cluster-level covariates, $\boldsymbol{x}_j$; Ind-level: individual-level covariates, $\boldsymbol{z}_{jk}$; Unadj: unadjusted test; Adj-C: test adjusted for the covariates on the horizontal axis (with individual-level covariates aggregated at the cluster level); Adj-I: test adjusted for the covariates on the horizontal axis (with actual individual-level covariates); Fully Adj-I: test adjusted for all four covariates (with actual individual-level covariates).
	}
\end{figure*}

\begin{figure*}[htbp]
	\begin{center}
		\includegraphics[trim=5 15 5 10, clip, width=0.85\linewidth]{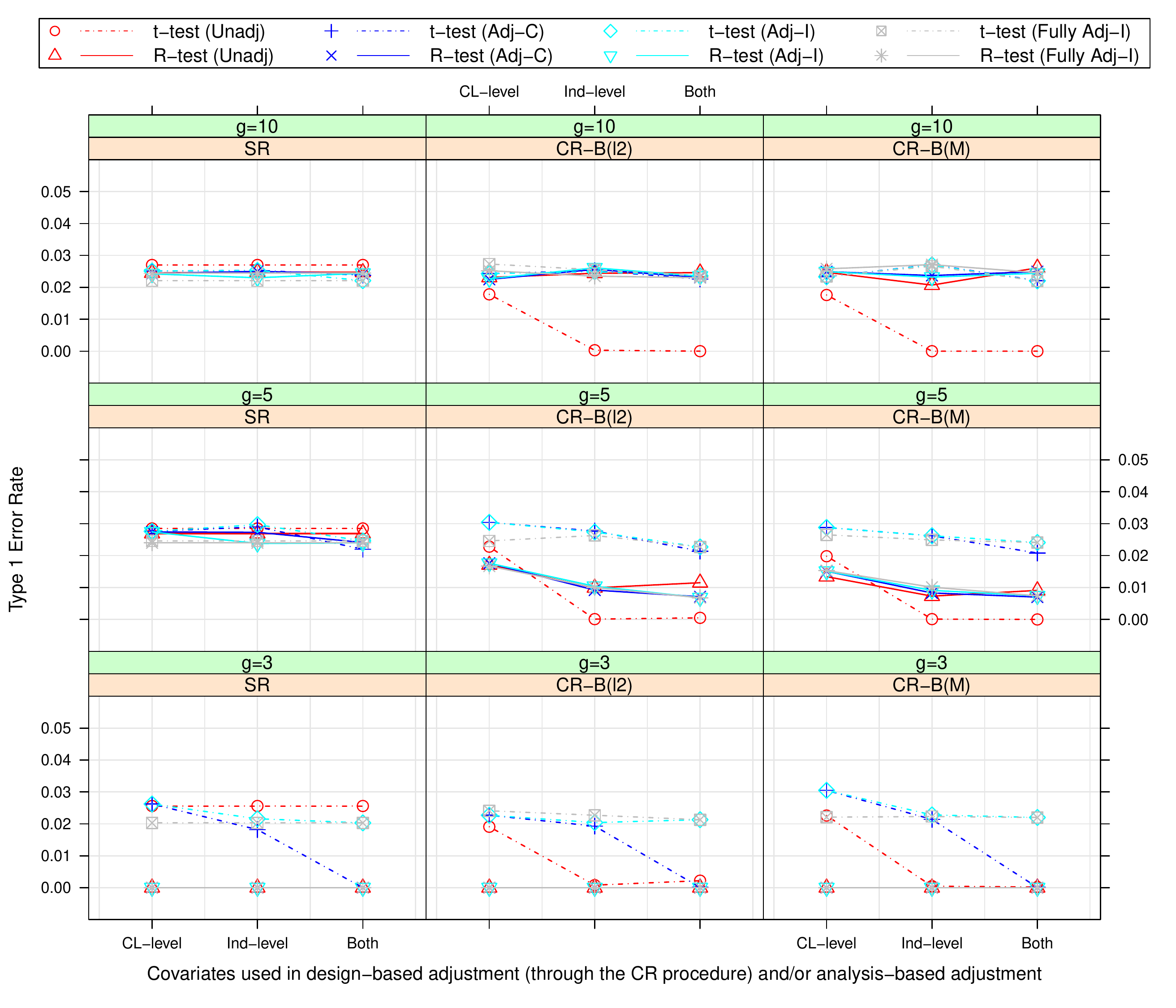}
	\end{center}
	\caption{\label{indi2_alpha_FWER_fig} Results under multiplicity adjustment: type I error rates for the pairwise hypothesis ($\mathcal{H}_{0}\text{: }\delta_{2}=0$) under simple randomization (SR) versus constrained randomization (CR) with 2 balance metrics $B_{(l2)}$ and $B_{(M)}$. CR implemented using covariates indicated on the horizontal axis; candidate set size $=10\%$ under CR; ICC $=0.05$; alpha level $=2.5\%$; R-test: randomization test; CL-level: cluster-level covariates, $\boldsymbol{x}_j$; Ind-level: individual-level covariates, $\boldsymbol{z}_{jk}$; Unadj: unadjusted test; Adj-C: test adjusted for the covariates on the horizontal axis (with individual-level covariates aggregated at the cluster level); Adj-I: test adjusted for the covariates on the horizontal axis (with actual individual-level covariates); Fully Adj-I: test adjusted for all four covariates (with actual individual-level covariates).
	}
\end{figure*}

\begin{figure*}[htbp]
	\begin{center}
		\includegraphics[trim=5 15 5 10, clip, width=0.85\linewidth]{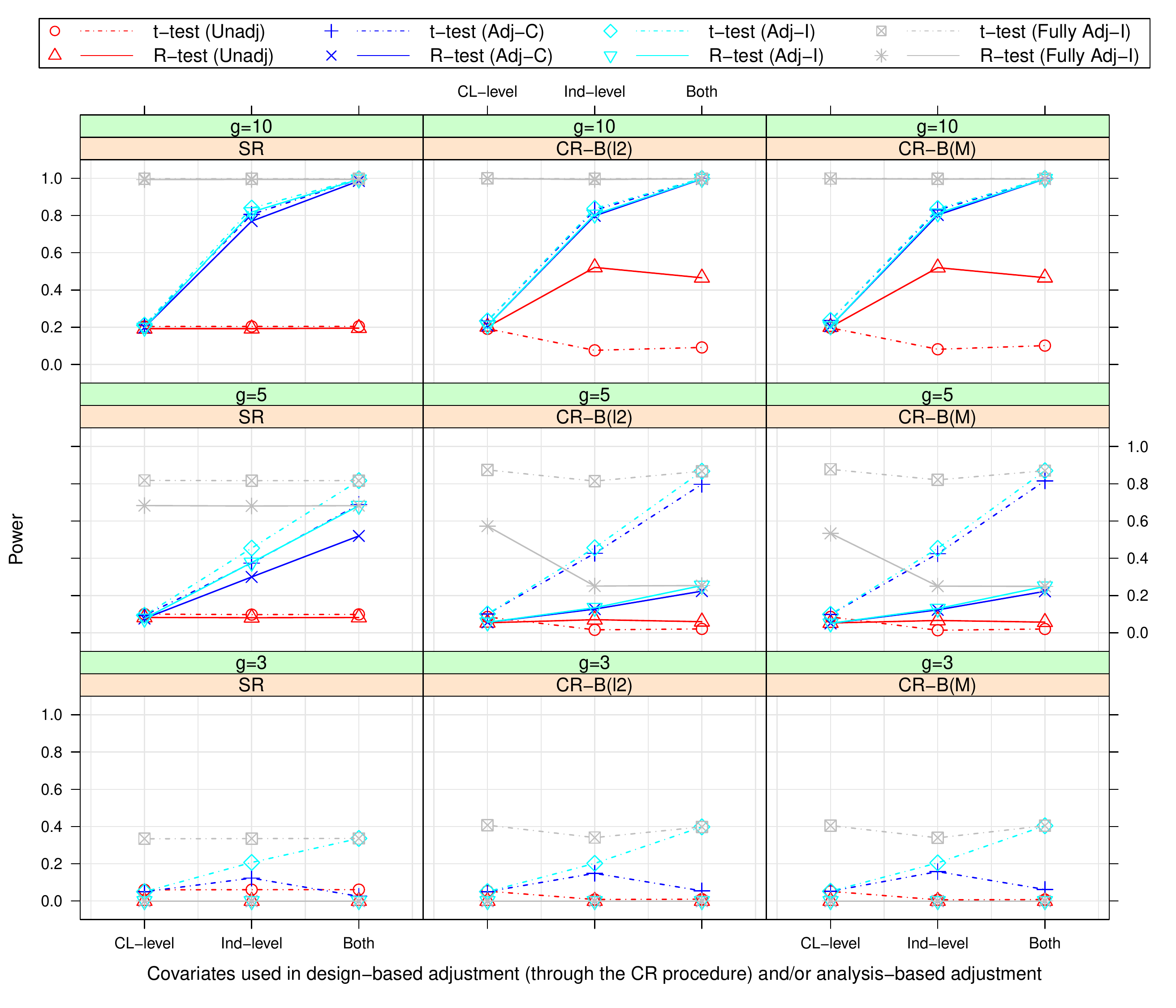}
	\end{center}
	\caption{\label{indi_power_FWER_fig} Results under  multiplicity adjustment: power for the pairwise hypothesis ($\mathcal{H}_{0}\text{: }\delta_{1}=0$) under simple randomization (SR) versus constrained randomization (CR) with 2 balance metrics $B_{(l2)}$ and $B_{(M)}$. CR implemented using covariates indicated on the horizontal axis; candidate set size $=10\%$ under CR; ICC $=0.05$; alpha level $=2.5\%$; R-test: randomization test; CL-level: cluster-level covariates, $\boldsymbol{x}_j$; Ind-level: individual-level covariates, $\boldsymbol{z}_{jk}$; Unadj: unadjusted test; Adj-C: test adjusted for the covariates on the horizontal axis (with individual-level covariates aggregated at the cluster level); Adj-I: test adjusted for the covariates on the horizontal axis (with actual individual-level covariates); Fully Adj-I: test adjusted for all four covariates (with actual individual-level covariates).
	}
\end{figure*}

\begin{figure*}[htbp]
	\begin{center}
		\includegraphics[trim=5 15 5 10, clip, width=0.85\linewidth]{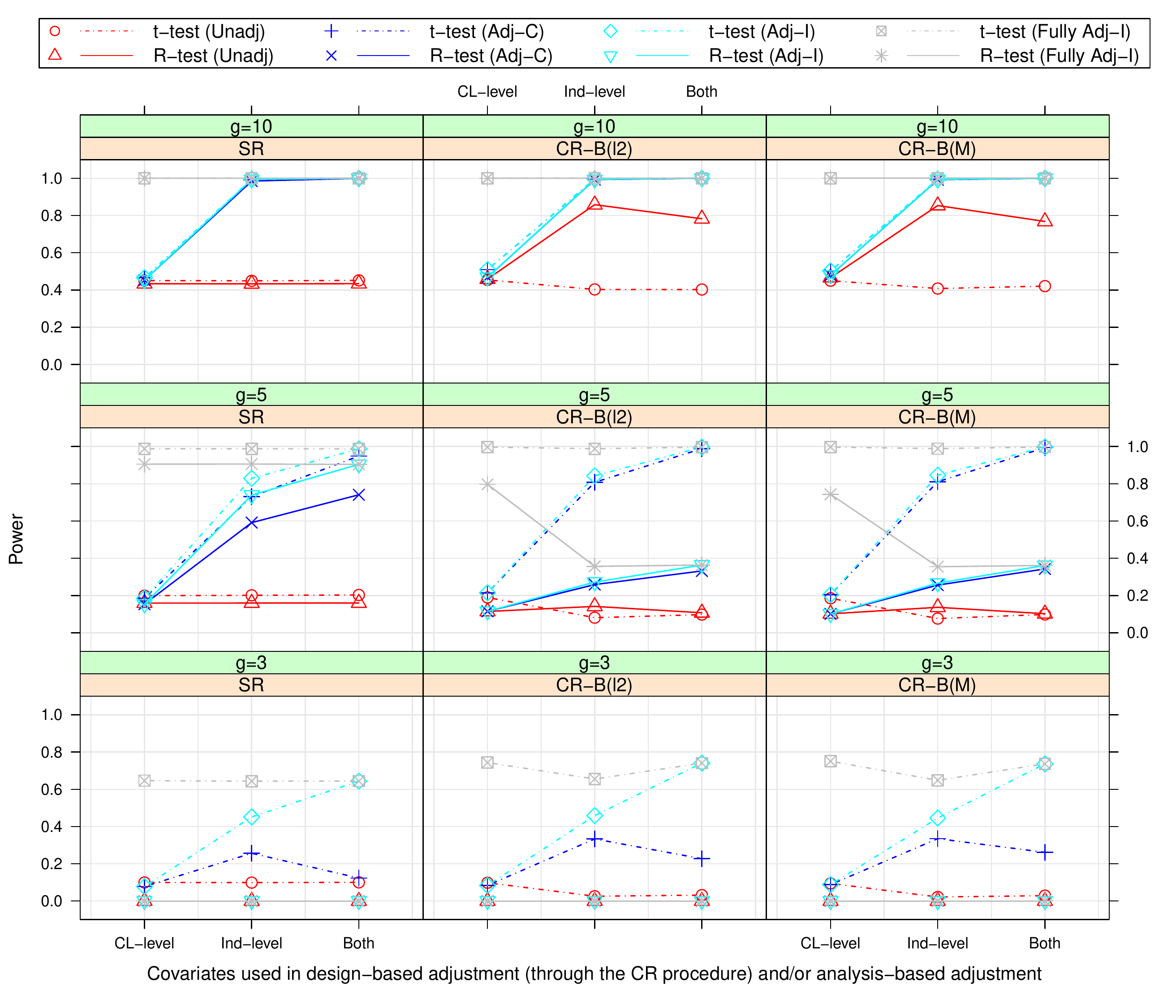}
	\end{center}
	\caption{\label{indi2_power_FWER_fig} Results under  multiplicity adjustment: power for the pairwise hypothesis ($\mathcal{H}_{0}\text{: }\delta_{2}=0$) under simple randomization (SR) versus constrained randomization (CR) with 2 balance metrics $B_{(l2)}$ and $B_{(M)}$. CR implemented using covariates indicated on the horizontal axis; candidate set size $=10\%$ under CR; ICC $=0.05$; alpha level $=2.5\%$; R-test: randomization test; CL-level: cluster-level covariates, $\boldsymbol{x}_j$; Ind-level: individual-level covariates, $\boldsymbol{z}_{jk}$; Unadj: unadjusted test; Adj-C: test adjusted for the covariates on the horizontal axis (with individual-level covariates aggregated at the cluster level); Adj-I: test adjusted for the covariates on the horizontal axis (with actual individual-level covariates); Fully Adj-I: test adjusted for all four covariates (with actual individual-level covariates).
	}
\end{figure*}

We performed a conservative Bonferroni adjustment \cite{aickin1996adjusting} for the tests of the two pairwise hypotheses, which is suggested for the scenario where the two active treatment arms consist of the same treatment given at different doses and the overall effectiveness will be concluded if any one of the treatment doses shows a significant effect relative to ‘standard of care’. In Web Table~\ref{alpha_FWER} and Web Figures~\ref{indi_alpha_FWER_fig}-\ref{indi2_alpha_FWER_fig}, we summarized the Monte Carlo type I error rates for the pairwise hypothesis ($\mathcal{H}_{0}\text{: }\delta_{1}=0$ and $\mathcal{H}_{0}\text{: }\delta_{2}=0$) under simple randomization (SR) and constrained randomization (CR). In Web Table~\ref{power_FWER} and Web Figures~\ref{indi_power_FWER_fig}-\ref{indi2_power_FWER_fig}, we summarized the results for power. Alpha level is held at 2.5\% throughout the comparison in this section.

\begin{table}[htbp]
	\caption{\label{alpha_FWER} Results under multiplicity adjustment: Type I error rates for the pairwise hypotheses ($\mathcal{H}_{0}\text{: }\delta_{1}=0$ and $\mathcal{H}_{0}\text{: }\delta_{2}=0$) under simple randomization (SR) versus constrained randomization (CR) \textcolor{black}{with candidate set sizes = 50\%, 10\%, and 100 of the randomization space}. All covariates were used in constrained randomization and the adjusted tests; constrained randomization was implemented using the $l2$ metric; ICC $=0.05$; alpha level $=2.5\%$. The nominal type I error rate is $0.025$, and the acceptance range for nominal type I error rate with $10,000$ replicates is $(0.0215,0.0285)$. } \centering
	\begin{tabular}{p{1.5cm}p{2.0cm}p{2.15cm}  p{0.62cm}p{0.62cm}p{0.62cm}p{0.8cm} p{0.62cm}p{0.62cm}p{0.62cm}p{0.8cm} }
		\toprule
		& & &  \multicolumn{4}{c}{\textbf{$t$-test}} & \multicolumn{4}{c}{\textbf{Randomization test}} \\
		\cmidrule(lr){4-7}\cmidrule(lr){8-11}
		$\mathcal{H}_{0}$ & \# of clusters per arm  & Analysis-based adjustment & SR & CR (50$\%$) & CR (10$\%$) & CR (100) & SR & CR (50$\%$) & CR (10$\%$) & CR (100) \\ 
		\hline
		$\delta_{1}=0$ & $g=10$ & Unadj & 0.028 & 0.003 & 0.000 & 0.000 & 0.023 & 0.026 & 0.029 & 0.024 \\ 
		& & Adj-C & 0.024 & 0.027 & 0.027 & 0.024 & 0.024 & 0.024 & 0.024 & 0.024  \\ 
		& & Adj-I & 0.026 & 0.027 & 0.025 & 0.024 & 0.024 & 0.026 & 0.022 & 0.025 \\ \addlinespace
		
		& $g=5$ & Unadj & 0.025 & 0.004 & 0.000 & 0.000 & 0.025 & 0.021 & 0.008 & 0.000 \\ 
		& & Adj-C & 0.020 & 0.021 & 0.021 & 0.022 & 0.022 & 0.022 & 0.007 & 0.000 \\ 
		& & Adj-I & 0.025 & 0.024 & 0.024 & 0.026 & 0.022 & 0.018 & 0.007 & 0.000 \\ \addlinespace
		
		& $g=3$ & Unadj & 0.027 & 0.006 & 0.001 & 0.001 & 0.000 & 0.000 & 0.000 & 0.000 \\ 
		& & Adj-C & 0.000 & 0.000 & 0.000 & 0.000 & 0.000 & 0.000 & 0.000 & 0.000 \\ 
		& & Adj-I & 0.024 & 0.024 & 0.020 & 0.026 & 0.000 & 0.000 & 0.000 & 0.000 \\  \addlinespace
		
		$\delta_{2}=0$ & $g=10$ & Unadj & 0.027 & 0.004 & 0.000 & 0.000 & 0.025 & 0.025 & 0.025 & 0.027 \\ 
		& & Adj-C & 0.024 & 0.023 & 0.023 & 0.025 & 0.024 & 0.023 & 0.023 & 0.024  \\ 
		& & Adj-I & 0.022 & 0.024 & 0.024 & 0.025 & 0.024 & 0.024 & 0.024 & 0.024 \\ \addlinespace
		
		& $g=5$ & Unadj & 0.028 & 0.005 & 0.000 & 0.000 & 0.027 & 0.020 & 0.011 & 0.000 \\ 
		& & Adj-C & 0.022 & 0.022 & 0.021 & 0.025 & 0.024 & 0.019 & 0.007 & 0.000 \\ 
		& & Adj-I & 0.025 & 0.026 & 0.023 & 0.027 & 0.024 & 0.018 & 0.007 & 0.000  \\ \addlinespace
		
		& $g=3$ & Unadj & 0.026 & 0.008 & 0.002 & 0.001 & 0.000 & 0.000 & 0.000 & 0.000 \\ 
		& & Adj-C & 0.000 & 0.000 & 0.000 & 0.000 & 0.000 & 0.000 & 0.000 & 0.000 \\ 
		& & Adj-I & 0.020 & 0.024 & 0.021 & 0.022 & 0.000 & 0.000 & 0.000 & 0.000 \\

		\bottomrule
	\end{tabular}
	\begin{flushleft}
	\end{flushleft}
\end{table}

\clearpage

\clearpage 

\clearpage 

\begin{table}[htbp]
	\caption{\label{power_FWER} Results under multiplicity adjustment: Power for the pairwise hypotheses ($\mathcal{H}_{0}\text{: }\delta_{1}=0$ and $\mathcal{H}_{0}\text{: }\delta_{2}=0$) under simple randomization (SR) versus constrained randomization (CR) \textcolor{black}{with candidate set sizes = 50\%, 10\%, and 100 of the randomization space}. All covariates were used in constrained randomization and the adjusted tests; constrained randomization was implemented using the $l2$ metric; ICC $=0.05$; alpha level $=2.5\%$; \textcolor{black}{power values corresponding to non-nominal type I errors are shaded out.} } \centering
	\begin{tabular}{p{1.5cm}p{2.0cm}p{2.15cm}  p{0.62cm}p{0.62cm}p{0.62cm}p{0.8cm} p{0.62cm}p{0.62cm}p{0.62cm}p{0.8cm} }
		\toprule
		& & &  \multicolumn{4}{c}{\textbf{$t$-test}} & \multicolumn{4}{c}{\textbf{Randomization test}} \\
		\cmidrule(lr){4-7}\cmidrule(lr){8-11}
		$\mathcal{H}_{0}$ & \# of clusters per arm  & Analysis-based adjustment & SR & CR (50$\%$) & CR (10$\%$) & CR (100) & SR & CR (50$\%$) & CR (10$\%$) & CR (100) \\ 
		\hline
		$\delta_{1}=0$ & $g=10$ & Unadj & 0.204 & \textcolor{gray}{0.155} & \textcolor{gray}{0.091} & \textcolor{gray}{0.042} & 0.195 & 0.313 & 0.466 & 0.699  \\ 
		& & Adj-C & 0.994 & 0.998 & 0.998 & 0.999 & 0.986 & 0.994 & 0.996 & 0.997 \\ 
		& & Adj-I & 0.997 & 0.998 & 0.999 & 0.999 & 0.994 & 0.997 & 0.998 & 0.998 \\  \addlinespace
		
		& $g=5$ & Unadj & 0.098 & \textcolor{gray}{0.055} & \textcolor{gray}{0.021} & \textcolor{gray}{0.005} & 0.082 & \textcolor{gray}{0.097} & \textcolor{gray}{0.059} & \textcolor{gray}{0.000} \\ 
		& & Adj-C & \textcolor{gray}{0.689} & \textcolor{gray}{0.748} & \textcolor{gray}{0.797} & 0.828 & 0.520 & 0.497 & \textcolor{gray}{0.223} & \textcolor{gray}{0.000} \\ 
		& & Adj-I & 0.817 & 0.845 & 0.867 & 0.876 & 0.682 & \textcolor{gray}{0.621} & \textcolor{gray}{0.253} & \textcolor{gray}{0.000} \\  \addlinespace
		
		& $g=3$ & Unadj & 0.061 & \textcolor{gray}{0.033} & \textcolor{gray}{0.010} & \textcolor{gray}{0.010} & \textcolor{gray}{0.000} & \textcolor{gray}{0.000} & \textcolor{gray}{0.000} & \textcolor{gray}{0.000} \\ 
		& & Adj-C & \textcolor{gray}{0.025} & \textcolor{gray}{0.040} & \textcolor{gray}{0.054} & \textcolor{gray}{0.059} & \textcolor{gray}{0.000} & \textcolor{gray}{0.000} & \textcolor{gray}{0.000} & \textcolor{gray}{0.000} \\ 
		& & Adj-I & 0.336 & 0.378 & \textcolor{gray}{0.398} & 0.409 & \textcolor{gray}{0.000} & \textcolor{gray}{0.000} & \textcolor{gray}{0.000} & \textcolor{gray}{0.000} \\  \addlinespace
		
		$\delta_{2}=0$ & $g=10$ & Unadj & 0.452 & \textcolor{gray}{0.436} & \textcolor{gray}{0.403} & \textcolor{gray}{0.378} & 0.434 & 0.605 & 0.782 & 0.949  \\ 
		& & Adj-C & 1.000 & 1.000 & 1.000 & 1.000 & 1.000 & 1.000 & 1.000 & 1.000 \\ 
		& & Adj-I & 1.000 & 1.000 & 1.000 & 1.000 & 1.000 & 1.000 & 1.000 & 1.000 \\  \addlinespace
		
		& $g=5$ & Unadj & 0.203 & \textcolor{gray}{0.151} & \textcolor{gray}{0.098} & \textcolor{gray}{0.044} & 0.160 & \textcolor{gray}{0.195} & \textcolor{gray}{0.107} & \textcolor{gray}{0.000} \\ 
		& & Adj-C & 0.949 & 0.978 & 0.989 & 0.995 & 0.741 & \textcolor{gray}{0.747} & \textcolor{gray}{0.332} & \textcolor{gray}{0.000} \\ 
		& & Adj-I & 0.987 & 0.995 & 0.997 & 0.998 & 0.904 & \textcolor{gray}{0.873} & \textcolor{gray}{0.364} & \textcolor{gray}{0.000} \\  \addlinespace
		
		& $g=3$ & Unadj & 0.100 & \textcolor{gray}{0.066} & \textcolor{gray}{0.032} & \textcolor{gray}{0.028} & \textcolor{gray}{0.000} & \textcolor{gray}{0.000} & \textcolor{gray}{0.000} & \textcolor{gray}{0.000} \\ 
		& & Adj-C & \textcolor{gray}{0.123} & \textcolor{gray}{0.164} & \textcolor{gray}{0.227} & \textcolor{gray}{0.239} & \textcolor{gray}{0.000} & \textcolor{gray}{0.000} & \textcolor{gray}{0.000} & \textcolor{gray}{0.000} \\ 
		& & Adj-I & \textcolor{gray}{0.644} & 0.701 & 0.741 & 0.744 & \textcolor{gray}{0.000} & \textcolor{gray}{0.000} & \textcolor{gray}{0.000} & \textcolor{gray}{0.000} \\  
		\bottomrule
	\end{tabular}
	\begin{flushleft}
	\end{flushleft}
\end{table}

\clearpage 

\clearpage 

\clearpage 

\section*{Appendix G: Additional simulation results under reduced effect sizes}

\begin{figure*}[htbp]
	\begin{center}
		\includegraphics[trim=5 15 5 10, clip, width=0.85\linewidth]{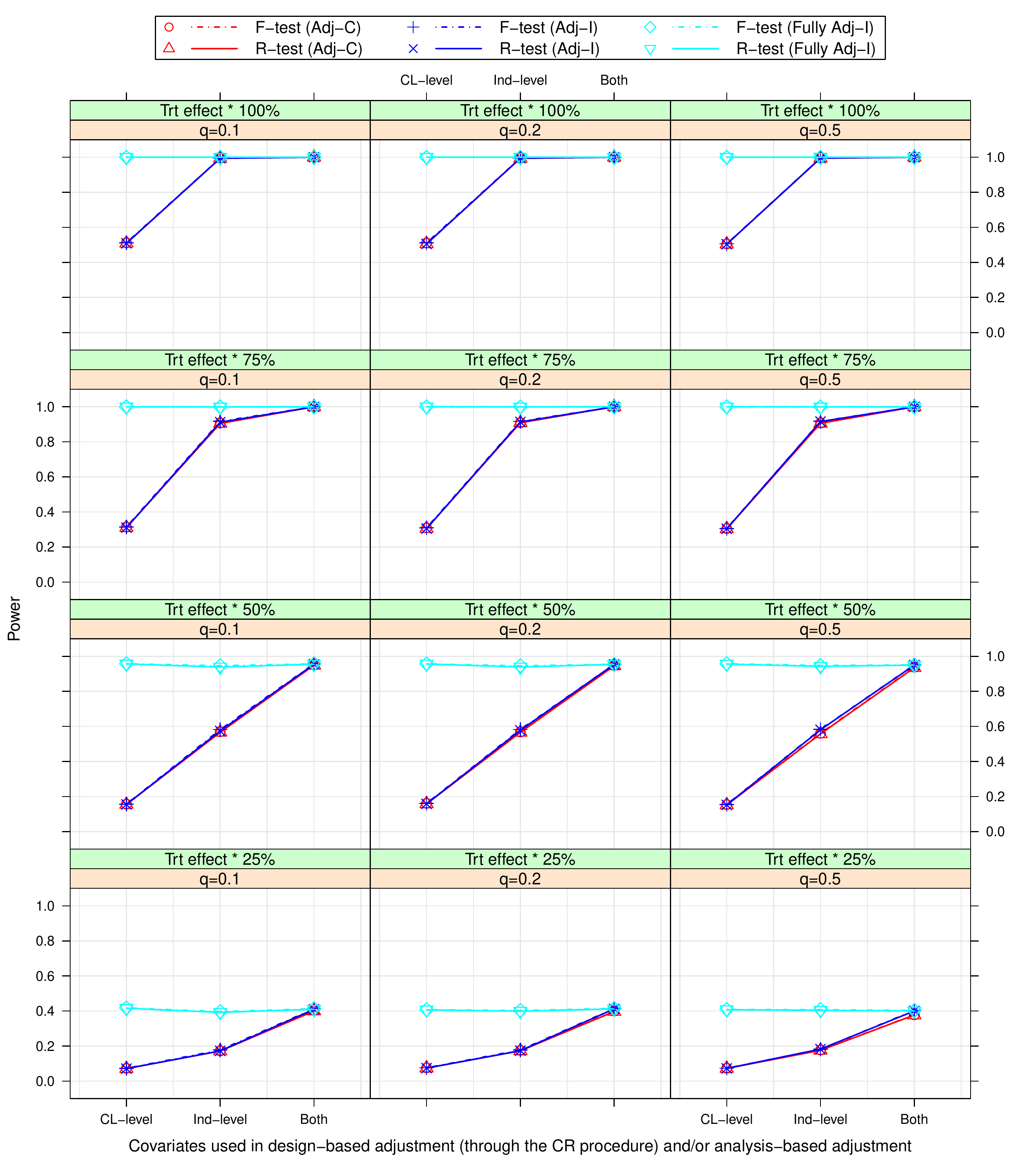}
	\end{center}
	\caption{\label{global_power_re_fig} Results under reduced effect sizes (75\%, 50\%, and 25\% of the original magnitude): power for the global hypothesis ($\mathcal{H}_{0}\text{: }\delta_{1}=\delta_{2}=0$) under constrained randomization (CR) with $B_{(M)}$ balance metric and $q$ = 0.1, 0.2, and 0.5. CR implemented using covariates indicated on the horizontal axis; ICC $=0.05$; alpha level $=5\%$; R-test: randomization test; CL-level: cluster-level covariates, $\boldsymbol{x}_j$; Ind-level: individual-level covariates, $\boldsymbol{z}_{jk}$; Unadj: unadjusted test; Adj-C: test adjusted for the covariates on the horizontal axis (with individual-level covariates aggregated at the cluster level); Adj-I: test adjusted for the covariates on the horizontal axis (with actual individual-level covariates); Fully Adj-I: test adjusted for all four covariates (with actual individual-level covariates).
	}
\end{figure*}

\textcolor{black}{We presented the simulation results with $g=3, 5, 10$ in the main text. It was shown that the adjusted model-based tests and the randomization-based tests (UMPR/LMPR tests) provided a similar level of power when the number of clusters per arm is large. However, with $g=10$, the power for both types of tests reaches the maximum ($100\%$ power), making the comparison uninformative. Therefore, we presented additional simulation results for power with effect sizes reduced to 75\%, 50\%, and 25\% of the original magnitude to better compare the power of the UMPR/LMPR tests with that of the model-based tests. We showed the power for the global hypothesis in Web Figure~\ref{global_power_re_fig}. Results for the pairwise hypotheses are qualitatively similar, thus omitted for presentation. These results demonstrate that the randomization test is asymptotically equivalent to the model-based test in terms of statistical power.}

\clearpage 

\section*{Appendix H: Additional simulation results under non-normal data generation processes}
\begin{figure*}[htbp]
	\begin{center}
		\includegraphics[trim=5 15 5 10, clip, width=0.85\linewidth]{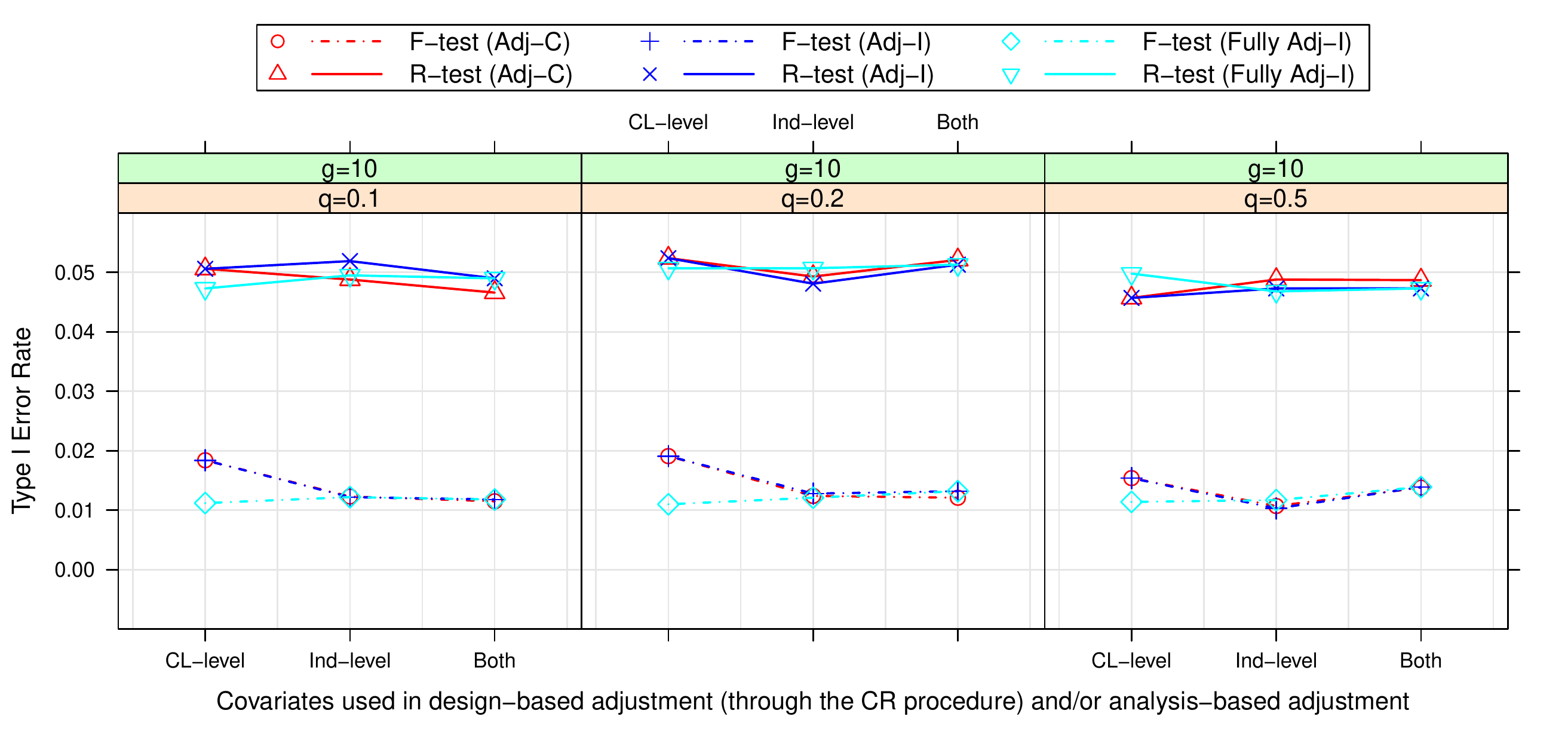}
	\end{center}
	\caption{\label{global_alpha_cauchy_error_fig} Results under Cauchy residual: type I error rate for the global hypothesis ($\mathcal{H}_{0}\text{: }\delta_{1}=\delta_{2}=0$) under constrained randomization (CR) with $B_{(M)}$ balance metric and $q$ = 0.1, 0.2, and 0.5. CR implemented using covariates indicated on the horizontal axis; alpha level $=5\%$; R-test: randomization test; R-test: randomization test; CL-level: cluster-level covariates, $\boldsymbol{x}_j$; Ind-level: individual-level covariates, $\boldsymbol{z}_{jk}$; Unadj: unadjusted test; Adj-C: test adjusted for the covariates on the horizontal axis (with individual-level covariates aggregated at the cluster level); Adj-I: test adjusted for the covariates on the horizontal axis (with actual individual-level covariates); Fully Adj-I: test adjusted for all four covariates (with actual individual-level covariates).
	}
	
\end{figure*}

\begin{figure*}[htbp]
	\begin{center}
		\includegraphics[trim=5 15 5 10, clip, width=0.85\linewidth]{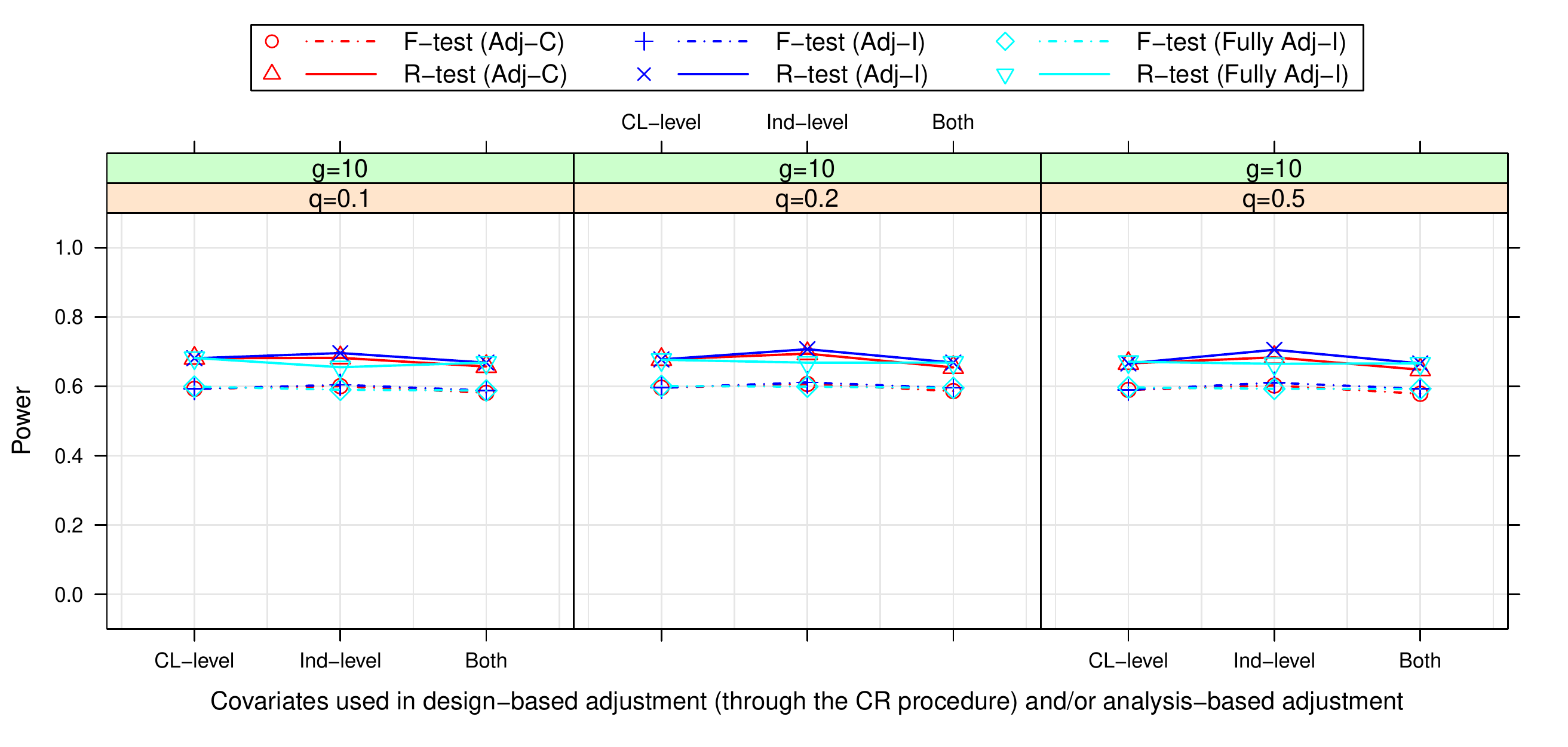}
	\end{center}
	\caption{\label{global_power_cauchy_error_fig} Results under Cauchy residual: power for the global hypothesis ($\mathcal{H}_{0}\text{: }\delta_{1}=\delta_{2}=0$) under constrained randomization (CR) with $B_{(M)}$ balance metric and $q$ = 0.1, 0.2, and 0.5. CR implemented using covariates indicated on the horizontal axis; alpha level $=5\%$; R-test: randomization test; CL-level: cluster-level covariates, $\boldsymbol{x}_j$; Ind-level: individual-level covariates, $\boldsymbol{z}_{jk}$; Unadj: unadjusted test; Adj-C: test adjusted for the covariates on the horizontal axis (with individual-level covariates aggregated at the cluster level); Adj-I: test adjusted for the covariates on the horizontal axis (with actual individual-level covariates); Fully Adj-I: test adjusted for all four covariates (with actual individual-level covariates).
	}
	
\end{figure*}

\begin{figure*}[htbp]
	\begin{center}
		\includegraphics[trim=5 15 5 10, clip, width=0.85\linewidth]{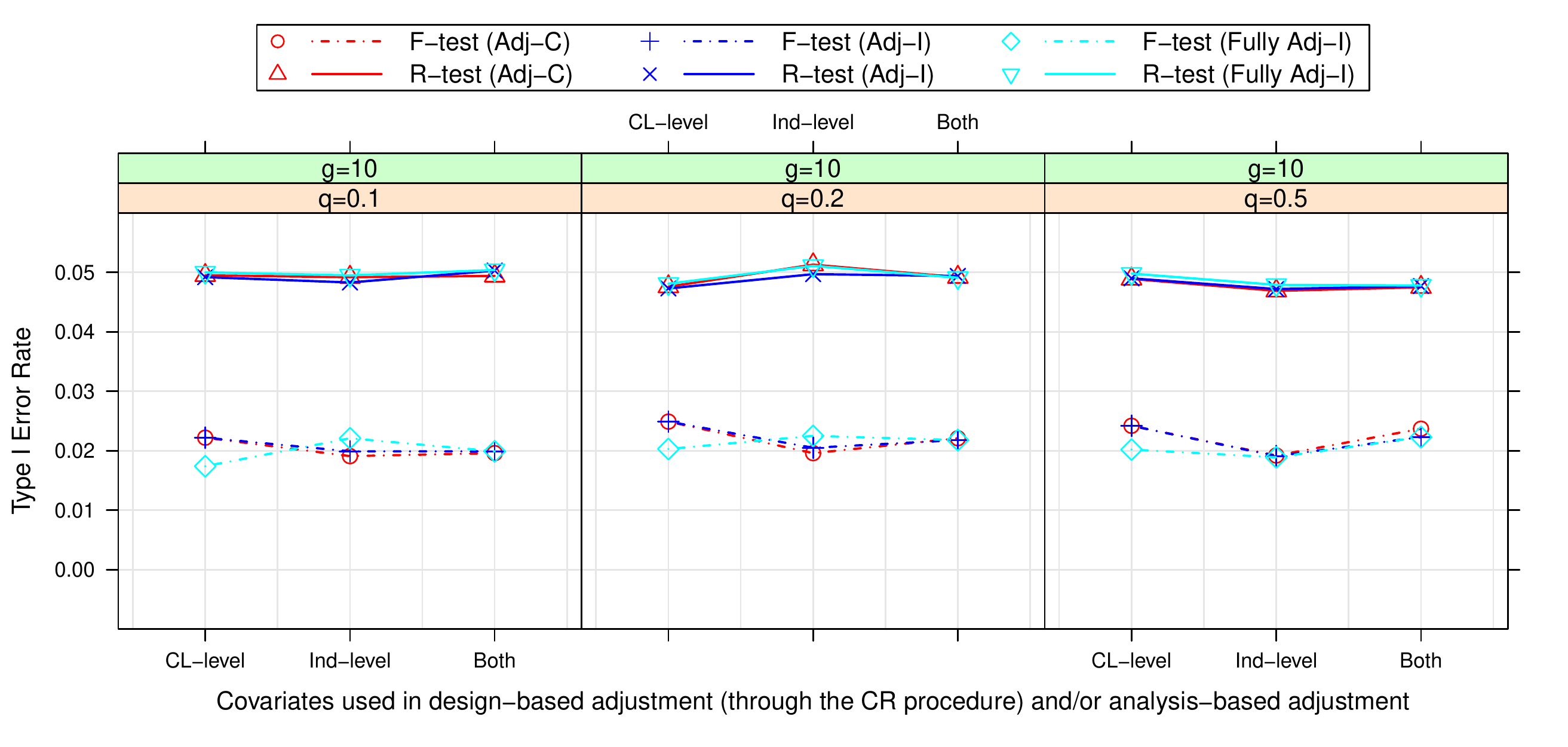}
	\end{center}
	\caption{\label{indi_alpha_cauchy_error_fig} Results under Cauchy residual: type I error rate for the pairwise hypothesis ($\mathcal{H}_{0}\text{: }\delta_{1}=0$) under constrained randomization (CR) with $B_{(M)}$ balance metric and $q$ = 0.1, 0.2, and 0.5. CR implemented using covariates indicated on the horizontal axis; alpha level $=5\%$; R-test: randomization test; CL-level: cluster-level covariates, $\boldsymbol{x}_j$; Ind-level: individual-level covariates, $\boldsymbol{z}_{jk}$; Unadj: unadjusted test; Adj-C: test adjusted for the covariates on the horizontal axis (with individual-level covariates aggregated at the cluster level); Adj-I: test adjusted for the covariates on the horizontal axis (with actual individual-level covariates); Fully Adj-I: test adjusted for all four covariates (with actual individual-level covariates).
	}
	
\end{figure*}

\begin{figure*}[htbp]
	\begin{center}
		\includegraphics[trim=5 15 5 10, clip, width=0.85\linewidth]{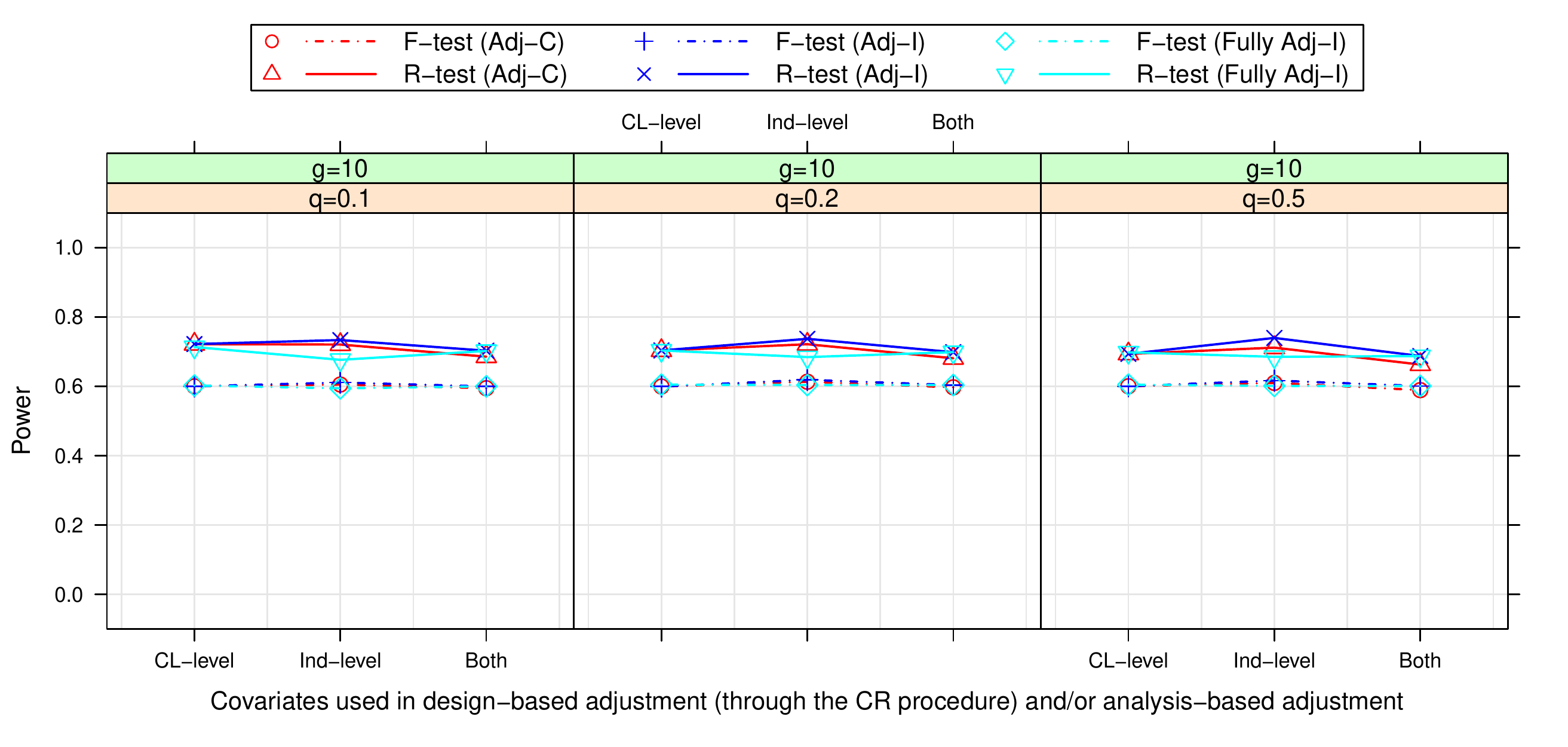}
	\end{center}
	\caption{\label{indi_power_cauchy_error_fig} Results under Cauchy residual: power for the pairwise hypothesis ($\mathcal{H}_{0}\text{: }\delta_{1}=0$) under constrained randomization (CR) with $B_{(M)}$ balance metric and $q$ = 0.1, 0.2, and 0.5. CR implemented using covariates indicated on the horizontal axis; alpha level $=5\%$; R-test: randomization test; CL-level: cluster-level covariates, $\boldsymbol{x}_j$; Ind-level: individual-level covariates, $\boldsymbol{z}_{jk}$; Unadj: unadjusted test; Adj-C: test adjusted for the covariates on the horizontal axis (with individual-level covariates aggregated at the cluster level); Adj-I: test adjusted for the covariates on the horizontal axis (with actual individual-level covariates); Fully Adj-I: test adjusted for all four covariates (with actual individual-level covariates).
	}
	
\end{figure*}

\begin{figure*}[htbp]
	\begin{center}
		\includegraphics[trim=5 15 5 10, clip, width=0.85\linewidth]{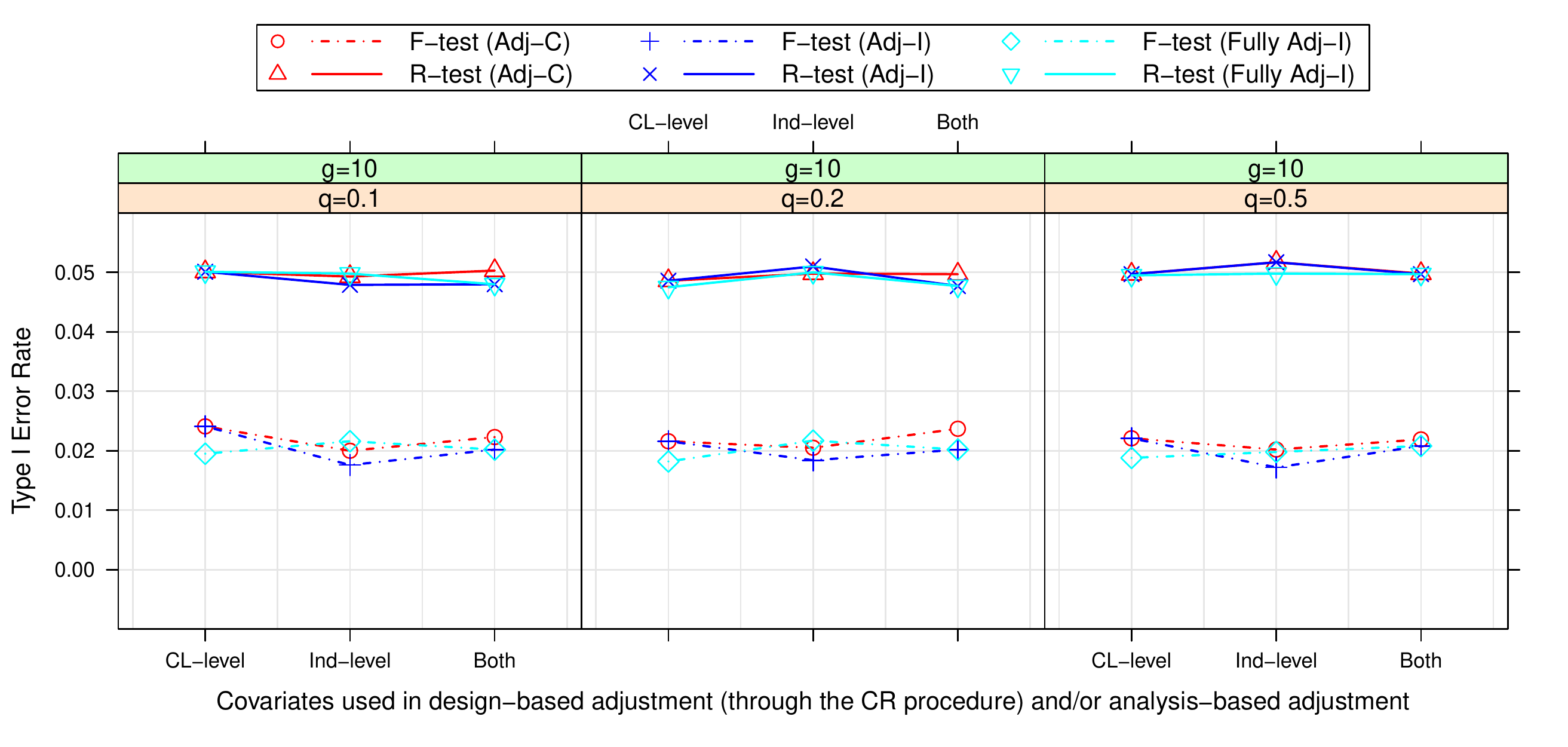}
	\end{center}
	\caption{\label{global_alpha_cauchy_cl_fig} Results under Cauchy random cluster effect: type I error rate for the global hypothesis ($\mathcal{H}_{0}\text{: }\delta_{1}=\delta_{2}=0$) under constrained randomization (CR) with $B_{(M)}$ balance metric and $q$ = 0.1, 0.2, and 0.5. CR implemented using covariates indicated on the horizontal axis; alpha level $=5\%$; R-test: randomization test; CL-level: cluster-level covariates, $\boldsymbol{x}_j$; Ind-level: individual-level covariates, $\boldsymbol{z}_{jk}$; Unadj: unadjusted test; Adj-C: test adjusted for the covariates on the horizontal axis (with individual-level covariates aggregated at the cluster level); Adj-I: test adjusted for the covariates on the horizontal axis (with actual individual-level covariates); Fully Adj-I: test adjusted for all four covariates (with actual individual-level covariates).
	}
	
\end{figure*}

\begin{figure*}[htbp]
	\begin{center}
		\includegraphics[trim=5 15 5 10, clip, width=0.85\linewidth]{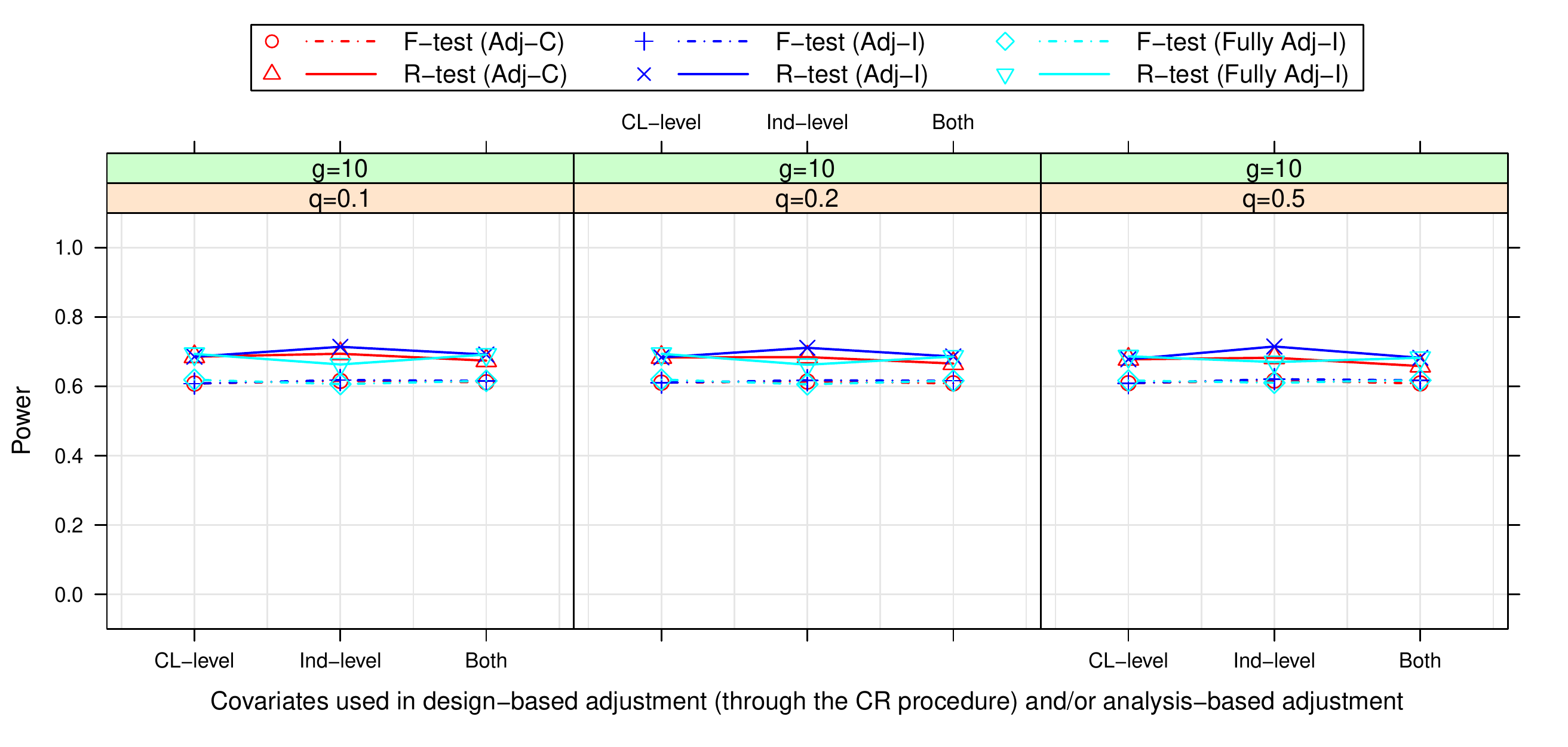}
	\end{center}
	\caption{\label{global_power_cauchy_cl_fig} Results under Cauchy random cluster effect: power for the global hypothesis ($\mathcal{H}_{0}\text{: }\delta_{1}=\delta_{2}=0$) under constrained randomization (CR) with $B_{(M)}$ balance metric and $q$ = 0.1, 0.2, and 0.5. CR implemented using covariates indicated on the horizontal axis; alpha level $=5\%$; R-test: randomization test; CL-level: cluster-level covariates, $\boldsymbol{x}_j$; Ind-level: individual-level covariates, $\boldsymbol{z}_{jk}$; Unadj: unadjusted test; Adj-C: test adjusted for the covariates on the horizontal axis (with individual-level covariates aggregated at the cluster level); Adj-I: test adjusted for the covariates on the horizontal axis (with actual individual-level covariates); Fully Adj-I: test adjusted for all four covariates (with actual individual-level covariates).
	}
	
\end{figure*}

\begin{figure*}[htbp]
	\begin{center}
		\includegraphics[trim=5 15 5 10, clip, width=0.85\linewidth]{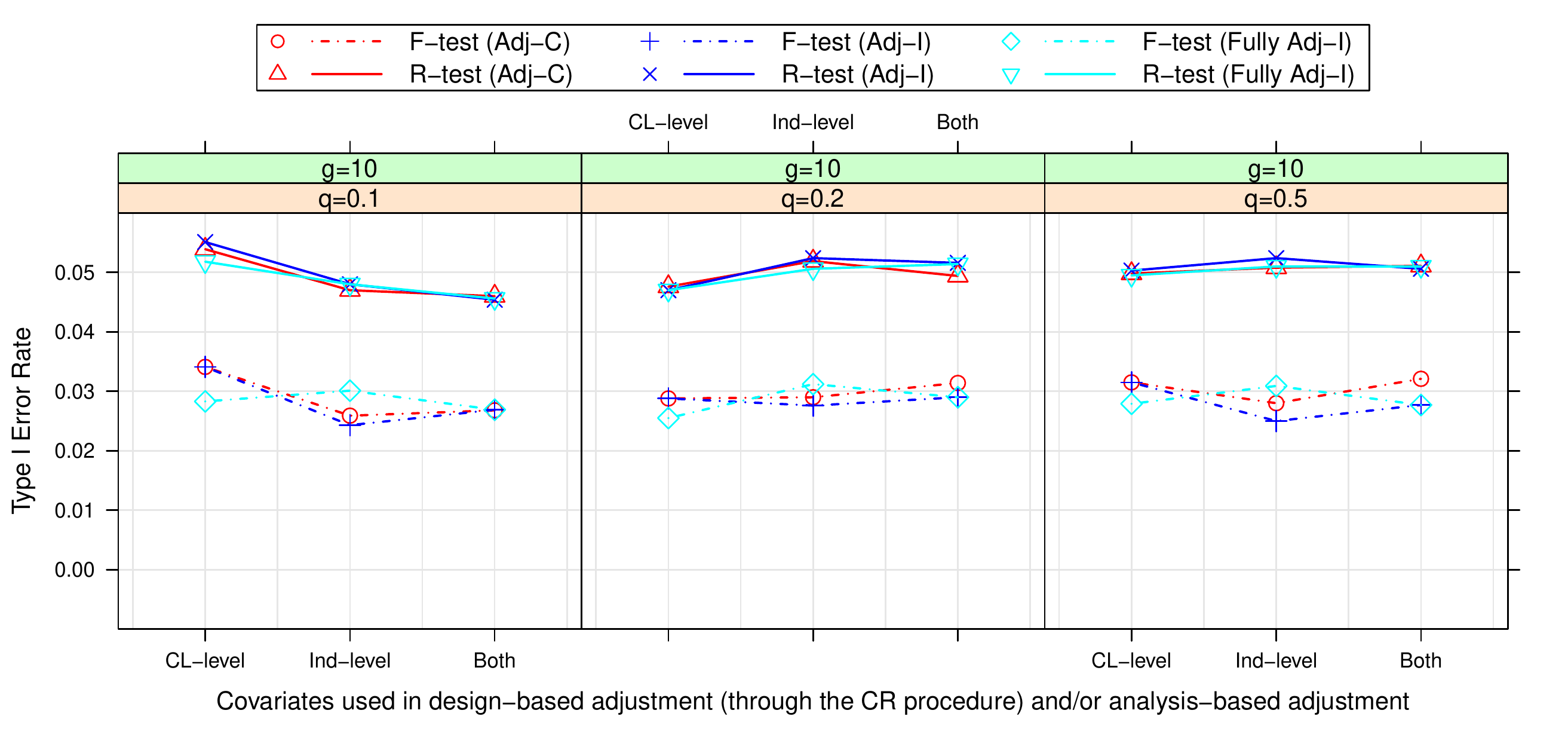}
	\end{center}
	\caption{\label{indi_alpha_cauchy_cl_fig} Results under Cauchy random cluster effect: type I error rate for the pairwise hypothesis ($\mathcal{H}_{0}\text{: }\delta_{1}=0$) under constrained randomization (CR) with $B_{(M)}$ balance metric and $q$ = 0.1, 0.2, and 0.5. CR implemented using covariates indicated on the horizontal axis; alpha level $=5\%$; R-test: randomization test; CL-level: cluster-level covariates, $\boldsymbol{x}_j$; Ind-level: individual-level covariates, $\boldsymbol{z}_{jk}$; Unadj: unadjusted test; Adj-C: test adjusted for the covariates on the horizontal axis (with individual-level covariates aggregated at the cluster level); Adj-I: test adjusted for the covariates on the horizontal axis (with actual individual-level covariates); Fully Adj-I: test adjusted for all four covariates (with actual individual-level covariates).
	}
	
\end{figure*}

\begin{figure*}[htbp]
	\begin{center}
		\includegraphics[trim=5 15 5 10, clip, width=0.85\linewidth]{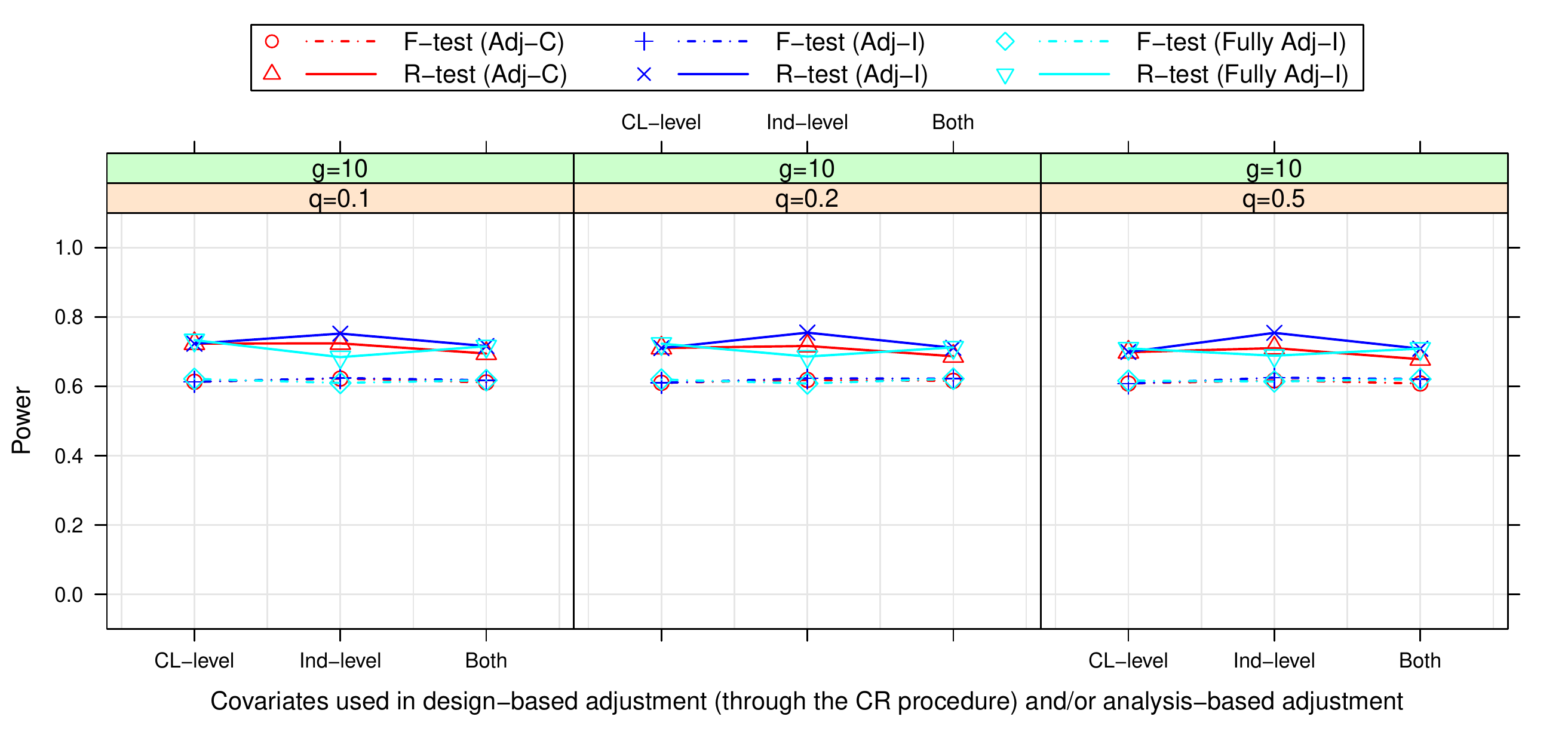}
	\end{center}
	\caption{\label{indi_power_cauchy_cl_fig} Results under Cauchy random cluster effect: power for the pairwise hypothesis ($\mathcal{H}_{0}\text{: }\delta_{1}=0$) under constrained randomization (CR) with $B_{(M)}$ balance metric and $q$ = 0.1, 0.2, and 0.5. CR implemented using covariates indicated on the horizontal axis; alpha level $=5\%$; R-test: randomization test; CL-level: cluster-level covariates, $\boldsymbol{x}_j$; Ind-level: individual-level covariates, $\boldsymbol{z}_{jk}$; Unadj: unadjusted test; Adj-C: test adjusted for the covariates on the horizontal axis (with individual-level covariates aggregated at the cluster level); Adj-I: test adjusted for the covariates on the horizontal axis (with actual individual-level covariates); Fully Adj-I: test adjusted for all four covariates (with actual individual-level covariates).
	}
	
\end{figure*}

\textcolor{black}{To evaluate the robustness of the analytical methods under comparison against violations of distributional assumptions, we generated non-normal data in a similar way to Small et al.\cite{small2008randomization}, specifying the random cluster effect $\gamma_j$ and error term $\epsilon_{jk}$ to follow the standard Cauchy distribution, respectively. Note that the standard Cauchy distribution has a thicker tail than the normal distributions. Treatment effects were set to $8$ for both treatment arms to evaluate power. Number of clusters per arm $g$ was held at 10. Results of type I error rates and power under Cauchy residual and random cluster effect are presented in Web Figures~\ref{global_alpha_cauchy_error_fig}-\ref{indi_power_cauchy_cl_fig}. In terms of type I error rates, the model-based tests are overly conservative in either scenario, while the randomization test (both UMPR and LMPR, depending on the hypothesis of interest) maintains the nominal test size. For power, both tests are negatively affected by the violations of normality assumptions. However, the randomization test can be more powerful than the model-based test. In summary, the randomization test is a more flexible and robust alternative to the model-based test when distributional assumptions are not guaranteed to hold.}

\clearpage 

\section*{References}

\begingroup
\renewcommand{\section}[2]{} 

\endgroup

\end{document}